\documentclass[sigconf]{acmart}
\usepackage{seqsplit}
\usepackage{booktabs} 
\usepackage{tikz}
\usepackage{booktabs} 
\usepackage{float}
\usepackage{multirow}
\usepackage{lipsum}
\usepackage{graphicx}
\usepackage{csquotes}
\usepackage{array}
\usepackage{xcolor}
\usepackage{soul}
\usepackage{multicol}
\usepackage{balance}

\usepackage{acmart-taps}

\usepackage[final]{pdfpages}
\newcolumntype{P}[1]{>{\centering\arraybackslash}p{#1}}
\newcolumntype{M}[1]{>{\centering\arraybackslash}m{#1}}
\usepackage{blindtext}
\usepackage[utf8]{inputenc}
\usepackage{subcaption}
\usepackage{xfrac}
\usepackage{multicol}
\newcommand{\ignore}[1]{}

\newcommand{\rev}[1]{\textcolor{black}{#1}}
\newcolumntype{P}[1]{>{\centering\arraybackslash}p{#1}}
\newcolumntype{M}[1]{>{\centering\arraybackslash}m{#1}}
\AtBeginDocument{%
  }

\copyrightyear{2026}
\acmYear{2026}
\setcopyright{cc}
\setcctype{by}
\acmConference[IMX '26]{ACM International Conference on Interactive Media Experiences}{June 09--11, 2026}{Athlone, Ireland}
\acmBooktitle{ACM International Conference on Interactive Media Experiences (IMX '26), June 09--11, 2026, Athlone, Ireland}
\acmDOI{10.1145/3788851.3805015}
\acmISBN{979-8-4007-2448-0/2026/06}

\begin{document}

\title[Human vs. AI Support in VRChat Discord: Engagement and Interaction Patterns]{Comparative Analysis of Human vs. AI-powered Support in VRChat Communities on Discord: User Engagement, Response Dynamics and Interaction Patterns}

%
\author{He Zhang}
\email{hpz5211@psu.edu}
\orcid{0000-0002-8169-1653}
\affiliation{%
  \institution{College of Information Sciences and Technology\\The Pennsylvania State University}
  \city{University Park}
  \state{Pennsylvania}
  \country{USA}
  \postcode{16802}
}
\author{Bumjin Kim}
\email{bqk5313@psu.edu}
\orcid{0009-0000-6568-6355}
\affiliation{%
  \institution{College of Information Sciences and Technology\\The Pennsylvania State University}
  \city{University Park}
  \state{Pennsylvania}
  \country{USA}
  \postcode{16802}
}

\author{John M. Carroll}
\orcid{0000-0001-5189-337X}
\email{jmc56@psu.edu}
\affiliation{%
  \institution{College of Information Sciences and Technology\\The Pennsylvania State University}
  \city{University Park}
  \state{Pennsylvania}
  \country{USA}
  \postcode{16802}
}
\author{Jie Cai}
\authornote{Corresponding author.}
\orcid{0000-0002-0582-555X}
\email{jie-cai@mail.tsinghua.edu.cn}
\affiliation{%
  \institution{Department of Computer Science and Technology\\Tsinghua University}
  \city{Beijing}
  \country{China}
  \postcode{100190}
}


\begin{abstract}
  The integration of AI-driven support systems within online communities has opened new avenues for enhancing user engagement and support efficiency in recent years. This study investigates the differences in user interactions and engagement within two distinct support channels on the VRChat Discord server: ``user support,'' where human users provide assistance to peers, and ``AI support,'' where an AI chatbot addresses user queries. By analyzing user engagement, response dynamics, and interaction patterns across these channels, we uncover different usage patterns and user attitudes toward each approach. Our research employs both quantitative and qualitative methods to explore the trends in the VRChat community when using AI and user support, highlighting the unique advantages and limitations of AI-driven support compared to traditional human assistance. The findings offer valuable insights into optimizing AI and human support systems, aiming to foster more effective support strategies and create more engaging online communities.
\end{abstract}



\begin{teaserfigure}
  \includegraphics[width=1\columnwidth]{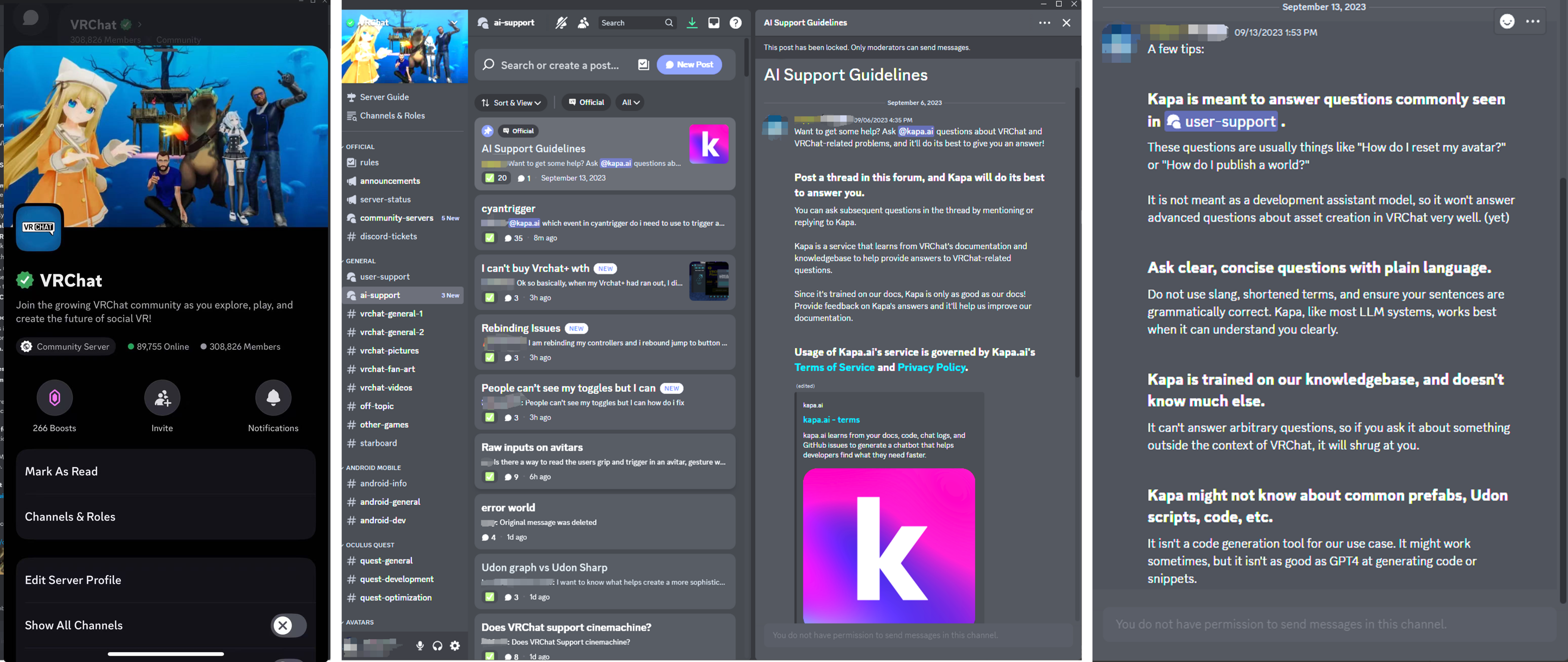}
  \caption{\textbf{Example of Discord VRChat Channel and AI-support Sub-channel. It shows the structure and layout of the support channels, highlighting the rules for using the VRChat Discord server and several tips about Kapa, the AI trained to answer questions in the AI support channel.}}
  \label{fig.teaserfigure}
\end{teaserfigure}


\maketitle

\section{Introduction}
In recent years, online communities have become essential spaces for social interaction, writing, and knowledge sharing~\cite{haythornthwaite2007social,faraj2015leading}. Within these communities, platforms play a vital role, with Discord~\footnote{https://discord.com} being one of the most popular, particularly among gaming and technology enthusiasts. Originally designed as a voice and text chat application for gamers, Discord has evolved into a multifunctional platform that supports a wide range of online communities, allowing like-minded users to create and join servers dedicated to specific topics.

Support channels within these communities are crucial components, serving as vital spaces where members can seek help, share expertise, enhance social connections, and contribute to community growth~\cite{cai2026twitchthirdpartydeveloperssupport}. The importance of these channels lies in their ability to facilitate collaboration and mutual assistance among community members, providing an environment for communication and problem-solving. As the significance of these channels has grown, the methods of providing support have diversified. Beyond traditional human assistance, which includes both official support and peer-to-peer help, artificial intelligence (AI) powered systems such as chatbots have also been introduced into these communities. With the development of large language models (LLMs) and Artificial General Intelligence (AGI), these chatbots can more effectively understand user queries, moving beyond predefined templates to offer more efficient solutions in support channels~\cite{10.1145/3449161}.

As Discord has expanded, it has attracted a large number of user communities centered around specific applications. The VRChat Discord server, created by the official VRChat operations team, boasts a large and diverse user base, with over 308,826 members as of January 2025~\footnote{\url{https://wiki.vrchat.com/wiki/Discord_servers}}. Its user-friendly design enables seamless engagement with the platform, emphasizing how real-time interactions shape the nature of support. As one of the most active servers on the Discord platform, VRChat offers a wealth of data for studying interaction patterns and support mechanisms. Moreover, VRChat~\footnote{https://hello.vrchat.com/}, as a virtual reality (VR) social platform, serves as a quintessential case for examining the effectiveness of support within online communities. Support needs in the VRChat ecosystem are particularly complex, as they often span multiple layers of practice and expertise, including hardware configuration, software setup, avatar creation, platform-specific troubleshooting, and community-specific norms and etiquette~\cite{chen2026understandingnewcomerpersistencesocial,10.1145/3706598.3713561}. Users may need help not only with technical issues such as VR headset compatibility, performance optimization, or account settings, but also with socially situated practices such as understanding behavioral expectations, navigating identity expression through avatars, and participating appropriately in community spaces~\cite{10.1145/3706599.3720120}. The VRChat Discord server provides a unique research environment, where both human and AI support systems coexist, offering similar levels of access to members. This complexity makes community-based peer support especially valuable in areas where formal documentation, static FAQs, or conventional customer service may be incomplete, difficult to contextualize, or too slow to respond to evolving user needs. In such settings, support is not merely about solving isolated technical problems, but also about translating community knowledge, sharing experiential strategies, and helping newcomers integrate into a highly social and technically layered environment~\cite{10.1145/3715336.3735825}. Within this setting, however, the AI system studied here, ``\textit{Kapa.ai}'', should be understood primarily as a functional, text-based support tool rather than as a full social participant. In the VRChat Discord server, its role is centered on answering user questions in a dedicated channel, providing relatively direct and utility-oriented responses. Unlike human supporters, who may contribute through broader conversational, relational, and community-building practices, Kapa.ai primarily operates as a question-answering agent. In addition, in the setting we examined, it did not meaningfully process media-based inputs such as images or other rich contextual artifacts, which further constrained the kinds of support it could provide. This setting presents a valuable opportunity to explore how human and AI support methods influence user engagement, satisfaction, and interaction patterns.

The primary objective of this study is to understand the dynamic relationship between AI and human support within the VRChat Discord server, with a particular focus on how AI systems represented by chatbots influence user engagement, the facilitation of community discussion, the performance of social roles, and response dynamics and interaction patterns. By analyzing the ``user-support'' and ``ai-support~\footnote{https://www.kapa.ai/content/terms-of-service}'' channels, we aim to answer the following key questions: 

\begin{itemize}

\item[\textbf{RQ1.}] How do the AI-support and User-support sub-channels differ in activity patterns, user engagement, and forms of participation?
\item[\textbf{RQ2.}] How do these two support channels contribute differently to collaborative support, including problem-solving, knowledge sharing, and the coordination of community participation?

\end{itemize}

Specifically, this research seeks to identify the unique strengths and limitations of each support method, as well as how functional AI assistance and socially embedded peer support may complement one another, providing deeper insights into how they can be optimized to enhance the overall support experience and foster more engaging and effective online communities.

\section{Related Work}

\subsection{VRChat Communities}
VR has triggered paradigm shifts across various fields, offering users an immersive interactive experience~\cite{zhang2024exploring}. In particular, VRChat, as a leading social virtual reality platform, has gained significant attention in recent years due to its unique integration of virtual reality technology and social networking features.~\cite{9417636,10.1145/3411764.3445426,wang2020social,10.1145/3492836}. This platform offers users the opportunity to create and inhabit avatar-based worlds, fostering a diverse and dynamic community~\cite{10.1145/3432938,dwivedi2022metaverse}. 

\subsubsection{Social Interactions in VR Environments}

Social interactions in VR environments like VRChat differ significantly from traditional online platforms~\cite{10.1145/3744736.3749361}. Freeman and Maloney~\cite{10.1145/3432938} highlight the importance of embodied presence in VR social interactions, noting that users experience a heightened sense of ``being there'' with others. Tanenbaum et al.~\cite{10.1145/3313831.3376606} explored how embodied presence in VR facilitates more nuanced communication, including non-verbal cues absent in text-based interactions. Their study revealed that avatar customization plays a crucial role in self-expression and social dynamics, which is consistent with the Embodied Social Presence Theory proposed by Mennecke et al~\cite{5428431}. McVeigh-Schultz et al.~\cite{10.1145/3197391.3205451} examined the unique affordances of social VR platforms, highlighting how VRChat's customizable environments contribute to users' sense of identity and belonging. Researchers found that the ability to create and share virtual spaces fostered stronger community ties~\cite{ABRAMCZUK2023103104,ABURUMMAN2022102819}. Moreover, the immersive nature of VR can lead to more intense social experiences. Research by Smith and Neff~\cite{10.1145/3173574.3173863} suggest that social interactions in VR can have psychological effects similar to those of face-to-face interactions, potentially leading to stronger emotional connections between users.

\subsubsection{Social Interactions Outside the VR Environment and Needs of VRChat Communities}

The distinct nature of the VRChat community gives rise to specific support needs, such as avatar creation, which serves as a key element of social interaction in the virtual world~\cite{5428431}. This often involves issues related to the integration of VR hardware and software, as well as design-related challenges~\cite{10.1145/3411764.3445335}. Additionally, users often extend their social interactions to other platforms to maintain connections outside the VR environment~\cite{10.1145/3579529}. Social interactions outside VR help provide continuity for the VRChat community. For example, platforms like Discord and Reddit are commonly used by VRChat users to communicate, share content, organize events, and support each other~\cite{10.1145/3411764.3445335}. This cross-platform socialization addresses limitations within VRChat, such as the need for asynchronous communication and accessibility for users who may not be in VR at all times~\cite{rzeszewski2024social}. External channels support the VRChat community by offering spaces for collaboration, troubleshooting, and fostering a sense of belonging~\cite{10.1145/3313831.3376606,10.1145/3706599.3720120}.

\subsection{Online Communities and Support Systems}
Online communities have evolved into essential support networks across various domains. Faraj et al.\cite{faraj2015leading} define online communities as new forms of organizing that facilitate knowledge exchange, problem-solving, and innovation. These communities enable users to request and provide assistance more effectively and responsively than traditional customer support systems\cite{10.1145/3313831.3376606}.

On the one hand, for developers, online communities serve as collaborative spaces where they can ask questions, share expertise, and receive feedback~\cite{10.1145/3651990}. Active engagement in such communities contributes to collective learning and the advancement of projects. Player-developer-official interactions can encourage developers to address issues promptly and align products with user needs~\cite{10.1145/3613904.3642787}. On the other hand, for players, online communities offer platforms for sharing strategies, tips, and experiences, fostering social bonds that can develop into lasting friendships~\cite{10.1145/985921.986080}. Engaging in these communities enhances the overall experience and contributes to a strong sense of community.

Unlike customer service, where limited experts control information, online communities distribute knowledge among a large base, enabling faster responses and diverse problem-solving approaches~\cite{faraj2015leading}. This democratization of knowledge empowers users and enhances the overall effectiveness of support systems.

\subsubsection{User Engagement and Interaction Patterns in Support Channels}

Research on human-AI collaboration patterns in online community support systems has gained significant attention in recent years~\cite{MALINEN2015228}. Studies have demonstrated that user engagement in online support channels exhibits multidimensional features, including problem-solving or supporting design orientation~\cite{BARCELLINI2008558}, community interaction~\cite{TSAI2013475,10.1145/3392863}, and knowledge sharing~\cite{faraj2011knowledge,HSU2007153}.

Early research primarily focused on user engagement patterns in traditional online support forums, offering insights such as the identification of core user groups through social network analysis~\cite{10.1145/3068777.3068781}, tracking the user behavior in online communities~\cite{10.1145/503376.503446}. Users' relationship often play a pivotal role in knowledge dissemination and problem resolution~\cite{10.1145/2187836.2187907}. Large-scale data analyses have highlighted that high-quality interactions are frequently linked to responsive mechanisms and detailed solutions~\cite{GHANI2019417}.

On platforms like Discord, user engagement in support channels exhibits unique characteristics~\cite{10.1145/3274319}. Interactions in these channels are not limited to technical problem-solving but also encompass rich elements of community building. Users foster a collaborative and trusting community atmosphere by sharing experiences, offering advice, and providing emotional support. Specific interaction paradigms emerge, including problem-oriented, experience-sharing, and collaborative problem-solving interactions~\cite{10.1145/3613904.3642787}.

Recent studies further reveal that user engagement in Discord support channels is influenced by multiple factors, such as response time~\cite{10.1145/3478431.3499385}, solution quality~\cite{10.1145/3359157}, and the community atmosphere~\cite{10.1145/3610053,10.1145/3610191}. Particularly in professional technical communities, the accuracy of knowledge and the feasibility of solutions are critical factors in maintaining user activity and engagement~\cite{10.1145/3613904.3642787}.

However, there are limitations in current research, especially in the comparative analysis of user interaction patterns across different types of support channels. By systematically analyzing the interaction patterns and user engagement characteristics of AI-supported and human-supported channels on Discord, this study aims to fill the research gaps. It seeks to provide new theoretical insights into online support systems in technology-mediated environments and offer valuable references for system optimization in practice.

\subsubsection{AI Bots in Online Communities}
With the advancement of AI technologies, particularly the widespread application of AI chatbots on social platforms, user support has shown new characteristics and trends~\cite{10.1145/3579505,10.1145/3706598.3713503,kim2025enhancing}. The debate over human versus AI support has intensified with advancements in artificial intelligence technologies. Human support is crucial when complex problem-solving or emotional intelligence is required~\cite{10.1145/3579505}. Human agents provide personalized, empathetic interactions that can effectively address users' frustrations and unique circumstances. AI support, on the other hand, offers rapid responses to routine queries, operates continuously without time constraints, and can learn from user interactions to improve over time~\cite{10.1145/2858036.2858288}. AI assistants can efficiently handle high volumes of straightforward inquiries, freeing human agents to focus on more complex tasks~\cite{zhang2024redefiningqualitativeanalysisai}. However, challenges with AI support include limitations in understanding nuanced human language, managing ambiguous queries, and providing emotionally sensitive responses~\cite{cui2017superagent}. Users may experience frustration if AI assistants fail to comprehend and address their issues~\cite{ZHANG2025100144} or lack the empathy provided by human support. Research suggests that a hybrid approach, integrating AI efficiency with human empathy, may offer the most effective support systems~\cite{hill2015real}. Combining the strengths of both can enhance user satisfaction and improve the overall support experience.

Currently, AI chatbots are widely applied on social platforms~\cite{ferrara2023social,RAPP2021102630} and have brought unique value to online communities~\cite{zhu2022me,kempt2020chatbots,davenport2018ai,10.1145/3290605.3300680}. During interactions with AI chatbots, users emphasize the bots' problem-solving abilities and express concerns about safety~\cite{10.1145/3593013.3594013,SEITZ2022102848,10.1145/3610191,10.1145/3686941}, empathy~\cite{blut2021understanding,10.1145/3637410}, and interaction patterns~\cite{mouronte2024patterns,MALLICK2024103355,10.1145/3678884.3681875}. Understanding the user experience with AI chatbots in online communities has become particularly important~\cite{HOORN2024103142}, as it can help bridge gaps in specific skills and knowledge~\cite{HOBERT2023103108,10.1145/3449083}, improve technical frameworks~\cite{JANSSEN2022102921}, and better support online communities~\cite{mouronte2024patterns}.

\section{\rev{Methods}}

\subsection{Data \rev{Collection,} Exploration and Pre-processing}

We collected data from publicly accessible VRChat Discord servers, specifically focusing on two sub-channels highly relevant to our research topic: ``user-support'' and ``ai-support''. The data collection period spans from August 15, 2023, to June 13, 2024 (the start date corresponds to the creation of the ``user support'' sub-channel on August 15, 2023, while the ``ai-support'' sub-channel was created on September 12, 2023), encompassing all posts and their respective replies within these sub-channels. The collected data includes the following fields: AuthorID, Author, Date (timestamp), Content (detailed post content), Attachments (file or image paths attached to replies), and Reactions (user responses to the content).

To maintain context and facilitate analysis, we implemented an organizational structure where we used a [post title] tag (``SourceFile'') to group all content from the same thread, indicating whether multiple Content'' entries belong to the same post. Entries were annotated in chronological order within each thread, ensuring that thread-level context was retained and that initial questions could be clearly distinguished from subsequent replies.

Our dataset comprises 81,784 entries from the ``user-support'' sub-channel [8,482 $(\approx 10.37\%)$ initial questions and 73,302 $(\approx 89.63\%)$ related replies\ and 28,261 entries from the ``ai-support'' sub-channel [2,905 $(\approx 10.28\%)$ initial questions and 25,356 $(\approx 89.72\%)$ related replies]. Notably, both sub-channels demonstrate a similar composition, with approximately 10\% of entries being initial questions and 90\% being follow-up replies. This distribution highlights the collaborative nature of these support channels, where a single question often generates multiple responses and discussions. %

\subsubsection{\rev{\textbf{Ethical Considerations for Data Collection and Use}}}
\rev{This study analyzes content from publicly accessible channels on the official VRChat Discord server, and it involves secondary analysis of publicly available data without any interaction or intervention with users. We restricted collection to channels visible to any server member, did not circumvent access controls, respected rate limits, and complied with Discord’s Terms of Service and community guidelines. All user handles were de-identified using a deterministic mapping function that replaces each username with an anonymized identifier, and the identifier of the system bot (kappa.ai\#2237) was retained to enable analysis of interactions. We followed established internet research ethics guidance (e.g., Association of Internet Researchers, 2019~\cite{heise2019internet}) and prioritized contextual integrity and harm minimization throughout. The original posts and replies remain hosted on Discord’s infrastructure and are governed by Discord’s Terms of Service and community policies. Our research copy is de-identified and cannot be used to directly locate or contact specific users. We acknowledge that, in principle, a reader could discover a specific user’s content on Discord via manual or programmatic deep search, but only if those posts remain available in fully public channels; if posts are removed, made private, or otherwise not publicly visible, such discovery would not be feasible through search of Discord’s platform. Any identification arising from a platform-side breach or unauthorized access is outside the scope and control of this study. }
\subsection{\rev{Quantitative Analysis Methods}}

\subsubsection{\rev{{\textbf{Temporal Analysis}}}}
Temporal analysis was performed to examine how user activity and content similarity evolved over time. We first converted the \textit{Date} column in the dataset to ISO8601 format, preserving the timezone information, and then grouped the data by day to compute key metrics: (1) \textbf{Daily Question and Answer Counts}: The total number of questions and answers posted each day was calculated. (2) \textbf{Average Similarity Over Time}: The average semantic similarity of replies to questions was computed for each day. A rolling 7-day window was applied to smooth fluctuations and better capture trends in the data over time. 

\rev{This method directly quantifies and compares activity trends and engagement differences across the support systems and answers RQ1 from a temporal perspective.}

\subsubsection{\textbf{User Profile Building}}
To further enhance our analysis, we constructed user profiles based on interaction patterns. For each user, we tracked the topics they engaged with, the number of questions they asked or answered, their average response similarity to the original questions they addressed, and the total number of reactions they received. To operationalize topical alignment, we used the Sentence-BERT model all-MiniLM-L6-v2 to generate embeddings and computed cosine similarity between each reply and the original question in the thread, rather than the immediately preceding message. We then plotted the relationship between average response quality and the average number of reactions per response for each user, where the x-axis represents average answer quality based on semantic similarity to the original question, and the y-axis represents the average number of reactions per answer as an engagement metric.

\subsubsection{\textbf{Sentiment Analysis and Cumulative Average Sentiment Calculation}}
To analyze the sentiment of each comment, we employed a BERT-based sentiment classification model~\cite{devlin2019bertpretrainingdeepbidirectional}. Specifically, we utilized the pre-trained multilingual sentiment model (``\textit{all-MiniLM-L6-v2}''), bert-base-multilingual-uncased-sentiment~\footnote{https://huggingface.co/nlptown/bert-base-multilingual-uncased-sentiment}, designed for multilingual sentiment classification. The model was fine-tuned on a sentiment classification task, predicting one of five sentiment labels (from 0: Very Negative to 4: Very Positive). For each text entry, we performed the following steps: (1) \textbf{Text Tokenization}: We tokenized the input text using the corresponding BERT tokenizer, ensuring the input length was capped at 512 tokens to fit within the model's processing capabilities. (2) \textbf{Sentiment Prediction}: We passed the tokenized text through the BERT model, which outputted a sentiment score as a probability distribution across the five possible sentiment classes. The class with the highest score was selected as the sentiment label. (3) \textbf{Sentiment Assignment}: The sentiment score was appended as a new column (Sentiment) to the original dataset for subsequent analysis.

To assess the sentiment evolution over time for each source, we calculated the cumulative average sentiment score. This measure allowed us to track how the general emotional tone of replies changed as the conversation progressed. Additionally, this method allows us to understand the response (quantity) situation in each question post. To achieve this, we performed the following steps: (1) \textbf{Sorting and Grouping}: The dataset was first sorted by each source file and the row number to maintain the chronological order of comments. (2) \textbf{Cumulative Sum and Count}: For each source file, we calculated the cumulative sum of sentiment scores and the number of comments at each point in the sequence. (3) \textbf{Cumulative Average Calculation}: The cumulative average sentiment score was then computed by dividing the cumulative sum by the cumulative count, providing a running average that updated with each new reply.

We apply this method to help distinguish differences between AI and human support in tone and conversational persistence and thereby reveals how users interact and the resulting usage patterns for RQ2.

\subsubsection{\textbf{Advanced Network Analysis}}
In addition to sentiment analysis, we performed an advanced network analysis~\cite{doi:10.1126/science.1165821,otte2002social} to investigate the interaction patterns within the ``User Support'' and ``AI Support'' communities. This section outlines the steps taken to construct and analyze these networks. To protect the privacy of the users, all usernames were anonymized. A mapping function was used to replace each username with an anonymized identifier. One exception was the system user (an AI bot in AI-support sub-channel) \textit{kappa.ai\#2237}, which was kept intact due to its relevance in AI Support interactions. This anonymization ensured that the analysis focused on interaction patterns rather than individual identities. 

We modeled the interaction between users as a graph \(G\), where nodes represent users, and edges represent interactions between a question author and a respondent. The edge weights were based on either a default value of 1 or the \textit{similarity\_to\_question} metric, when available, which provided a quantitative measure of how closely a reply was related to the original question. 

For each comment that was part of a thread, an edge was created between the author of the original question and the respondent. This process was repeated across both the User Support and AI Support datasets, resulting in two distinct networks: one for user-to-user interactions and another for user-to-AI interactions. To understand the structure and dynamics of these networks, we applied the Girvan-Newman community detection algorithm \cite{girvan2002community}. This method iteratively removes edges with the highest betweenness centrality, which naturally reveals tightly-knit communities within the graph. Additionally, we calculated three centrality measures for each user: (1) \textbf{Degree Centrality}: A measure of how many direct connections a user has, indicating their level of interaction. (2) \textbf{Betweenness Centrality}:  A measure of the extent to which a user lies on paths between other users, indicating their role as a potential mediator. (3) \textbf{Eigenvector Centrality}: A measure of a user's influence within the network, calculated based on their connections to highly connected users. In cases where eigenvector centrality failed to converge, we used the PageRank algorithm~\cite{brin1998anatomy} as an alternative.

To provide a visual representation of the interaction patterns and detected communities, we generated network visualizations for both the User Support and AI Support networks. Nodes were colored according to their community affiliation, with larger nodes representing users with higher degrees of centrality. A spring layout algorithm was applied to position the nodes, producing a visually coherent network structure. 

This method reveals whether interactions revolve around AI mediated one to one answers or form multi topic collaborative communities and answers how users interact and whether communities and brokers emerge for RQ2.

\subsection{Qualitative Analysis Methods}
We further applied qualitative method in analysis to answer RQ2, which we could further validate the findings from the quantitative analysis and provided more detailed explanations through qualitative analysis~\cite{10.1145/3640457.3688069}. Additionally, it is worth noting that the quantitative and qualitative analyses were initially conducted independently. It was only after obtaining preliminary concepts and identifying some trends that we integrated the results of both analyses. This approach helped researchers avoid preconceived notions at the start of the qualitative analysis and remain open to the nuances and patterns emerging from the content. In this case, our qualitative analysis is mainly based on grounded theory, primarily employs two main methods: Inductive Coding~\cite{thomas2003general} and Constant Comparative Method~\cite{glaser1965constant}, and also enhanced by Topic Modeling~\cite{jelodar2019latent} and AI-powered Qualitative Analysis Method~\cite{zhang2024redefiningqualitativeanalysisai}. These methods do not follow a strictly linear process but are interactive and iterative. In this section, we will describe how we processed these methods in detail. 

\subsubsection{\textbf{Inductive Coding and Constant Comparative Method}}

In conducting this study, we drew on the analytical methods and processes used by Sparrow et al.~\cite{10.1145/3411764.3445363} and Zhang et al.~\cite{10535527}. Overall, we followed these steps: (1) familiarizing ourselves with the data, (2) using open coding~\cite{holton2007coding} to identify preliminary concepts and categories, (3) generating coding themes, (4) reviewing the themes and codes, (5) discussing the themes and reaching consensus, and (6) writing up the results. The first author and senior researcher prioritized discussing the data and the initial coding approach at the start of the project, and then, we began by using open coding to gain a preliminary understanding of the data and identify initial concepts and categories.

Specifically, we randomly selected around 100 ``question'' entries from our dataset for open coding to categorize the issues discussed by users in the two sub-channels. Subsequently, we reviewed the replies (including responses from other users and the AI bot) to these ``question'' posts in both sub-channels and applied open coding to those replies. The primary coders held several meetings to thoroughly discuss the data (both questions and replies), focusing on the initial codes and concepts, resolving potential disagreements and ambiguities, and refining the themes to ensure they accurately represented the data. The coders then discussed the coding results with senior researchers. Based on the collective input and consensus reached, the themes were iteratively revised.

To provide more detail, in inductive coding, researchers coded the data at both the semantic and latent levels and used the comparative analysis method~\cite{glaser1965constant}, continuously reflecting on why users acted as they did and comparing the content of the two sub-channels. This led to the gradual formation of concepts and insights into the differences in user behavior patterns and attitudes between AI-support and user-support. As coding progressed, we continually compared newly emerging data, generating new themes or refining the codes accordingly.

By applying this method, the resulting semantic and latent themes explain how users interact and why particular patterns arise, which addresses RQ2, and this method also helps interpret activity and engagement differences noted in RQ1 through triangulation with quantitative trends.

\subsubsection{\textbf{Topic Modeling}}
\begin{figure}[ht]
  \centering
  \includegraphics[width=1\linewidth]{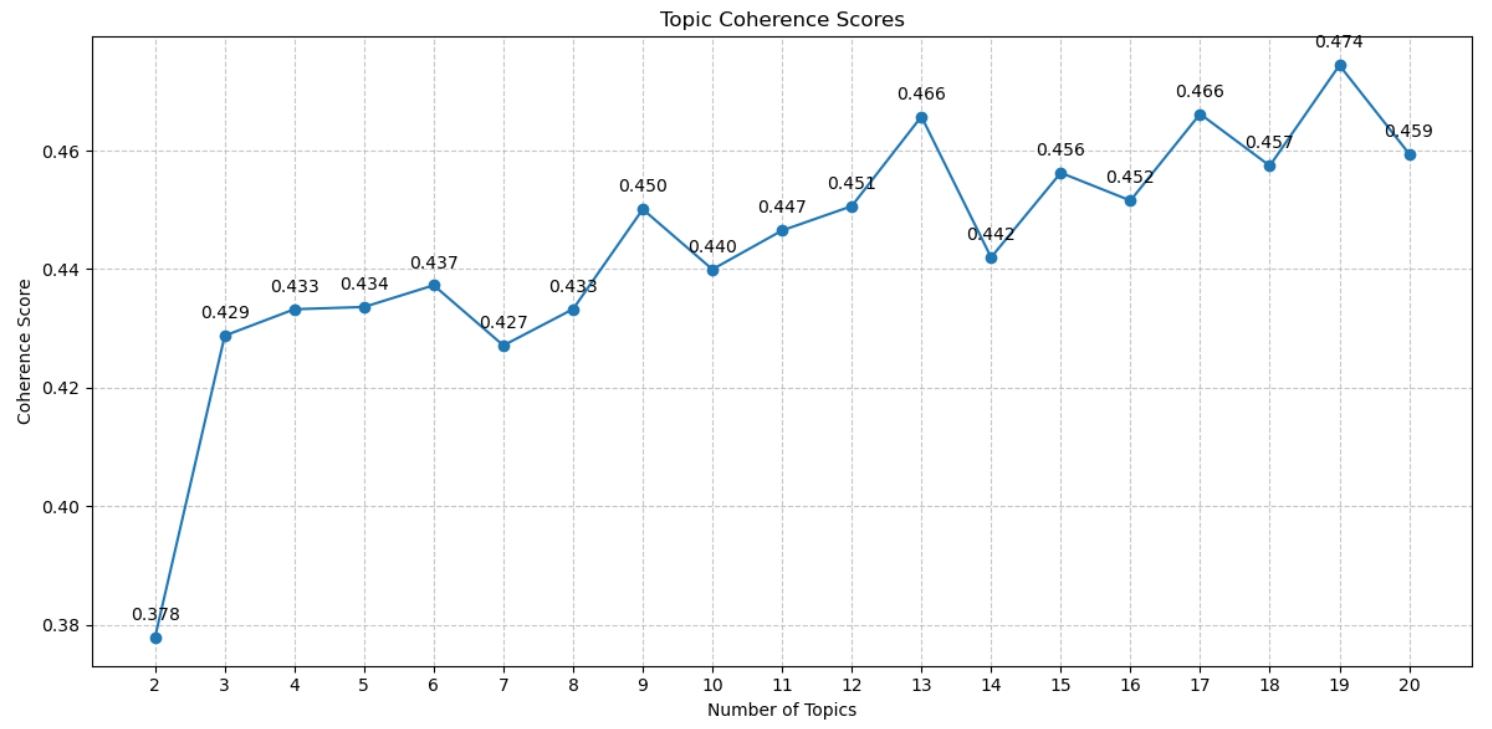}
  \caption{\textbf{The results of the coherence model for sub-channel ``user support''. The x-axis represents the number of topics, and the y-axis represents the coherence score. We calculated the coherence scores for 2 topics to 20 topics. In this case, for ``User Support'', the topic number has been considered as ``19'' with the highest coherence score, ``0.474''.}}
  \label{fig.coherence-scores-user}
\end{figure}

\begin{figure}[ht]
  \centering
  \includegraphics[width=1\linewidth]{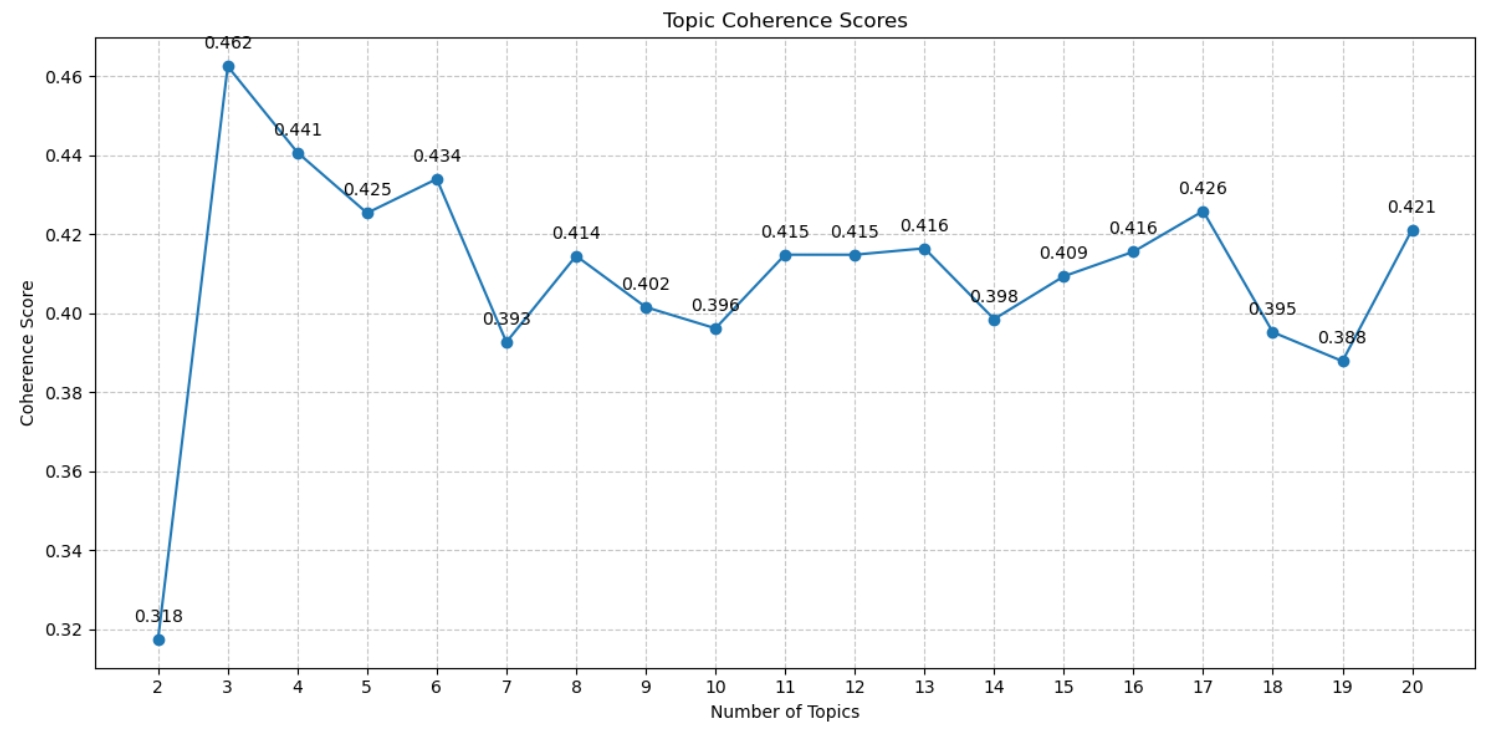}
  \caption{\textbf{The results of the coherence model for sub-channel ``AI support''. The x-axis represents the number of topics, and the y-axis represents the coherence score. We calculated the coherence scores for 2 topics to 20 topics. In this case, for ``AI Support'', the topic number has been considered as ``3'' with the highest coherence score, ``0.462''.}}
  \label{fig.coherence-scores-ai}
\end{figure}

After completing the aforementioned coding and comparisons, we have obtained some interesting results to answer our RQs and reinforce the findings from our quantitative analysis. However, we aim to further introduce topic modeling, primarily due to the scale of the dataset we are dealing with. For large-scale datasets, incorporating topic modeling provides more opportunities to uncover hidden patterns or themes within the data, potentially offering new perspectives for understanding the dataset. Additionally, Discord comments are short texts that contain sparse, noisy, and uninformative words, making it challenging to generate accurate topics. Previous studies have highlighted the effectiveness of topic modeling in extracting thematic structures from such short-text datasets, making it a valuable method for identifying underlying topics and patterns \cite{5416713,likhitha2019detailed,albalawi2020using}. Also, topic modeling is a common and effective method for exploring hidden thematic structures~\cite{10.1145/2133806.2133826}, and it does not solely rely on keyword searches~\cite{jelodar2019latent}. Specifically in this study, we used Latent Dirichlet Allocation (LDA) for topic modeling on our social media data, a method that has been proven effective by several previous studies in exploring topics within social media data~\cite{zhao2011comparing,10.1145/3613904.3642787,egger2022topic}. The LDA, alpha, and beta parameters were set to the default value of 1.0 divided by the number of topics. We conducted a cluster analysis on the pre-processed data and identified a potentially dominant cluster quantity in the results (Figure~\ref{fig.coherence-scores-user} and~\ref{fig.coherence-scores-ai}), which led us to determine the number of thematic trends. The coherence score is considered one of the most effective metrics for topic modeling~\cite{10.5555/2390948.2391052}. Thus, the number of topics was determined based on the coherence score~\cite{stevens2012exploring,10.1145/3613904.3642787} of the extracted topics, with various possible topic numbers ranging from 2 to 20. Figure~\ref{fig.coherence-scores-user} and~\ref{fig.coherence-scores-ai} show that the coherence model suggests using 19 topics as the optimal number for sub-channel ``user support'' and 3 topics for sub-channel ``AI support''. The top 10 keywords for each topic of both user-support and AI-support sub-channels have been shown in Table.~\ref{tab:top10wordstopics}.

\aptLtoX[graphic=no,type=html]{\begin{table*}[ht]
\centering
\begin{tabular}{c|ccccccccccc}
\hline
\hline
\multicolumn{1}{l|}{\,} &
  \multicolumn{11}{c}{User-Support} \\
\hline
Categories &   \multicolumn{1}{c|}{Number  of  Topics} &
  Word 1 &
  Word 2 &
  Word 3 &
  Word 4 &
  Word 5 &
  Word 6 &
  Word 7 &
  Word 8 &
  Word 9 &
  Word 10 \\ \hline
\multirow{19}{*}{User-Support} &
  \multicolumn{1}{c|}{Topic 1} &
  \begin{tabular}[c]{@{}c@{}}vr \break (321.837)\end{tabular} &
  \begin{tabular}[c]{@{}c@{}}steam \break (304.929)\end{tabular} &
  \begin{tabular}[c]{@{}c@{}}vrchat \break (296.832)\end{tabular} &
  \begin{tabular}[c]{@{}c@{}}quest \break (226.877)\end{tabular} &
  \begin{tabular}[c]{@{}c@{}}pc \break (152.881)\end{tabular} &
  \begin{tabular}[c]{@{}c@{}}desktop \break (121.104)\end{tabular} &
  \begin{tabular}[c]{@{}c@{}}use \break (106.223)\end{tabular} &
  \begin{tabular}[c]{@{}c@{}}cotroller \break (87.759)\end{tabular} &
  \begin{tabular}[c]{@{}c@{}}oculus \break (87.670)\end{tabular} &
  \begin{tabular}[c]{@{}c@{}}open \break (80.761)\end{tabular} \\
 &
  \multicolumn{1}{c|}{2} &
  \begin{tabular}[c]{@{}c@{}}help \break (346.783)\end{tabular} &
  \begin{tabular}[c]{@{}c@{}}ve \break (194.796)\end{tabular} &
  \begin{tabular}[c]{@{}c@{}}don \break (130.333)\end{tabular} &
  \begin{tabular}[c]{@{}c@{}}need \break (122.352)\end{tabular} &
  \begin{tabular}[c]{@{}c@{}}know \break (117.426)\end{tabular} &
  \begin{tabular}[c]{@{}c@{}}work \break (69.475)\end{tabular} &
  \begin{tabular}[c]{@{}c@{}}tried \break (58.088)\end{tabular} &
  \begin{tabular}[c]{@{}c@{}}fix \break (56.310)\end{tabular} &
  \begin{tabular}[c]{@{}c@{}}avatar \break (55.113)\end{tabular} &
  \begin{tabular}[c]{@{}c@{}}app \break (48.666)\end{tabular} \\
 &
  \multicolumn{1}{c|}{3} &
  \begin{tabular}[c]{@{}c@{}}gpu \break (151.767)\end{tabular} &
  \begin{tabular}[c]{@{}c@{}}ram \break (133.037)\end{tabular} &
  \begin{tabular}[c]{@{}c@{}}cpu \break (125.835)\end{tabular} &
  \begin{tabular}[c]{@{}c@{}}fps \break (105.007)\end{tabular} &
  \begin{tabular}[c]{@{}c@{}}specs \break (96.906)\end{tabular} &
  \begin{tabular}[c]{@{}c@{}}10 \break (89.260)\end{tabular} &
  \begin{tabular}[c]{@{}c@{}}and \break (79.789)\end{tabular} &
  \begin{tabular}[c]{@{}c@{}}windows \break (78.798)\end{tabular} &
  \begin{tabular}[c]{@{}c@{}}ryzen \break (77.053)\end{tabular} &
  \begin{tabular}[c]{@{}c@{}}rtx \break (76.053)\end{tabular} \\
 &
  \multicolumn{1}{c|}{4} &
  \begin{tabular}[c]{@{}c@{}}avatar \break (648.182)\end{tabular} &
  \begin{tabular}[c]{@{}c@{}}unity \break (615.649)\end{tabular} &
  \begin{tabular}[c]{@{}c@{}}upload \break (438.129)\end{tabular} &
  \begin{tabular}[c]{@{}c@{}}sdk \break (328.840)\end{tabular} &
  \begin{tabular}[c]{@{}c@{}}project \break (262.311)\end{tabular} &
  \begin{tabular}[c]{@{}c@{}}new \break (224.787)\end{tabular} &
  \begin{tabular}[c]{@{}c@{}}error \break (213.507)\end{tabular} &
  \begin{tabular}[c]{@{}c@{}}version \break (212.403)\end{tabular} &
  \begin{tabular}[c]{@{}c@{}}just \break (205.141)\end{tabular} &
  \begin{tabular}[c]{@{}c@{}}trying \break (175.953)\end{tabular} \\
 &
  \multicolumn{1}{c|}{5} &
  \begin{tabular}[c]{@{}c@{}}vrchat \break (339.991)\end{tabular} &
  \begin{tabular}[c]{@{}c@{}}support \break (158.038)\end{tabular} &
  \begin{tabular}[c]{@{}c@{}}ve \break (105.956)\end{tabular} &
  \begin{tabular}[c]{@{}c@{}}just \break (97.963)\end{tabular} &
  \begin{tabular}[c]{@{}c@{}}user \break (97.200)\end{tabular} &
  \begin{tabular}[c]{@{}c@{}}problem \break (90.999)\end{tabular} &
  \begin{tabular}[c]{@{}c@{}}like \break (78.017)\end{tabular} &
  \begin{tabular}[c]{@{}c@{}}issue \break (74.428)\end{tabular} &
  \begin{tabular}[c]{@{}c@{}}time \break (74.047)\end{tabular} &
  \begin{tabular}[c]{@{}c@{}}https \break (60.404)\end{tabular} \\
 &
  \multicolumn{1}{c|}{6} &
  \begin{tabular}[c]{@{}c@{}}fix \break (136.374)\end{tabular} &
  \begin{tabular}[c]{@{}c@{}}does \break (100.794)\end{tabular} &
  \begin{tabular}[c]{@{}c@{}}issue \break (90.827)\end{tabular} &
  \begin{tabular}[c]{@{}c@{}}update \break (87.512)\end{tabular} &
  \begin{tabular}[c]{@{}c@{}}ve \break (83.661)\end{tabular} &
  \begin{tabular}[c]{@{}c@{}}open \break (73.357)\end{tabular} &
  \begin{tabular}[c]{@{}c@{}}reinstalled \break (72.027)\end{tabular} &
  \begin{tabular}[c]{@{}c@{}}completely \break (62.678)\end{tabular} &
  \begin{tabular}[c]{@{}c@{}}vrchat \break (60.538)\end{tabular} &
  \begin{tabular}[c]{@{}c@{}}working \break (56.474)\end{tabular} \\
 &
  \multicolumn{1}{c|}{7} &
  \begin{tabular}[c]{@{}c@{}}trackers \break (182.289)\end{tabular} &
  \begin{tabular}[c]{@{}c@{}}tracking \break (136.505)\end{tabular} &
  \begin{tabular}[c]{@{}c@{}}base \break (115.098)\end{tabular} &
  \begin{tabular}[c]{@{}c@{}}vive \break (108.880)\end{tabular} &
  \begin{tabular}[c]{@{}c@{}}just \break (98.470)\end{tabular} &
  \begin{tabular}[c]{@{}c@{}}im \break (97.658)\end{tabular} &
  \begin{tabular}[c]{@{}c@{}}headset \break (93.874)\end{tabular} &
  \begin{tabular}[c]{@{}c@{}}index \break (87.272)\end{tabular} &
  \begin{tabular}[c]{@{}c@{}}use \break (84.440)\end{tabular} &
  \begin{tabular}[c]{@{}c@{}}body \break (84.216)\end{tabular} \\
 &
  \multicolumn{1}{c|}{8} &
  \begin{tabular}[c]{@{}c@{}}fine \break (232.487)\end{tabular} &
  \begin{tabular}[c]{@{}c@{}}vrchat \break (198.225)\end{tabular} &
  \begin{tabular}[c]{@{}c@{}}just \break (165.866)\end{tabular} &
  \begin{tabular}[c]{@{}c@{}}world \break (131.792)\end{tabular} &
  \begin{tabular}[c]{@{}c@{}}loading \break (115.750)\end{tabular} &
  \begin{tabular}[c]{@{}c@{}}game \break (114.210)\end{tabular} &
  \begin{tabular}[c]{@{}c@{}}works \break (113.001)\end{tabular} &
  \begin{tabular}[c]{@{}c@{}}music \break (92.046)\end{tabular} &
  \begin{tabular}[c]{@{}c@{}}fix \break (91.943)\end{tabular} &
  \begin{tabular}[c]{@{}c@{}}desktop \break (87.511)\end{tabular} \\
 &
  \multicolumn{1}{c|}{9} &
  \begin{tabular}[c]{@{}c@{}}world \break (1647.967)\end{tabular} &
  \begin{tabular}[c]{@{}c@{}}load \break (913.888)\end{tabular} &
  \begin{tabular}[c]{@{}c@{}}home \break (642.023)\end{tabular} &
  \begin{tabular}[c]{@{}c@{}}vrchat \break (571.413)\end{tabular} &
  \begin{tabular}[c]{@{}c@{}}join \break (561.209)\end{tabular} &
  \begin{tabular}[c]{@{}c@{}}worlds \break (527.275)\end{tabular} &
  \begin{tabular}[c]{@{}c@{}}just \break (483.589)\end{tabular} &
  \begin{tabular}[c]{@{}c@{}}try \break (438.388)\end{tabular} &
  \begin{tabular}[c]{@{}c@{}}game \break (408.361)\end{tabular} &
  \begin{tabular}[c]{@{}c@{}}time \break (304.105)\end{tabular} \\
 &
  \multicolumn{1}{c|}{10} &
  vr (342.837) &
  \begin{tabular}[c]{@{}c@{}}chat \break (296.011)\end{tabular} &
  \begin{tabular}[c]{@{}c@{}}like \break (99.145)\end{tabular} &
  \begin{tabular}[c]{@{}c@{}}cheat \break (73.516)\end{tabular} &
  \begin{tabular}[c]{@{}c@{}}anti \break (72.199)\end{tabular} &
  \begin{tabular}[c]{@{}c@{}}help \break (68.017)\end{tabular} &
  \begin{tabular}[c]{@{}c@{}}ve \break (66.689)\end{tabular} &
  \begin{tabular}[c]{@{}c@{}}just \break (64.797)\end{tabular} &
  \begin{tabular}[c]{@{}c@{}}fix \break (59.281)\end{tabular} &
  \begin{tabular}[c]{@{}c@{}}run \break (58.644)\end{tabular} \\
 &
  \multicolumn{1}{c|}{11} &
  \begin{tabular}[c]{@{}c@{}}video \break (305.712)\end{tabular} &
  \begin{tabular}[c]{@{}c@{}}quest \break (280.592)\end{tabular} &
  \begin{tabular}[c]{@{}c@{}}play \break (176.282)\end{tabular} &
  \begin{tabular}[c]{@{}c@{}}audio \break (174.103)\end{tabular} &
  \begin{tabular}[c]{@{}c@{}}link \break (172.285)\end{tabular} &
  \begin{tabular}[c]{@{}c@{}}vrchat \break (163.896)\end{tabular} &
  \begin{tabular}[c]{@{}c@{}}pc \break (147.820)\end{tabular} &
  \begin{tabular}[c]{@{}c@{}}players \break (125.293)\end{tabular} &
  \begin{tabular}[c]{@{}c@{}}just \break (124.656)\end{tabular} &
  \begin{tabular}[c]{@{}c@{}}use \break (122.902)\end{tabular} \\
 &
  \multicolumn{1}{c|}{12} &
  \begin{tabular}[c]{@{}c@{}}avatar \break (91.683)\end{tabular} &
  \begin{tabular}[c]{@{}c@{}}log \break (69.835)\end{tabular} &
  \begin{tabular}[c]{@{}c@{}}instance \break (59.458)\end{tabular} &
  \begin{tabular}[c]{@{}c@{}}00 \break (57.569)\end{tabular} &
  \begin{tabular}[c]{@{}c@{}}2024 \break (21.810)\end{tabular} &
  \begin{tabular}[c]{@{}c@{}}45 \break (48.847)\end{tabular} &
  \begin{tabular}[c]{@{}c@{}}time \break (46.764)\end{tabular} &
  \begin{tabular}[c]{@{}c@{}}make \break (42.720)\end{tabular} &
  \begin{tabular}[c]{@{}c@{}}04 \break (40.053)\end{tabular} &
  \begin{tabular}[c]{@{}c@{}}way \break (37.202)\end{tabular} \\
 &
  \multicolumn{1}{c|}{13} &
  \begin{tabular}[c]{@{}c@{}}tried \break (785.504)\end{tabular} &
  \begin{tabular}[c]{@{}c@{}}game \break (441.978)\end{tabular} &
  \begin{tabular}[c]{@{}c@{}}ve \break (325.189)\end{tabular} &
  \begin{tabular}[c]{@{}c@{}}vrchat \break (314.706)\end{tabular} &
  \begin{tabular}[c]{@{}c@{}}reinstalling \break (309.920)\end{tabular} &
  \begin{tabular}[c]{@{}c@{}}files \break (244.672)\end{tabular} &
  \begin{tabular}[c]{@{}c@{}}restarting \break (184.611)\end{tabular} &
  \begin{tabular}[c]{@{}c@{}}help \break (170.613)\end{tabular} &
  \begin{tabular}[c]{@{}c@{}}work \break (168.193)\end{tabular} &
  \begin{tabular}[c]{@{}c@{}}cache \break (166.075)\end{tabular} \\
 &
  \multicolumn{1}{c|}{14} &
  \begin{tabular}[c]{@{}c@{}}avatar\break (5244.573)\end{tabular} &
  \begin{tabular}[c]{@{}c@{}}avatars \break (360.375)\end{tabular} &
  \begin{tabular}[c]{@{}c@{}}know \break (275.632)\end{tabular} &
  \begin{tabular}[c]{@{}c@{}}just \break (255.617)\end{tabular} &
  \begin{tabular}[c]{@{}c@{}}like \break (214.170)\end{tabular} &
  \begin{tabular}[c]{@{}c@{}}don \break (210.795)\end{tabular} &
  \begin{tabular}[c]{@{}c@{}}use \break (101.575)\end{tabular} &
  \begin{tabular}[c]{@{}c@{}}menu \break (173.396)\end{tabular} &
  \begin{tabular}[c]{@{}c@{}}people \break (158.334)\end{tabular} &
  \begin{tabular}[c]{@{}c@{}}make \break (141.335)\end{tabular} \\
 &
  \multicolumn{1}{c|}{15} &
  \begin{tabular}[c]{@{}c@{}}account \break (717.042)\end{tabular} &
  \begin{tabular}[c]{@{}c@{}}vrchat \break (325.855)\end{tabular} &
  \begin{tabular}[c]{@{}c@{}}email \break (283.773)\end{tabular} &
  \begin{tabular}[c]{@{}c@{}}log \break (202.409)\end{tabular} &
  \begin{tabular}[c]{@{}c@{}}group \break (187.535)\end{tabular} &
  \begin{tabular}[c]{@{}c@{}}website \break (183.376)\end{tabular} &
  \begin{tabular}[c]{@{}c@{}}login \break (171.803)\end{tabular} &
  \begin{tabular}[c]{@{}c@{}}vrc \break (164.059)\end{tabular} &
  \begin{tabular}[c]{@{}c@{}}know \break (162.430)\end{tabular} &
  \begin{tabular}[c]{@{}c@{}}password \break (144.053)\end{tabular} \\
 &
  \multicolumn{1}{c|}{16} &
  \begin{tabular}[c]{@{}c@{}}issue \break (214.474)\end{tabular} &
  \begin{tabular}[c]{@{}c@{}}having \break (173.880)\end{tabular} &
  \begin{tabular}[c]{@{}c@{}}vrchat \break (156.870)\end{tabular} &
  \begin{tabular}[c]{@{}c@{}}game\break (149.041)\end{tabular} &
  \begin{tabular}[c]{@{}c@{}}like \break (142.168)\end{tabular} &
  \begin{tabular}[c]{@{}c@{}}fix \break (131.167)\end{tabular} &
  \begin{tabular}[c]{@{}c@{}}just \break (118.605)\end{tabular} &
  \begin{tabular}[c]{@{}c@{}}mic \break (107.547)\end{tabular} &
  \begin{tabular}[c]{@{}c@{}}issues \break (96.722)\end{tabular} &
  \begin{tabular}[c]{@{}c@{}}open \break (89.929)\end{tabular} \\
 &
  \multicolumn{1}{c|}{17} &
  \begin{tabular}[c]{@{}c@{}}game \break (227.488)\end{tabular} &
  \begin{tabular}[c]{@{}c@{}}loading \break (177.679)\end{tabular} &
  \begin{tabular}[c]{@{}c@{}}screen \break (164.680)\end{tabular} &
  \begin{tabular}[c]{@{}c@{}}stuck \break (156.816)\end{tabular} &
  \begin{tabular}[c]{@{}c@{}}vrchat \break (149.110)\end{tabular} &
  \begin{tabular}[c]{@{}c@{}}just \break (144.147)\end{tabular} &
  \begin{tabular}[c]{@{}c@{}}keeps \break (109.879)\end{tabular} &
  \begin{tabular}[c]{@{}c@{}}avatar \break (95.775)\end{tabular} &
  \begin{tabular}[c]{@{}c@{}}fix \break (93.361)\end{tabular} &
  \begin{tabular}[c]{@{}c@{}}world \break (88.431)\end{tabular} \\
 &
  \multicolumn{1}{c|}{18} &
  \begin{tabular}[c]{@{}c@{}}vrc \break (153.214)\end{tabular} &
  \begin{tabular}[c]{@{}c@{}}vrchat \break (139.806)\end{tabular} &
  \begin{tabular}[c]{@{}c@{}}com \break (136.211)\end{tabular} &
  \begin{tabular}[c]{@{}c@{}}editor \break (128.521)\end{tabular} &
  \begin{tabular}[c]{@{}c@{}}https \break (75.425)\end{tabular} &
  \begin{tabular}[c]{@{}c@{}}object \break (72.715)\end{tabular} &
  \begin{tabular}[c]{@{}c@{}}api \break (67.945)\end{tabular} &
  \begin{tabular}[c]{@{}c@{}}unitypengine \break (66.053)\end{tabular} &
  \begin{tabular}[c]{@{}c@{}}packages \break (58.254)\end{tabular} &
  \begin{tabular}[c]{@{}c@{}}cs \break (50.053)\end{tabular} \\
 &
  \multicolumn{1}{c|}{19} &
  \begin{tabular}[c]{@{}c@{}}error \break (308.338)\end{tabular} &
  \begin{tabular}[c]{@{}c@{}}world \break (218.128)\end{tabular} &
  \begin{tabular}[c]{@{}c@{}}issue \break (131.130)\end{tabular} &
  \begin{tabular}[c]{@{}c@{}}getting \break (128.688)\end{tabular} &
  \begin{tabular}[c]{@{}c@{}}message \break (114.969)\end{tabular} &
  \begin{tabular}[c]{@{}c@{}}vrchat \break (73.726)\end{tabular} &
  \begin{tabular}[c]{@{}c@{}}im \break (73.211)\end{tabular} &
  \begin{tabular}[c]{@{}c@{}}tried \break (66.493)\end{tabular} &
  \begin{tabular}[c]{@{}c@{}}able \break (59.820)\end{tabular} &
  \begin{tabular}[c]{@{}c@{}}vrc \break (56.902)\end{tabular} \\ \hline
\multicolumn{1}{l|}{} &
  \multicolumn{11}{c}{AI-Support} \\ \hline
 &
  \multicolumn{1}{c|}{Number  of  Topics} &
  Word 1 &
  Word 2 &
  Word 3 &
  Word 4 &
  Word 5 &
  Word 6 &
  Word 7 &
  Word 8 &
  Word 9 &
  Word 10 \\ \hline
\multirow{3}{*}{AI-Support} &
  \multicolumn{1}{c|}{1} &
  \begin{tabular}[c]{@{}c@{}}try \break (377.166)\end{tabular} &
  \begin{tabular}[c]{@{}c@{}}ai \break (338.308)\end{tabular} &
  \begin{tabular}[c]{@{}c@{}}kapa \break (338.306)\end{tabular} &
  \begin{tabular}[c]{@{}c@{}}hi \break (304.259)\end{tabular} &
  \begin{tabular}[c]{@{}c@{}}support \break (278.305)\end{tabular} &
  \begin{tabular}[c]{@{}c@{}}question \break (267.320)\end{tabular} &
  \begin{tabular}[c]{@{}c@{}}base \break (263.023)\end{tabular} &
  \begin{tabular}[c]{@{}c@{}}answer \break (260.318)\end{tabular} &
  \begin{tabular}[c]{@{}c@{}}bot \break (258.332)\end{tabular} &
  \begin{tabular}[c]{@{}c@{}}knowledge \break (252.332)\end{tabular} \\
 &
  \multicolumn{1}{c|}{2} &
  \begin{tabular}[c]{@{}c@{}}vrc \break (396.116)\end{tabular} &
  \begin{tabular}[c]{@{}c@{}}editor \break (308.327)\end{tabular} &
  \begin{tabular}[c]{@{}c@{}}vrchat \break (259.467)\end{tabular} &
  \begin{tabular}[c]{@{}c@{}}object \break (177.503)\end{tabular} &
  \begin{tabular}[c]{@{}c@{}}api \break (161.298)\end{tabular} &
  \begin{tabular}[c]{@{}c@{}}unityengine \break (157.320)\end{tabular} &
  \begin{tabular}[c]{@{}c@{}}cs \break (156.332)\end{tabular} &
  \begin{tabular}[c]{@{}c@{}}com \break (152.293)\end{tabular} &
  \begin{tabular}[c]{@{}c@{}}packages \break (139.330)\end{tabular} &
  \begin{tabular}[c]{@{}c@{}}string \break (135.299)\end{tabular} \\
 &
  \multicolumn{1}{c|}{3} &
  \begin{tabular}[c]{@{}c@{}}vrchat \break (574.893)\end{tabular} &
  \begin{tabular}[c]{@{}c@{}}avatar \break (496.540)\end{tabular} &
  \begin{tabular}[c]{@{}c@{}}game \break (254.913)\end{tabular} &
  \begin{tabular}[c]{@{}c@{}}world \break (245.501)\end{tabular} &
  \begin{tabular}[c]{@{}c@{}}quest \break (228.182)\end{tabular} &
  \begin{tabular}[c]{@{}c@{}}make \break (215.784)\end{tabular} &
  \begin{tabular}[c]{@{}c@{}}just \break (207.854)\end{tabular} &
  \begin{tabular}[c]{@{}c@{}}fix \break (190.657)\end{tabular} &
  \begin{tabular}[c]{@{}c@{}}like \break (184.708)\end{tabular} &
  \begin{tabular}[c]{@{}c@{}}work \break (176.842)\end{tabular} \\ \hline
		  \hline
\end{tabular}%
\caption{Top 10 Words for Each Topic in User-Support and AI-Support Sub-channels. There are 19 main topics for user-support sub-channel and 3 main topics for AI-support based on the result of topic coherence scores.}
\label{tab:top10wordstopics}
\end{table*}}{\begin{table*}[ht]
\centering
\resizebox{\textwidth}{!}{%
\begin{tabular}{c|ccccccccccc}
\toprule\hline
\multicolumn{1}{l|}{} &
  \multicolumn{11}{c}{User-Support} \\ \hline
Categories &
  \multicolumn{1}{c|}{\begin{tabular}[c]{@{}c@{}}Number \\ of \\ Topics\end{tabular}} &
  Word 1 &
  Word 2 &
  Word 3 &
  Word 4 &
  Word 5 &
  Word 6 &
  Word 7 &
  Word 8 &
  Word 9 &
  Word 10 \\ \hline
\multirow{19}{*}{User-Support} &
  \multicolumn{1}{c|}{Topic 1} &
  \begin{tabular}[c]{@{}c@{}}vr \\ (321.837)\end{tabular} &
  \begin{tabular}[c]{@{}c@{}}steam \\ (304.929)\end{tabular} &
  \begin{tabular}[c]{@{}c@{}}vrchat \\ (296.832)\end{tabular} &
  \begin{tabular}[c]{@{}c@{}}quest \\ (226.877)\end{tabular} &
  \begin{tabular}[c]{@{}c@{}}pc \\ (152.881)\end{tabular} &
  \begin{tabular}[c]{@{}c@{}}desktop \\ (121.104)\end{tabular} &
  \begin{tabular}[c]{@{}c@{}}use \\ (106.223)\end{tabular} &
  \begin{tabular}[c]{@{}c@{}}cotroller \\ (87.759)\end{tabular} &
  \begin{tabular}[c]{@{}c@{}}oculus \\ (87.670)\end{tabular} &
  \begin{tabular}[c]{@{}c@{}}open \\ (80.761)\end{tabular} \\
 &
  \multicolumn{1}{c|}{2} &
  \begin{tabular}[c]{@{}c@{}}help \\ (346.783)\end{tabular} &
  \begin{tabular}[c]{@{}c@{}}ve \\ (194.796)\end{tabular} &
  \begin{tabular}[c]{@{}c@{}}don \\ (130.333)\end{tabular} &
  \begin{tabular}[c]{@{}c@{}}need \\ (122.352)\end{tabular} &
  \begin{tabular}[c]{@{}c@{}}know \\ (117.426)\end{tabular} &
  \begin{tabular}[c]{@{}c@{}}work \\ (69.475)\end{tabular} &
  \begin{tabular}[c]{@{}c@{}}tried \\ (58.088)\end{tabular} &
  \begin{tabular}[c]{@{}c@{}}fix \\ (56.310)\end{tabular} &
  \begin{tabular}[c]{@{}c@{}}avatar \\ (55.113)\end{tabular} &
  \begin{tabular}[c]{@{}c@{}}app \\ (48.666)\end{tabular} \\
 &
  \multicolumn{1}{c|}{3} &
  \begin{tabular}[c]{@{}c@{}}gpu \\ (151.767)\end{tabular} &
  \begin{tabular}[c]{@{}c@{}}ram \\ (133.037)\end{tabular} &
  \begin{tabular}[c]{@{}c@{}}cpu \\ (125.835)\end{tabular} &
  \begin{tabular}[c]{@{}c@{}}fps \\ (105.007)\end{tabular} &
  \begin{tabular}[c]{@{}c@{}}specs \\ (96.906)\end{tabular} &
  \begin{tabular}[c]{@{}c@{}}10 \\ (89.260)\end{tabular} &
  \begin{tabular}[c]{@{}c@{}}and \\ (79.789)\end{tabular} &
  \begin{tabular}[c]{@{}c@{}}windows \\ (78.798)\end{tabular} &
  \begin{tabular}[c]{@{}c@{}}ryzen \\ (77.053)\end{tabular} &
  \begin{tabular}[c]{@{}c@{}}rtx \\ (76.053)\end{tabular} \\
 &
  \multicolumn{1}{c|}{4} &
  \begin{tabular}[c]{@{}c@{}}avatar \\ (648.182)\end{tabular} &
  \begin{tabular}[c]{@{}c@{}}unity \\ (615.649)\end{tabular} &
  \begin{tabular}[c]{@{}c@{}}upload \\ (438.129)\end{tabular} &
  \begin{tabular}[c]{@{}c@{}}sdk \\ (328.840)\end{tabular} &
  \begin{tabular}[c]{@{}c@{}}project \\ (262.311)\end{tabular} &
  \begin{tabular}[c]{@{}c@{}}new \\ (224.787)\end{tabular} &
  \begin{tabular}[c]{@{}c@{}}error \\ (213.507)\end{tabular} &
  \begin{tabular}[c]{@{}c@{}}version \\ (212.403)\end{tabular} &
  \begin{tabular}[c]{@{}c@{}}just \\ (205.141)\end{tabular} &
  \begin{tabular}[c]{@{}c@{}}trying \\ (175.953)\end{tabular} \\
 &
  \multicolumn{1}{c|}{5} &
  \begin{tabular}[c]{@{}c@{}}vrchat \\ (339.991)\end{tabular} &
  \begin{tabular}[c]{@{}c@{}}support \\ (158.038)\end{tabular} &
  \begin{tabular}[c]{@{}c@{}}ve \\ (105.956)\end{tabular} &
  \begin{tabular}[c]{@{}c@{}}just \\ (97.963)\end{tabular} &
  \begin{tabular}[c]{@{}c@{}}user \\ (97.200)\end{tabular} &
  \begin{tabular}[c]{@{}c@{}}problem \\ (90.999)\end{tabular} &
  \begin{tabular}[c]{@{}c@{}}like \\ (78.017)\end{tabular} &
  \begin{tabular}[c]{@{}c@{}}issue \\ (74.428)\end{tabular} &
  \begin{tabular}[c]{@{}c@{}}time \\ (74.047)\end{tabular} &
  \begin{tabular}[c]{@{}c@{}}https \\ (60.404)\end{tabular} \\
 &
  \multicolumn{1}{c|}{6} &
  \begin{tabular}[c]{@{}c@{}}fix \\ (136.374)\end{tabular} &
  \begin{tabular}[c]{@{}c@{}}does \\ (100.794)\end{tabular} &
  \begin{tabular}[c]{@{}c@{}}issue \\ (90.827)\end{tabular} &
  \begin{tabular}[c]{@{}c@{}}update \\ (87.512)\end{tabular} &
  \begin{tabular}[c]{@{}c@{}}ve \\ (83.661)\end{tabular} &
  \begin{tabular}[c]{@{}c@{}}open \\ (73.357)\end{tabular} &
  \begin{tabular}[c]{@{}c@{}}reinstalled \\ (72.027)\end{tabular} &
  \begin{tabular}[c]{@{}c@{}}completely \\ (62.678)\end{tabular} &
  \begin{tabular}[c]{@{}c@{}}vrchat \\ (60.538)\end{tabular} &
  \begin{tabular}[c]{@{}c@{}}working \\ (56.474)\end{tabular} \\
 &
  \multicolumn{1}{c|}{7} &
  \begin{tabular}[c]{@{}c@{}}trackers \\ (182.289)\end{tabular} &
  \begin{tabular}[c]{@{}c@{}}tracking \\ (136.505)\end{tabular} &
  \begin{tabular}[c]{@{}c@{}}base \\ (115.098)\end{tabular} &
  \begin{tabular}[c]{@{}c@{}}vive \\ (108.880)\end{tabular} &
  \begin{tabular}[c]{@{}c@{}}just \\ (98.470)\end{tabular} &
  \begin{tabular}[c]{@{}c@{}}im \\ (97.658)\end{tabular} &
  \begin{tabular}[c]{@{}c@{}}headset \\ (93.874)\end{tabular} &
  \begin{tabular}[c]{@{}c@{}}index \\ (87.272)\end{tabular} &
  \begin{tabular}[c]{@{}c@{}}use \\ (84.440)\end{tabular} &
  \begin{tabular}[c]{@{}c@{}}body \\ (84.216)\end{tabular} \\
 &
  \multicolumn{1}{c|}{8} &
  \begin{tabular}[c]{@{}c@{}}fine \\ (232.487)\end{tabular} &
  \begin{tabular}[c]{@{}c@{}}vrchat \\ (198.225)\end{tabular} &
  \begin{tabular}[c]{@{}c@{}}just \\ (165.866)\end{tabular} &
  \begin{tabular}[c]{@{}c@{}}world \\ (131.792)\end{tabular} &
  \begin{tabular}[c]{@{}c@{}}loading \\ (115.750)\end{tabular} &
  \begin{tabular}[c]{@{}c@{}}game \\ (114.210)\end{tabular} &
  \begin{tabular}[c]{@{}c@{}}works \\ (113.001)\end{tabular} &
  \begin{tabular}[c]{@{}c@{}}music \\ (92.046)\end{tabular} &
  \begin{tabular}[c]{@{}c@{}}fix \\ (91.943)\end{tabular} &
  \begin{tabular}[c]{@{}c@{}}desktop \\ (87.511)\end{tabular} \\
 &
  \multicolumn{1}{c|}{9} &
  \begin{tabular}[c]{@{}c@{}}world \\ (1647.967)\end{tabular} &
  \begin{tabular}[c]{@{}c@{}}load \\ (913.888)\end{tabular} &
  \begin{tabular}[c]{@{}c@{}}home \\ (642.023)\end{tabular} &
  \begin{tabular}[c]{@{}c@{}}vrchat \\ (571.413)\end{tabular} &
  \begin{tabular}[c]{@{}c@{}}join \\ (561.209)\end{tabular} &
  \begin{tabular}[c]{@{}c@{}}worlds \\ (527.275)\end{tabular} &
  \begin{tabular}[c]{@{}c@{}}just \\ (483.589)\end{tabular} &
  \begin{tabular}[c]{@{}c@{}}try \\ (438.388)\end{tabular} &
  \begin{tabular}[c]{@{}c@{}}game \\ (408.361)\end{tabular} &
  \begin{tabular}[c]{@{}c@{}}time \\ (304.105)\end{tabular} \\
 &
  \multicolumn{1}{c|}{10} &
  vr (342.837) &
  \begin{tabular}[c]{@{}c@{}}chat \\ (296.011)\end{tabular} &
  \begin{tabular}[c]{@{}c@{}}like \\ (99.145)\end{tabular} &
  \begin{tabular}[c]{@{}c@{}}cheat \\ (73.516)\end{tabular} &
  \begin{tabular}[c]{@{}c@{}}anti \\ (72.199)\end{tabular} &
  \begin{tabular}[c]{@{}c@{}}help \\ (68.017)\end{tabular} &
  \begin{tabular}[c]{@{}c@{}}ve \\ (66.689)\end{tabular} &
  \begin{tabular}[c]{@{}c@{}}just \\ (64.797)\end{tabular} &
  \begin{tabular}[c]{@{}c@{}}fix \\ (59.281)\end{tabular} &
  \begin{tabular}[c]{@{}c@{}}run \\ (58.644)\end{tabular} \\
 &
  \multicolumn{1}{c|}{11} &
  \begin{tabular}[c]{@{}c@{}}video \\ (305.712)\end{tabular} &
  \begin{tabular}[c]{@{}c@{}}quest \\ (280.592)\end{tabular} &
  \begin{tabular}[c]{@{}c@{}}play \\ (176.282)\end{tabular} &
  \begin{tabular}[c]{@{}c@{}}audio \\ (174.103)\end{tabular} &
  \begin{tabular}[c]{@{}c@{}}link \\ (172.285)\end{tabular} &
  \begin{tabular}[c]{@{}c@{}}vrchat \\ (163.896)\end{tabular} &
  \begin{tabular}[c]{@{}c@{}}pc \\ (147.820)\end{tabular} &
  \begin{tabular}[c]{@{}c@{}}players \\ (125.293)\end{tabular} &
  \begin{tabular}[c]{@{}c@{}}just \\ (124.656)\end{tabular} &
  \begin{tabular}[c]{@{}c@{}}use \\ (122.902)\end{tabular} \\
 &
  \multicolumn{1}{c|}{12} &
  \begin{tabular}[c]{@{}c@{}}avatar \\ (91.683)\end{tabular} &
  \begin{tabular}[c]{@{}c@{}}log \\ (69.835)\end{tabular} &
  \begin{tabular}[c]{@{}c@{}}instance \\ (59.458)\end{tabular} &
  \begin{tabular}[c]{@{}c@{}}00 \\ (57.569)\end{tabular} &
  \begin{tabular}[c]{@{}c@{}}2024 \\ (21.810)\end{tabular} &
  \begin{tabular}[c]{@{}c@{}}45 \\ (48.847)\end{tabular} &
  \begin{tabular}[c]{@{}c@{}}time \\ (46.764)\end{tabular} &
  \begin{tabular}[c]{@{}c@{}}make \\ (42.720)\end{tabular} &
  \begin{tabular}[c]{@{}c@{}}04 \\ (40.053)\end{tabular} &
  \begin{tabular}[c]{@{}c@{}}way \\ (37.202)\end{tabular} \\
 &
  \multicolumn{1}{c|}{13} &
  \begin{tabular}[c]{@{}c@{}}tried \\ (785.504)\end{tabular} &
  \begin{tabular}[c]{@{}c@{}}game \\ (441.978)\end{tabular} &
  \begin{tabular}[c]{@{}c@{}}ve \\ (325.189)\end{tabular} &
  \begin{tabular}[c]{@{}c@{}}vrchat \\ (314.706)\end{tabular} &
  \begin{tabular}[c]{@{}c@{}}reinstalling \\ (309.920)\end{tabular} &
  \begin{tabular}[c]{@{}c@{}}files \\ (244.672)\end{tabular} &
  \begin{tabular}[c]{@{}c@{}}restarting \\ (184.611)\end{tabular} &
  \begin{tabular}[c]{@{}c@{}}help \\ (170.613)\end{tabular} &
  \begin{tabular}[c]{@{}c@{}}work \\ (168.193)\end{tabular} &
  \begin{tabular}[c]{@{}c@{}}cache \\ (166.075)\end{tabular} \\
 &
  \multicolumn{1}{c|}{14} &
  \begin{tabular}[c]{@{}c@{}}avatar\\ (5244.573)\end{tabular} &
  \begin{tabular}[c]{@{}c@{}}avatars \\ (360.375)\end{tabular} &
  \begin{tabular}[c]{@{}c@{}}know \\ (275.632)\end{tabular} &
  \begin{tabular}[c]{@{}c@{}}just \\ (255.617)\end{tabular} &
  \begin{tabular}[c]{@{}c@{}}like \\ (214.170)\end{tabular} &
  \begin{tabular}[c]{@{}c@{}}don \\ (210.795)\end{tabular} &
  \begin{tabular}[c]{@{}c@{}}use \\ (101.575)\end{tabular} &
  \begin{tabular}[c]{@{}c@{}}menu \\ (173.396)\end{tabular} &
  \begin{tabular}[c]{@{}c@{}}people \\ (158.334)\end{tabular} &
  \begin{tabular}[c]{@{}c@{}}make \\ (141.335)\end{tabular} \\
 &
  \multicolumn{1}{c|}{15} &
  \begin{tabular}[c]{@{}c@{}}account \\ (717.042)\end{tabular} &
  \begin{tabular}[c]{@{}c@{}}vrchat \\ (325.855)\end{tabular} &
  \begin{tabular}[c]{@{}c@{}}email \\ (283.773)\end{tabular} &
  \begin{tabular}[c]{@{}c@{}}log \\ (202.409)\end{tabular} &
  \begin{tabular}[c]{@{}c@{}}group \\ (187.535)\end{tabular} &
  \begin{tabular}[c]{@{}c@{}}website \\ (183.376)\end{tabular} &
  \begin{tabular}[c]{@{}c@{}}login \\ (171.803)\end{tabular} &
  \begin{tabular}[c]{@{}c@{}}vrc \\ (164.059)\end{tabular} &
  \begin{tabular}[c]{@{}c@{}}know \\ (162.430)\end{tabular} &
  \begin{tabular}[c]{@{}c@{}}password \\ (144.053)\end{tabular} \\
 &
  \multicolumn{1}{c|}{16} &
  \begin{tabular}[c]{@{}c@{}}issue \\ (214.474)\end{tabular} &
  \begin{tabular}[c]{@{}c@{}}having \\ (173.880)\end{tabular} &
  \begin{tabular}[c]{@{}c@{}}vrchat \\ (156.870)\end{tabular} &
  \begin{tabular}[c]{@{}c@{}}game\\ (149.041)\end{tabular} &
  \begin{tabular}[c]{@{}c@{}}like \\ (142.168)\end{tabular} &
  \begin{tabular}[c]{@{}c@{}}fix \\ (131.167)\end{tabular} &
  \begin{tabular}[c]{@{}c@{}}just \\ (118.605)\end{tabular} &
  \begin{tabular}[c]{@{}c@{}}mic \\ (107.547)\end{tabular} &
  \begin{tabular}[c]{@{}c@{}}issues \\ (96.722)\end{tabular} &
  \begin{tabular}[c]{@{}c@{}}open \\ (89.929)\end{tabular} \\
 &
  \multicolumn{1}{c|}{17} &
  \begin{tabular}[c]{@{}c@{}}game \\ (227.488)\end{tabular} &
  \begin{tabular}[c]{@{}c@{}}loading \\ (177.679)\end{tabular} &
  \begin{tabular}[c]{@{}c@{}}screen \\ (164.680)\end{tabular} &
  \begin{tabular}[c]{@{}c@{}}stuck \\ (156.816)\end{tabular} &
  \begin{tabular}[c]{@{}c@{}}vrchat \\ (149.110)\end{tabular} &
  \begin{tabular}[c]{@{}c@{}}just \\ (144.147)\end{tabular} &
  \begin{tabular}[c]{@{}c@{}}keeps \\ (109.879)\end{tabular} &
  \begin{tabular}[c]{@{}c@{}}avatar \\ (95.775)\end{tabular} &
  \begin{tabular}[c]{@{}c@{}}fix \\ (93.361)\end{tabular} &
  \begin{tabular}[c]{@{}c@{}}world \\ (88.431)\end{tabular} \\
 &
  \multicolumn{1}{c|}{18} &
  \begin{tabular}[c]{@{}c@{}}vrc \\ (153.214)\end{tabular} &
  \begin{tabular}[c]{@{}c@{}}vrchat \\ (139.806)\end{tabular} &
  \begin{tabular}[c]{@{}c@{}}com \\ (136.211)\end{tabular} &
  \begin{tabular}[c]{@{}c@{}}editor \\ (128.521)\end{tabular} &
  \begin{tabular}[c]{@{}c@{}}https \\ (75.425)\end{tabular} &
  \begin{tabular}[c]{@{}c@{}}object \\ (72.715)\end{tabular} &
  \begin{tabular}[c]{@{}c@{}}api \\ (67.945)\end{tabular} &
  \begin{tabular}[c]{@{}c@{}}unitypengine \\ (66.053)\end{tabular} &
  \begin{tabular}[c]{@{}c@{}}packages \\ (58.254)\end{tabular} &
  \begin{tabular}[c]{@{}c@{}}cs \\ (50.053)\end{tabular} \\
 &
  \multicolumn{1}{c|}{19} &
  \begin{tabular}[c]{@{}c@{}}error \\ (308.338)\end{tabular} &
  \begin{tabular}[c]{@{}c@{}}world \\ (218.128)\end{tabular} &
  \begin{tabular}[c]{@{}c@{}}issue \\ (131.130)\end{tabular} &
  \begin{tabular}[c]{@{}c@{}}getting \\ (128.688)\end{tabular} &
  \begin{tabular}[c]{@{}c@{}}message \\ (114.969)\end{tabular} &
  \begin{tabular}[c]{@{}c@{}}vrchat \\ (73.726)\end{tabular} &
  \begin{tabular}[c]{@{}c@{}}im \\ (73.211)\end{tabular} &
  \begin{tabular}[c]{@{}c@{}}tried \\ (66.493)\end{tabular} &
  \begin{tabular}[c]{@{}c@{}}able \\ (59.820)\end{tabular} &
  \begin{tabular}[c]{@{}c@{}}vrc \\ (56.902)\end{tabular} \\ \hline
\multicolumn{1}{l|}{} &
  \multicolumn{11}{c}{AI-Support} \\ \hline
 &
  \multicolumn{1}{c|}{\begin{tabular}[c]{@{}c@{}}Number \\ of \\ Topics\end{tabular}} &
  Word 1 &
  Word 2 &
  Word 3 &
  Word 4 &
  Word 5 &
  Word 6 &
  Word 7 &
  Word 8 &
  Word 9 &
  Word 10 \\ \hline
\multirow{3}{*}{AI-Support} &
  \multicolumn{1}{c|}{1} &
  \begin{tabular}[c]{@{}c@{}}try \\ (377.166)\end{tabular} &
  \begin{tabular}[c]{@{}c@{}}ai \\ (338.308)\end{tabular} &
  \begin{tabular}[c]{@{}c@{}}kapa \\ (338.306)\end{tabular} &
  \begin{tabular}[c]{@{}c@{}}hi \\ (304.259)\end{tabular} &
  \begin{tabular}[c]{@{}c@{}}support \\ (278.305)\end{tabular} &
  \begin{tabular}[c]{@{}c@{}}question \\ (267.320)\end{tabular} &
  \begin{tabular}[c]{@{}c@{}}base \\ (263.023)\end{tabular} &
  \begin{tabular}[c]{@{}c@{}}answer \\ (260.318)\end{tabular} &
  \begin{tabular}[c]{@{}c@{}}bot \\ (258.332)\end{tabular} &
  \begin{tabular}[c]{@{}c@{}}knowledge \\ (252.332)\end{tabular} \\
 &
  \multicolumn{1}{c|}{2} &
  \begin{tabular}[c]{@{}c@{}}vrc \\ (396.116)\end{tabular} &
  \begin{tabular}[c]{@{}c@{}}editor \\ (308.327)\end{tabular} &
  \begin{tabular}[c]{@{}c@{}}vrchat \\ (259.467)\end{tabular} &
  \begin{tabular}[c]{@{}c@{}}object \\ (177.503)\end{tabular} &
  \begin{tabular}[c]{@{}c@{}}api \\ (161.298)\end{tabular} &
  \begin{tabular}[c]{@{}c@{}}unityengine \\ (157.320)\end{tabular} &
  \begin{tabular}[c]{@{}c@{}}cs \\ (156.332)\end{tabular} &
  \begin{tabular}[c]{@{}c@{}}com \\ (152.293)\end{tabular} &
  \begin{tabular}[c]{@{}c@{}}packages \\ (139.330)\end{tabular} &
  \begin{tabular}[c]{@{}c@{}}string \\ (135.299)\end{tabular} \\
 &
  \multicolumn{1}{c|}{3} &
  \begin{tabular}[c]{@{}c@{}}vrchat \\ (574.893)\end{tabular} &
  \begin{tabular}[c]{@{}c@{}}avatar \\ (496.540)\end{tabular} &
  \begin{tabular}[c]{@{}c@{}}game \\ (254.913)\end{tabular} &
  \begin{tabular}[c]{@{}c@{}}world \\ (245.501)\end{tabular} &
  \begin{tabular}[c]{@{}c@{}}quest \\ (228.182)\end{tabular} &
  \begin{tabular}[c]{@{}c@{}}make \\ (215.784)\end{tabular} &
  \begin{tabular}[c]{@{}c@{}}just \\ (207.854)\end{tabular} &
  \begin{tabular}[c]{@{}c@{}}fix \\ (190.657)\end{tabular} &
  \begin{tabular}[c]{@{}c@{}}like \\ (184.708)\end{tabular} &
  \begin{tabular}[c]{@{}c@{}}work \\ (176.842)\end{tabular} \\ \hline\bottomrule
\end{tabular}%
}
\caption{Top 10 Words for Each Topic in User-Support and AI-Support Sub-channels. There are 19 main topics for user-support sub-channel and 3 main topics for AI-support based on the result of topic coherence scores.}
\label{tab:top10wordstopics}
\end{table*}}

From the results of the topic modeling, we found that the thematic diversity in the user-support sub-channel is far greater than in AI-support (with user-support covering a broader range of issues). At the same time, the vocabulary in user-support seems to revolve more around specific technical issues and products, such as VR devices (``Steam'', ``quest'', ``desktop'', ``controller'', ``oculus''), hardware (``gpu'', ``ram'', ``cpu'', ``ryzen'', ``rtx''), software technologies (``unity'', ``sdk'', ``project'', ``version''), and technical development issues (``tracking'', ``log'', ``cache'', ``menu''). In contrast, the vocabulary in AI-support tends to focus more on general support terms like ``support'', ``question'', ``vrc'' (VRChat), ``fix'', and others. This indicates that the trends in thematic diversity and the range of user issues align with the trends we observed in our quantitative analysis (complementing the quantitative comparisons for RQ1). At the same time, the specific categories shown in the topic modeling broadly address some of RQ2.

\section{Results of Quantitative Analysis}
We applied quantitative analysis methods to analyze the data in order to answer RQ1 and part of RQ2.

\subsection{\textbf{Results of} Temporal Analysis}

\begin{figure*}[ht]
  \centering
  \includegraphics[width=1\linewidth]{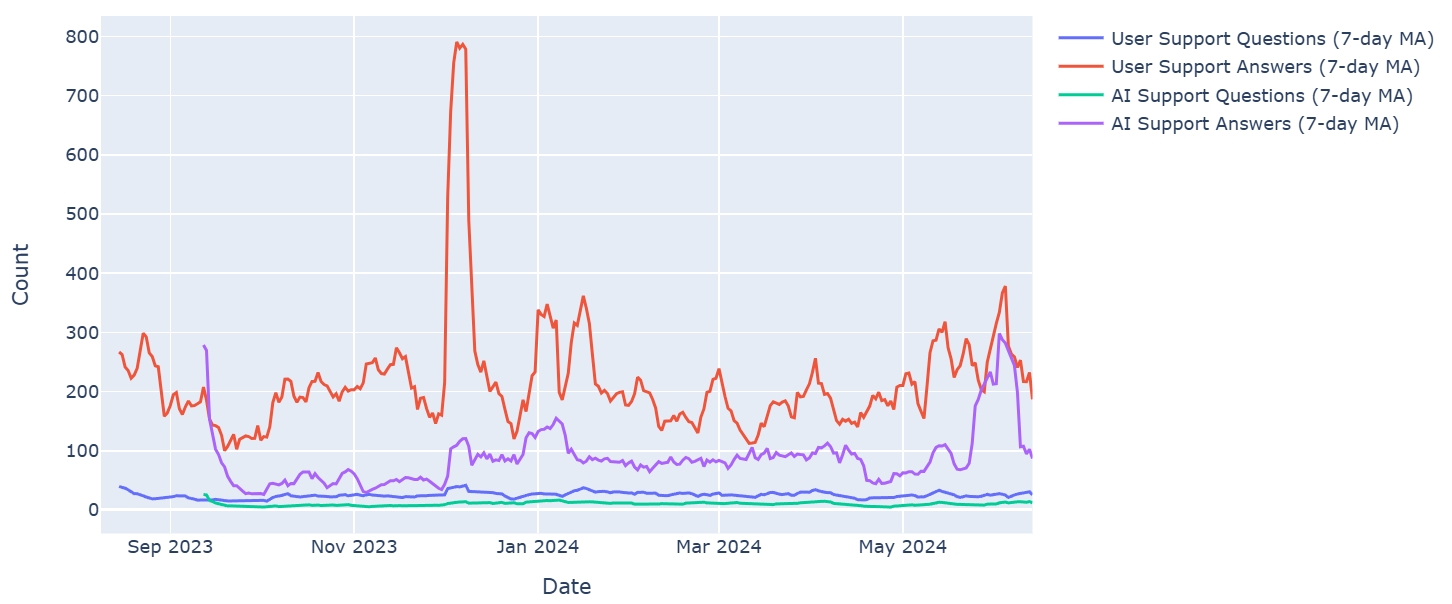}
  \caption{\textbf{Comparative Temporal Trends in Two Sub-Channels. Blue line represents User Support Questions, red line represents User Support Answers, light purple line represents AI Support Questions, and purple line represents AI Support Answers. All categories smmothed using a 7-day moving average (7-day MA). It shows temporal trends in the VRChat discord community’s support channels.}}
  \label{fig.Temporal-Trends}
\end{figure*}

The temporal trends of question and answer counts for both sub-channels. As shown in Fig.~\ref{fig.Temporal-Trends}, the trend in the number of questions and answers in the two sub-channels over time highlights differences in user activity patterns. Compared to AI-support, we found that the number of questions in the user-support sub-channel is similar to, but slightly higher than, that in the AI-support sub-channel. However, in terms of the number of replies and follow-up posts, user engagement in the user-support sub-channel is generally higher than in AI-support, with noticeable peaks in late December 2023 and mid-2024. These peaks suggest events or discussions within the community that triggered above-average activity. In contrast, the AI-support sub-channel shows relatively low and stable activity levels. Additionally, AI-support initially attracted significant user attention when it was first launched, but activity quickly declined over time, followed by a slight increase after a period. Overall, users tend to rely more on human responses than on AI assistance, with higher user engagement in the user-support sub-channel. Over time, as the AI bot support system iterated and user habits evolved, participation in AI-support increased, particularly in the later stages of the data collection period. The following sections will further explore these insights using both quantitative and qualitative analyses.

\begin{figure}[ht]
  \centering
  \includegraphics[width=1\linewidth]{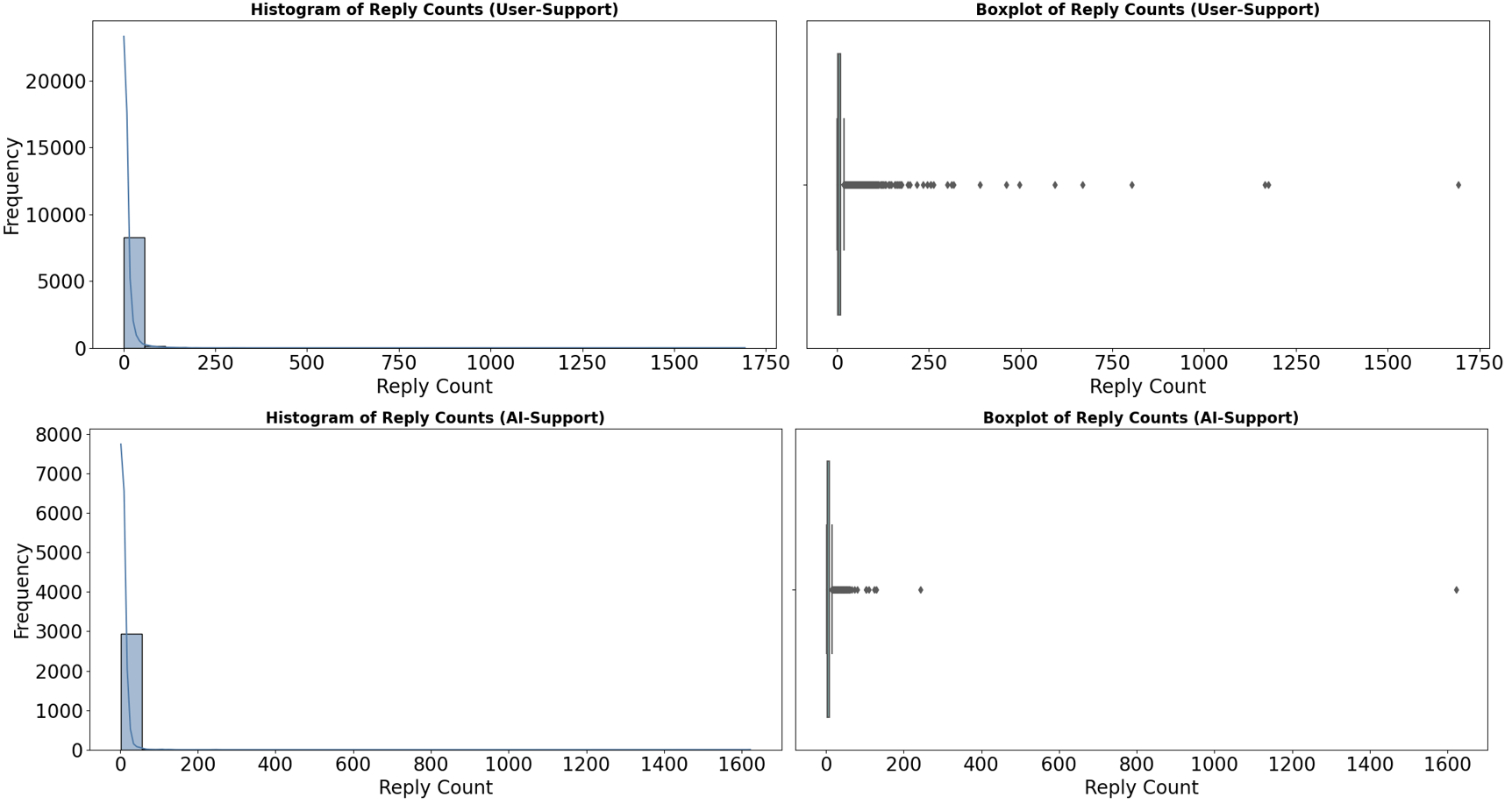}
  \caption{\textbf{Histograms of Reply Counts for the Two Sub-Channels (User-support on top, AI-support on the bottom). This figure shows the distribution of reply counts in the User-support and AI-support channels. It indicates that the range of reply counts in User-support threads is generally broader than that in AI-support threads.}}
  \label{fig.Histogram}
\end{figure}

The analysis of reply counts in Fig.~\ref{fig.Histogram} also reveals substantial differences between the two sub-channels. The histograms of reply counts show a skewed distribution for both the user support and AI support channels, but the user support channel demonstrates a much wider range of reply counts, with some threads accumulating over 1,700 replies. This contrasts sharply with the AI support channel, where most threads had fewer than 200 replies. The boxplots further illustrate this distribution, with the user support channel showing more frequent outliers, indicative of threads with significantly higher engagement compared to the median.


\subsection{\textbf{Results of User Profile Building}}

\begin{figure}[ht]
  \centering
  \includegraphics[width=.95\linewidth]{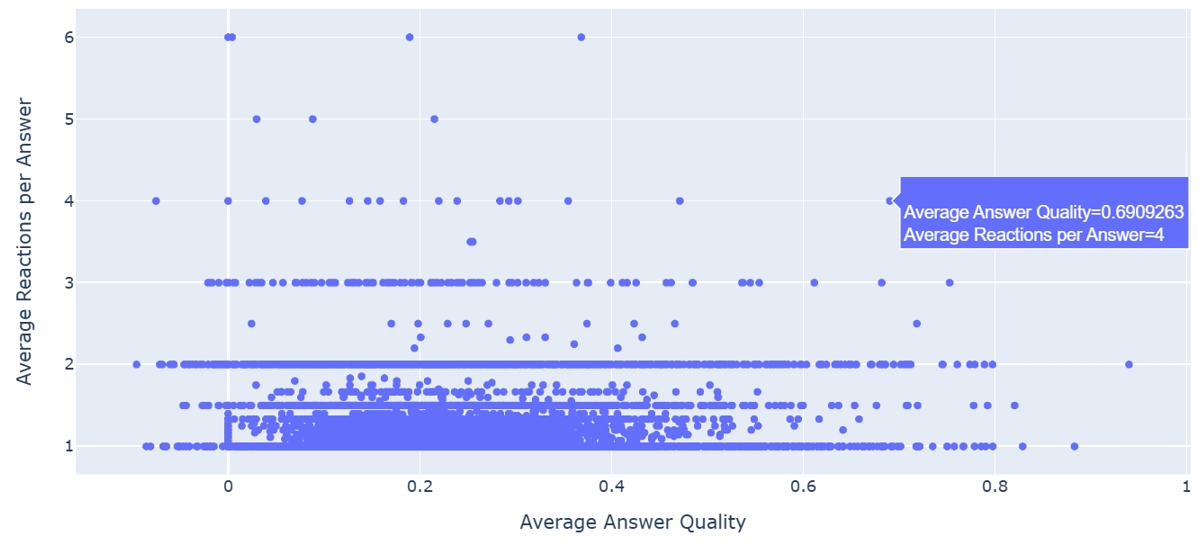}
  \caption{\textbf{The Relationship between Average Topical Alignment and the Average Number of Reactions per Response for User-support Sub-channel.}}
  \label{fig.Answer-Quality-vs-User-Reactions-Use}
\end{figure}

In the user-support sub-channel, as shown in Fig.~\ref{fig.Answer-Quality-vs-User-Reactions-Use}, most points are concentrated on the left side of the X-axis, with the average topical alignment mostly ranging between 0.0 and 0.4. This indicates that the average similarity of most user responses to the questions is relatively low, suggesting that while users did reply to the questions, their responses were not strongly related to the original queries. Additionally, although it could be theoretically assumed that higher-quality responses would receive more reactions, the graph shows that responses with higher quality did not significantly receive more reactions. In fact, some less relevant responses garnered a higher number of reactions. This suggests that the number of user reactions is not solely influenced by topical alignment but may also be affected by other factors (e.g., the topic of the response, users' interests, or emotions). We will further explain this finding using qualitative analysis methods in the following sections. The highlighted ``user'' shows a medium-to-high-quality response (quality close to 0.7), which also received a high number of reactions (four reactions). This may represent a highly relevant and well-received response.

\begin{figure}[ht]
  \centering
  \includegraphics[width=.95\linewidth]{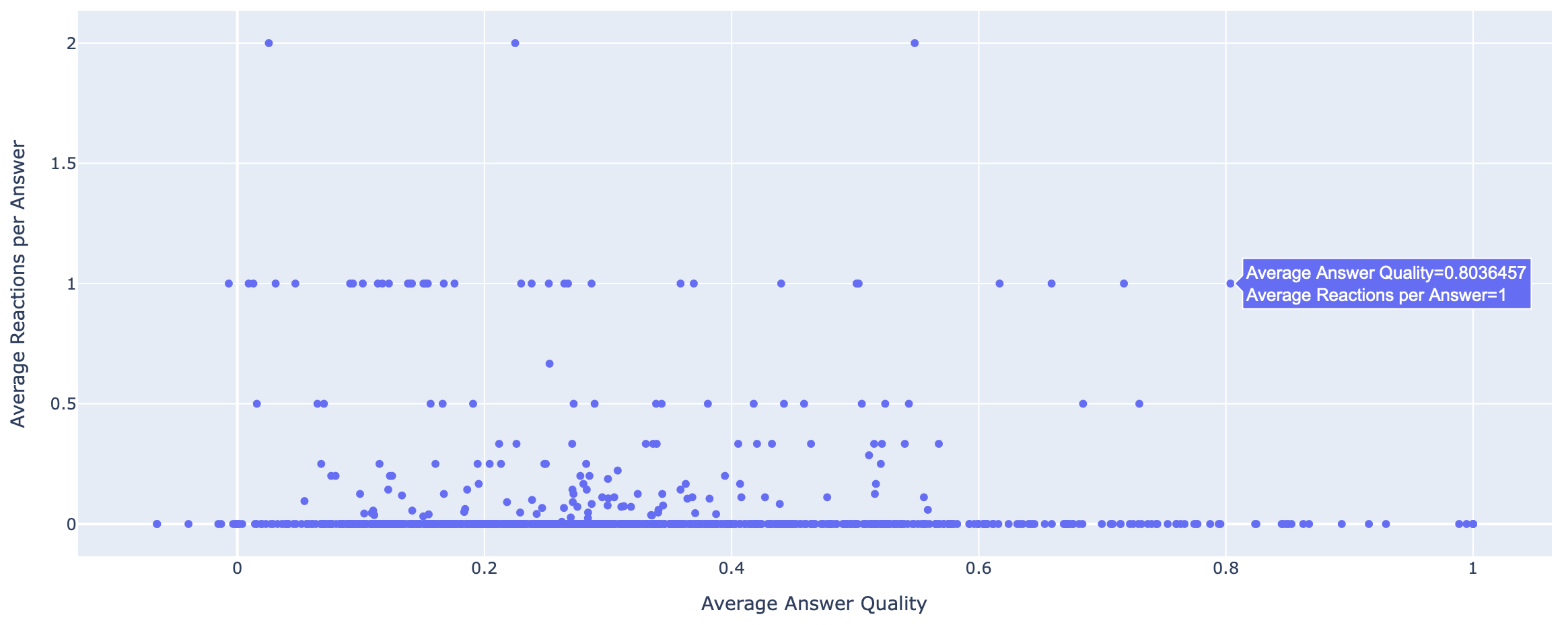}
  \caption{\textbf{The Relationship between Average Topical Alignment and the Average Number of Reactions per Response for AI-support Sub-channel.}}
  \label{fig.Answer-Quality-vs-User-Reactions-AI}
\end{figure}

In contrast, as shown in Fig.~\ref{fig.Answer-Quality-vs-User-Reactions-AI}, the AI-support sub-channel exhibits a different pattern: most points are concentrated near the bottom of the Y-axis, indicating that AI responses generally received very few reactions, often close to zero, regardless of their topical alignment. Although AI responses are more broadly distributed along the X-axis and include many relatively high-quality answers, this higher relevance did not translate into substantially more user reactions. Even the most reacted AI responses received at most two reactions, suggesting a weaker relationship between Topical Alignment and community feedback in the AI-support sub-channel. This pattern was further confirmed in the subsequent analyses.

\subsection{\textbf{Results of Sentiment Analysis and Cumulative Average Sentiment Calculation}}
We visualized the changes in cumulative average sentiment over time by plotting the cumulative average sentiment score against the reply number for each source. The visualization provided an intuitive view of how sentiment trends evolved across different sources.

The line plot was generated using the Seaborn library, with each source file represented by a unique line. To maintain clarity, we opted for a clean, focusing on the sentiment trajectory of each individual source. The x-axis represented the reply number, while the y-axis depicted the cumulative average sentiment score, ranging from 0 (Very Negative) to 4 (Very Positive).

From Fig.~\ref{fig.cumulative-average-sentiment-change-user} and~\ref{fig.cumulative-average-sentiment-change-ai}, we can observe that the cumulative average sentiment changes in both the user-support and AI-support sub-channels show significant fluctuations in sentiment scores, particularly during the early stages of responses. This indicates that the initial sentiments of many posts are highly random and may be extreme. However, as the number of replies accumulates, the volatility of the cumulative averages decreases in both user-support and AI-support. Despite this overall trend, there are subtle differences in sentiment fluctuations between the two sub-channels: (1) Conversations in user-support exhibit greater emotional investment, with more significant fluctuations in sentiment during the early stages. This reflects the emotional engagement users experience when solving problems through discussion. We acknowledge that the significant fluctuations observed in the early stages of the cumulative average sentiment are indeed a natural characteristic of this calculation method. However, as the number of replies increases, the cumulative average becomes more stable and less influenced by individual posts, thereby providing a clearer representation of the overall sentiment trend. This indicates that only a majority of later posts with more positive or neutral sentiments can shift the overall curve from a negative direction toward a more positive one, and vice versa. This suggests that (2) the cumulative sentiment trajectory is heavily influenced by the balance of sentiments in later interactions, reflecting the prevailing tone and resolution dynamics of the discussion over time. Hence, we concluded that the AI bot is more neutral and consistent in discussions. Although some conversations start with negative sentiment (scores <2), the final sentiment tends to be closer to neutral (score = 2) compared to user-support, where the final sentiment hovers around 1.5. In AI-support, users likely receive more functional but emotionally neutral feedback. The more stable nature of the sentiment curve suggests that users are more likely to accept AI-provided solutions or continue asking questions without needing much emotional investment. (3) AI-support does not exhibit the larger-scale and longer-duration discussions observed in user-support. This suggests that users have different expectations for AI-support and user-support. In AI-support, users are likely focused on obtaining quick answers without engaging in multiple rounds (or deeper, more emotional) discussions. In contrast, in user-support, resolving a problem might take longer and require the involvement of more participants.

\aptLtoX[graphic=no,type=html]{\begin{figure}[htbp]
    \centering
        \includegraphics[width=\linewidth]{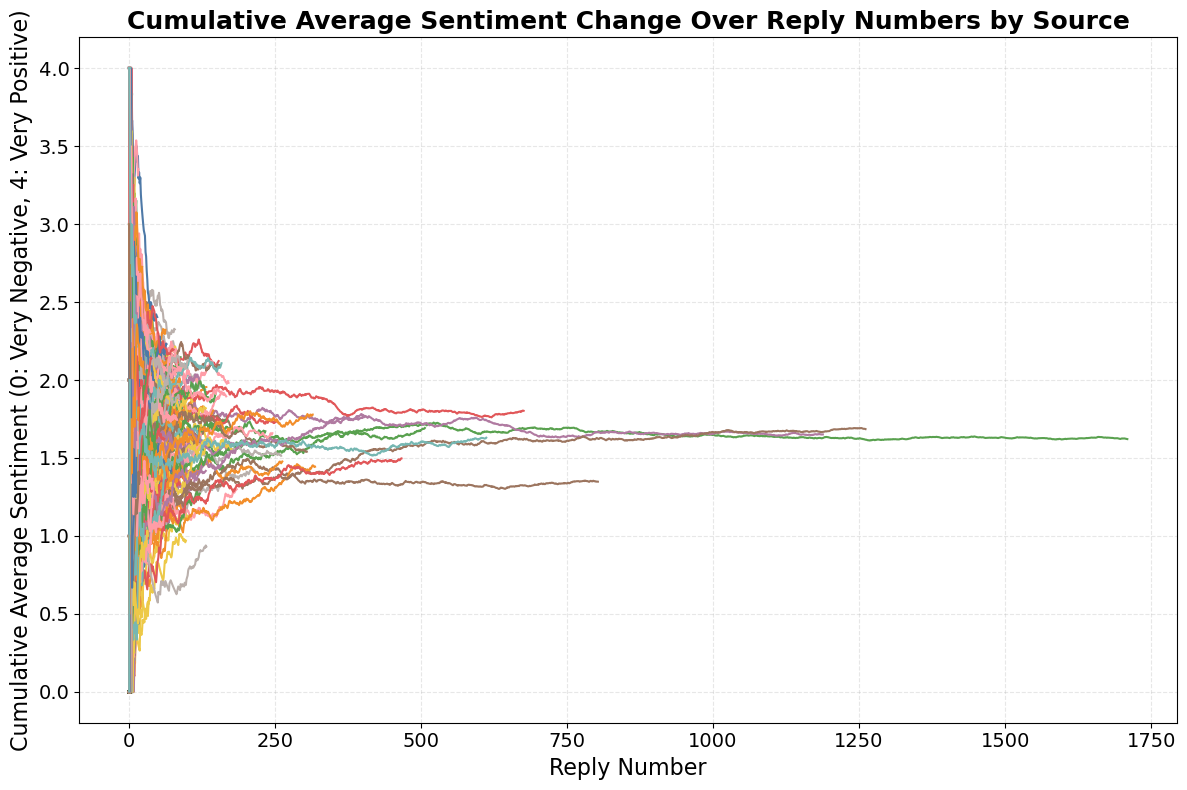}
        \caption{\textbf{Cumulative average sentiment change for user-support sub-channel over replies.}}
        \label{fig.cumulative-average-sentiment-change-user}
    \end{figure}
 
    \begin{figure}
        \centering
        \includegraphics[width=\linewidth]{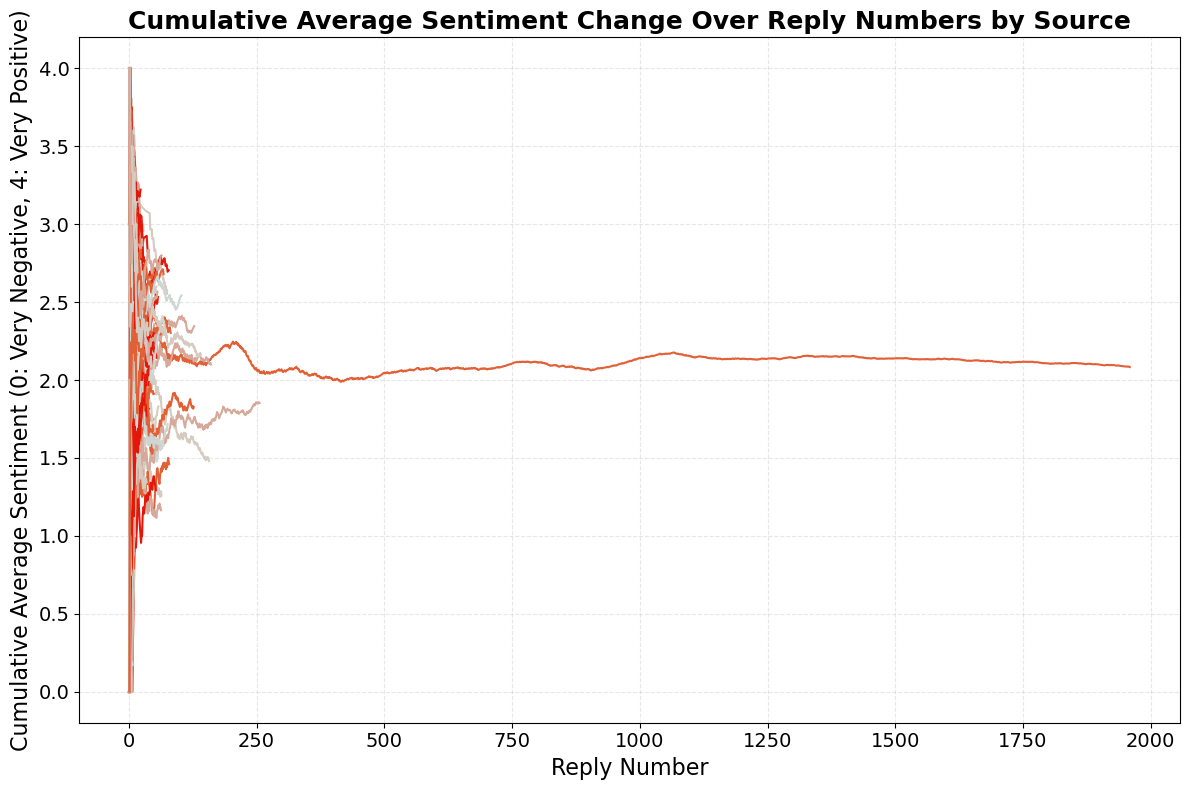}
        \caption{\textbf{Cumulative average sentiment change for AI-support sub-channel over replies.}}
        \label{fig.cumulative-average-sentiment-change-ai}
\end{figure}}{\begin{figure}[htbp]
    \centering
    \begin{minipage}{0.48\textwidth}
        \centering
        \includegraphics[width=\linewidth]{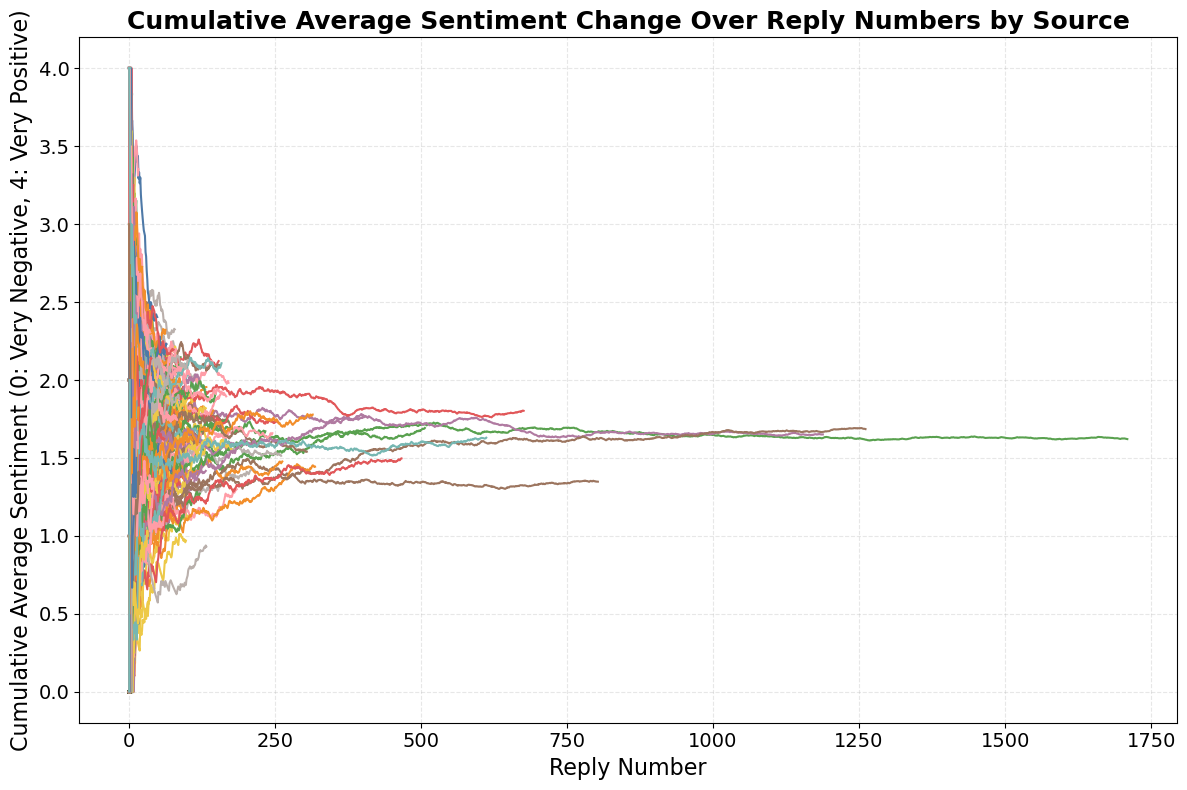}
        \caption{\textbf{Cumulative average sentiment change for user-support sub-channel over replies.}}
        \label{fig.cumulative-average-sentiment-change-user}
    \end{minipage}
    \hfill
    \begin{minipage}{0.48\textwidth}
        \centering
        \includegraphics[width=\linewidth]{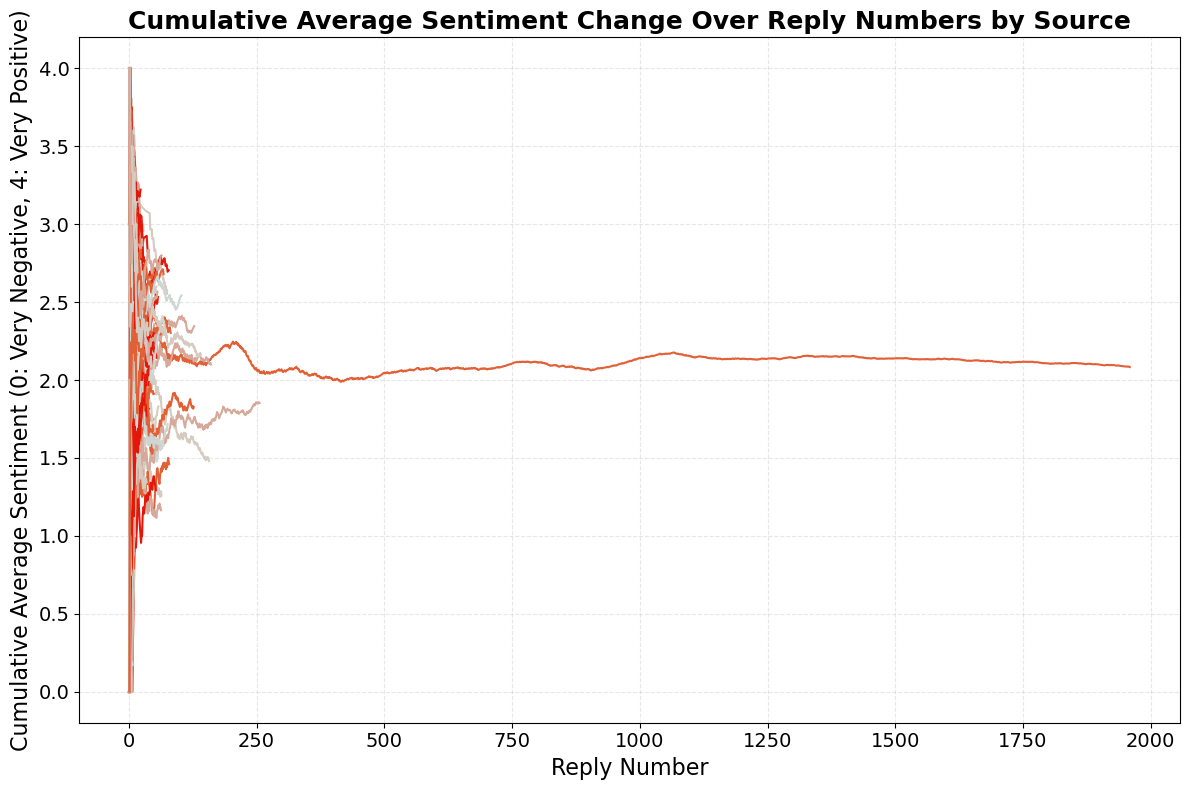}
        \caption{\textbf{Cumulative average sentiment change for AI-support sub-channel over replies.}}
        \label{fig.cumulative-average-sentiment-change-ai}
    \end{minipage}
\end{figure}}

\begin{figure}[!h]
  \centering
  \includegraphics[width=.95\linewidth]{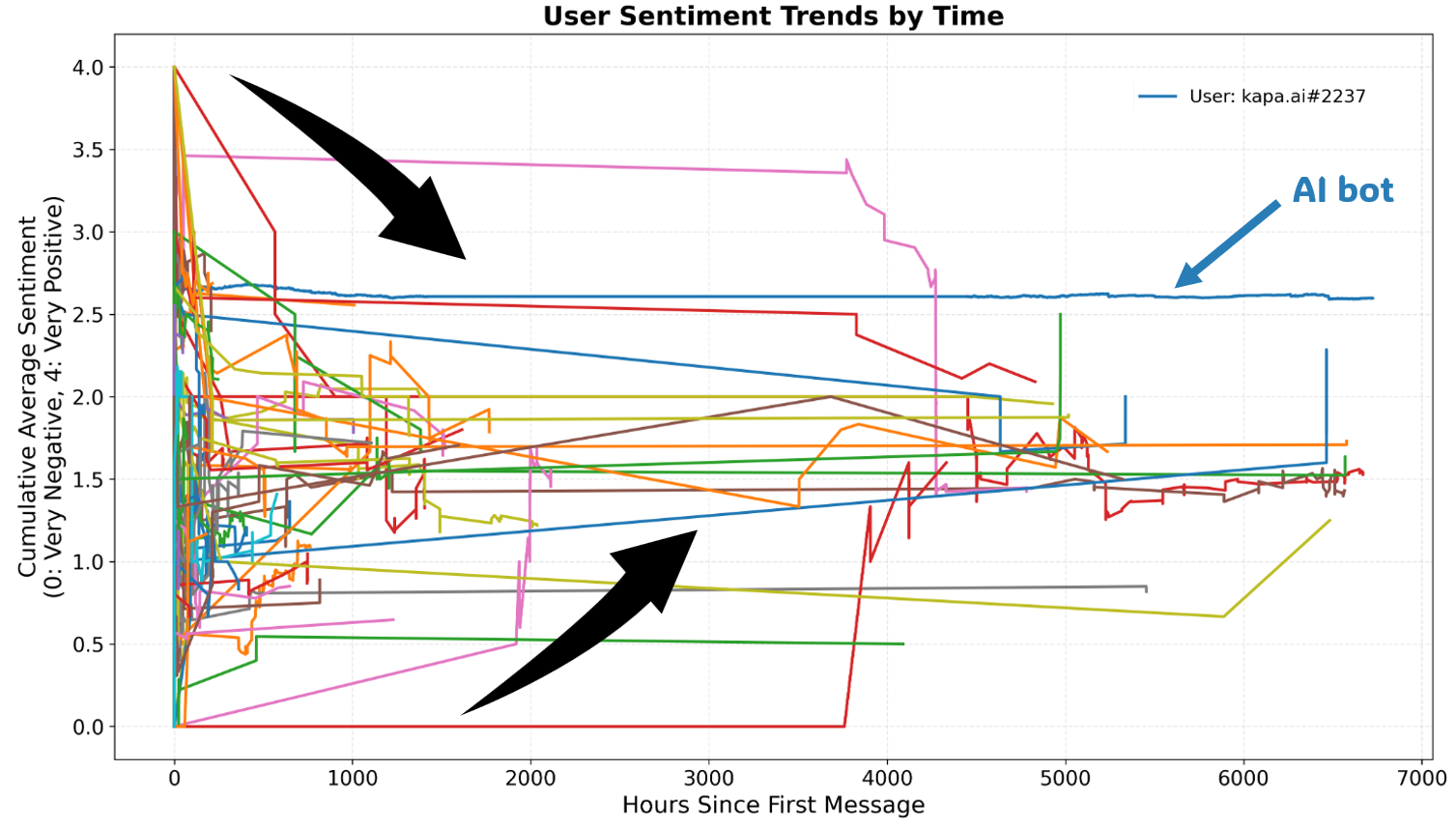}
        \caption{\textbf{User Sentiment Trends by Time. This figure shows individual user sentiment trends over time within AI-support sub-channel. The data indicates that sentiment gradually becomes neutral as users interact with the AI bot, especially those initially expressing strong emotions. AI responses consistently remained in a neutral to slightly positive state.}}
        \label{fig.User-Sentiment-Trends-by-Time}
\end{figure}

We also create Fig.~\ref{fig.User-Sentiment-Trends-by-Time} with specific users by time changes for AI-support sub-channel. Specifically, we presented the trend of the average sentiment score of each user's post over time. We found that the AI bot's responses consistently remained in a neutral to slightly positive state, while user sentiment gradually shifted towards neutrality as the conversation with the AI bot progressed, particularly for posts that started with more extreme emotions. In this case, we only displayed the top 100 users with the highest number of interactions (including an AI bot, \textit{kapa.ai\#2237}). 

\begin{figure}[!h]
  \centering
  \includegraphics[width=.95\linewidth]{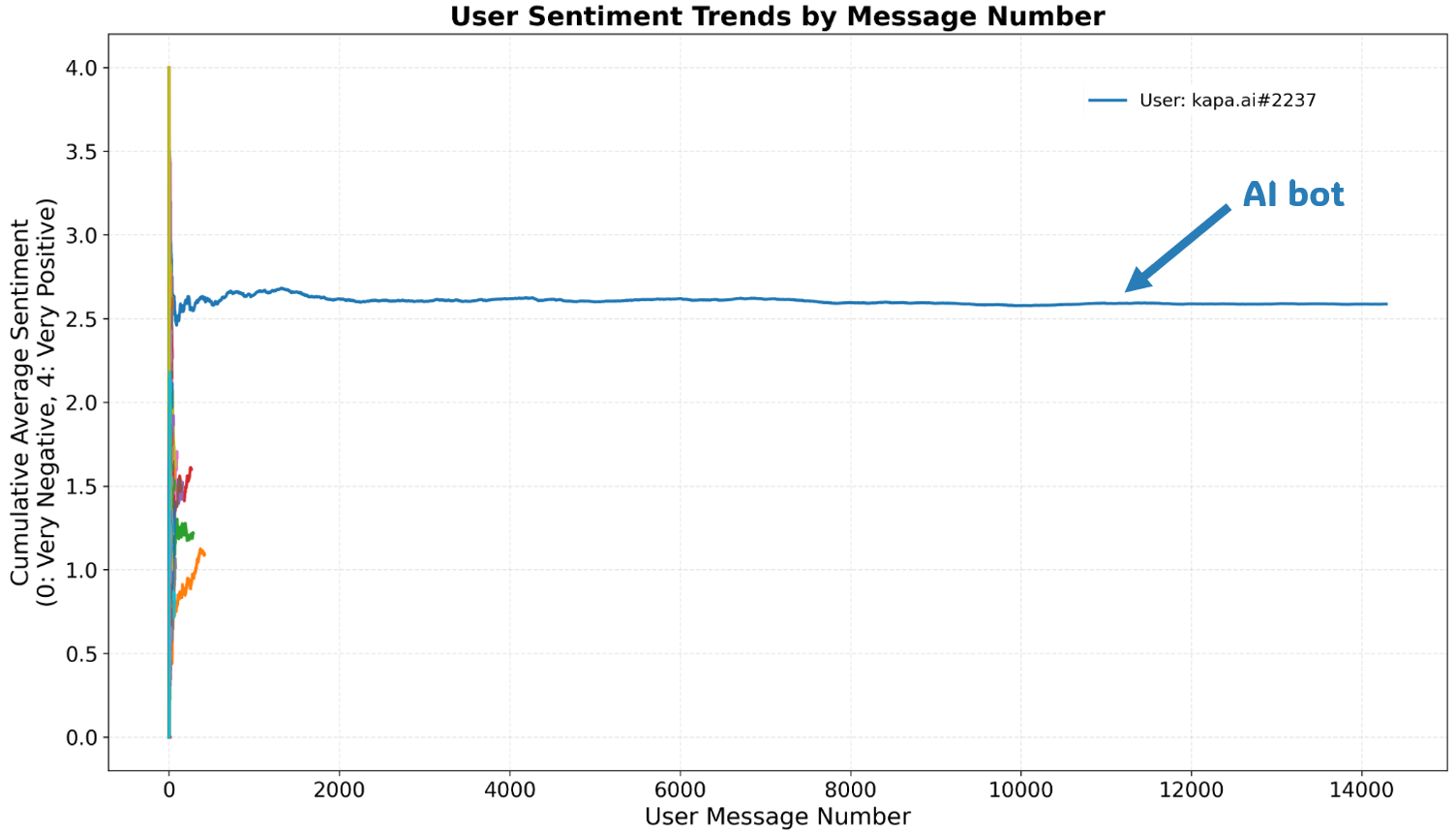}
        \caption{\textbf{User Sentiment Trends by Message Number. The number of messages from users is significantly lower than the AI bot. As same in Fig.~\ref{fig.User-Sentiment-Trends-by-Time}, AI responses consistently remained in a neutral to slightly positive state. Users’ sentiment trends also change slightly toward neutral as the responses increase.}}
        \label{fig.User-Sentiment-Trends-by-Message}
\end{figure}

\begin{figure}[!h]
  \centering
  \includegraphics[width=.98\linewidth]{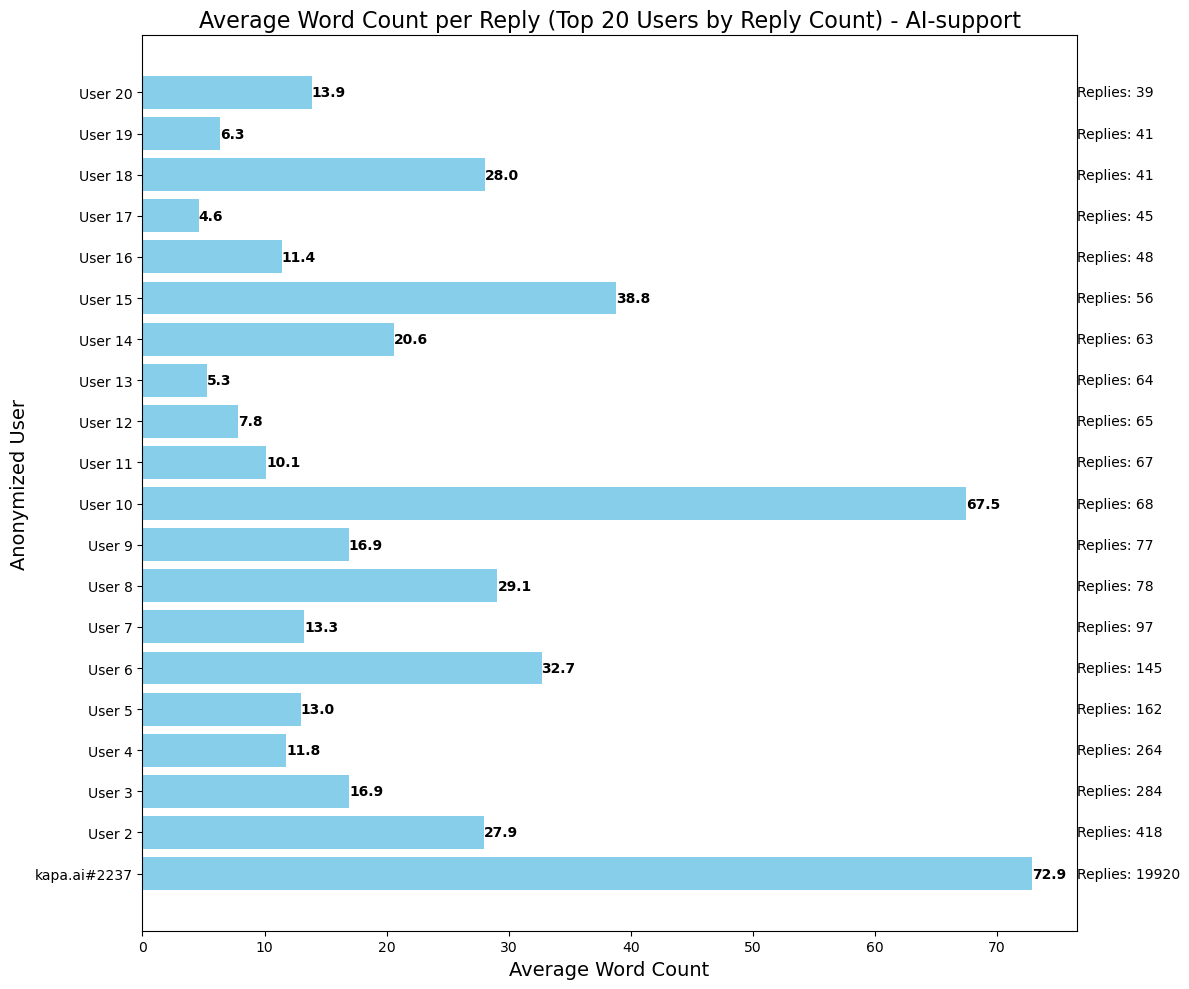}
  \caption{\textbf{Average Word Count per Reply (Top 20 Users by Reply Count) for AI-support, ranked by number of replies (the rank 1 user is AI bot with 19,920 replies). AI responses tend to have higher word counts due to the structured response format ($\approx$ 72.9 words), while users’ responses vary in length (from 4.6 words to 67.5 words).}}
        \label{fig.Average-word-count-AI-support}
\end{figure}

\begin{figure}[!h]
  \centering
  \includegraphics[width=.95\linewidth]{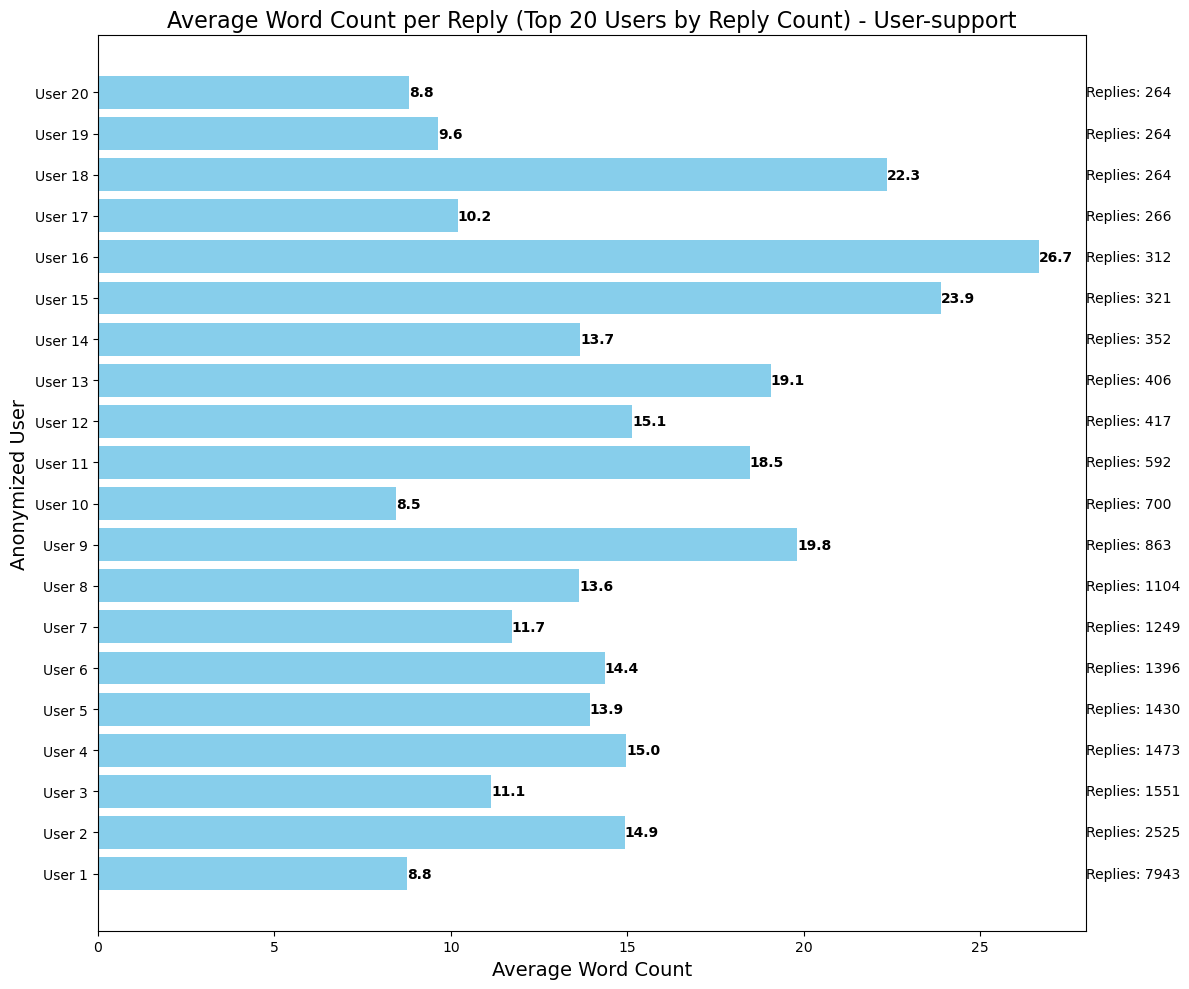}
        \caption{\textbf{Average Word Count per Reply (Top 20 Users by Reply Count) for User-support, ranked by number of replies (the rank 1 user has 7.943 replies). The responses vary in length but have replied more words (the average word count $\times$ the number of replies) compared to user's replies in the AI support sub-channel.}}
        \label{fig.Average-word-count-user-support}
\end{figure}

On the other hand, a similar trend can be observed in Fig.~\ref{fig.User-Sentiment-Trends-by-Message} (User Sentiment Trends with Response Count). Additionally, the number of responses from users in a single question is significantly lower than that of the AI bot. This is because, in this Discord sub-channel, the AI bot splits a complete reply into three separate messages to avoid exceeding the maximum content length for a single message. Fig.~\ref{fig.Average-word-count-AI-support}, which represents the ``average word count per reply'' for the AI-support sub-channel, shows that the AI bot's responses tend to be much longer on average compared to users' initial questions or follow-up inquiries. Additionally, in the ``User Support'' sub-channel, the average length of replies per user exhibits a trend similar to that observed in the ``AI Support'' sub-channel (Fig.~\ref{fig.Average-word-count-user-support}). However, the total reply length of the most active human users (the top 19 human users $\times$ the average word count $\times$ the number of replies = 6,794,120 words) is significantly higher than in the ``AI Support'' sub-channel (where the top 19 users collectively contributed approximately 797,659.8 words).

\subsection{\textbf{\rev{Results of} Advanced Network Analysis}}

\begin{figure}[htbp]
  \centering
  \includegraphics[width=0.95\linewidth]{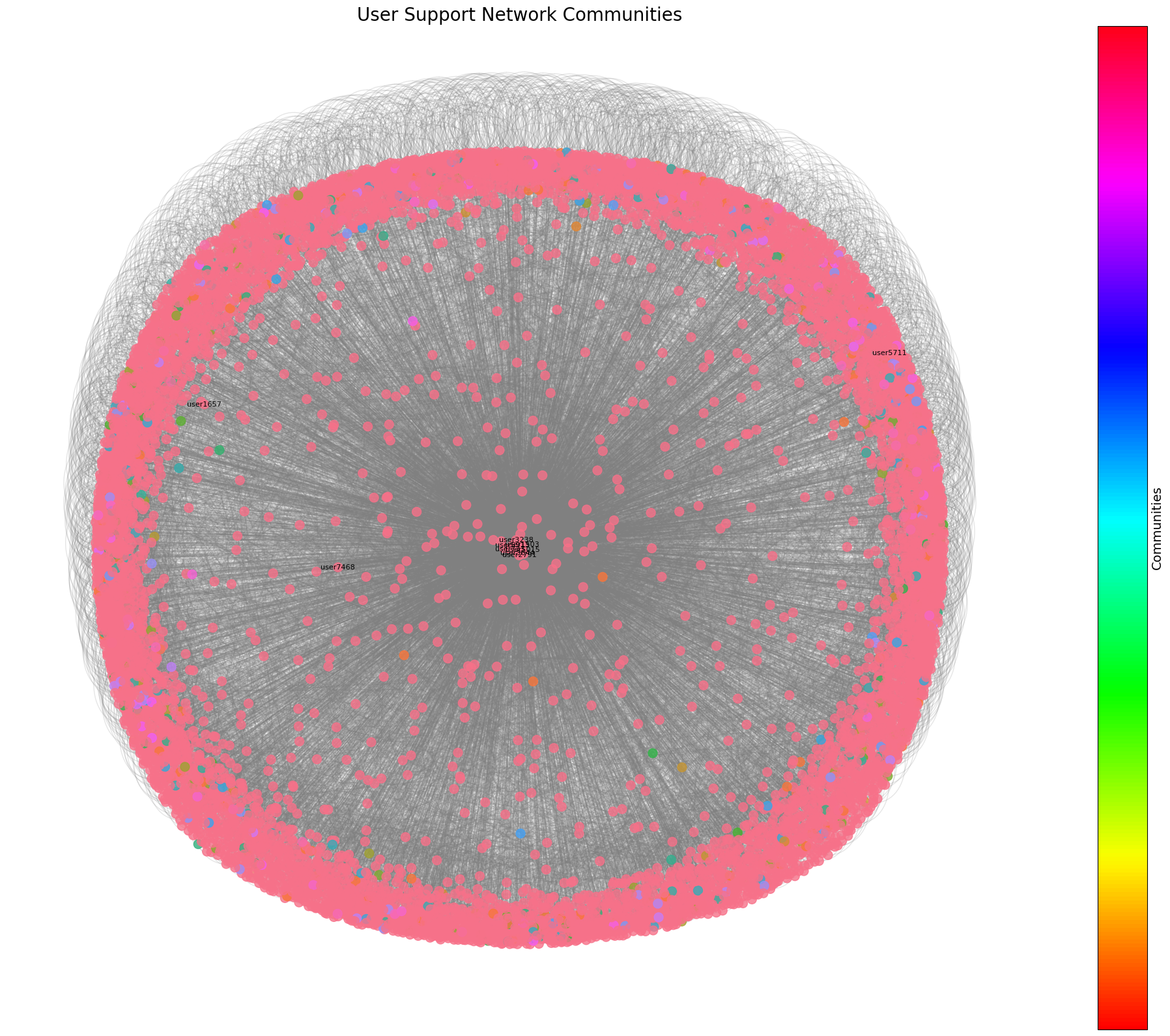}
  \caption{\textbf{Discord Community Network for User-support Sub-channel. This figure shows the community network for the user support channel, highlighting clusters of users that interact frequently. The user support network consists of numerous distinct communities which illustrates that the users are engaged in discussions on multiple topics. The variety of colors shows that the users are not engaging in a single topic but are instead participating in different conversations. Users who are listed in the central nodes indicate that they act as key responders and help the community across multiple topics.}}
  \label{fig.Discord-Community-Network-User}
\end{figure}

\begin{figure}[htbp]
  \centering
  \includegraphics[width=0.95\linewidth]{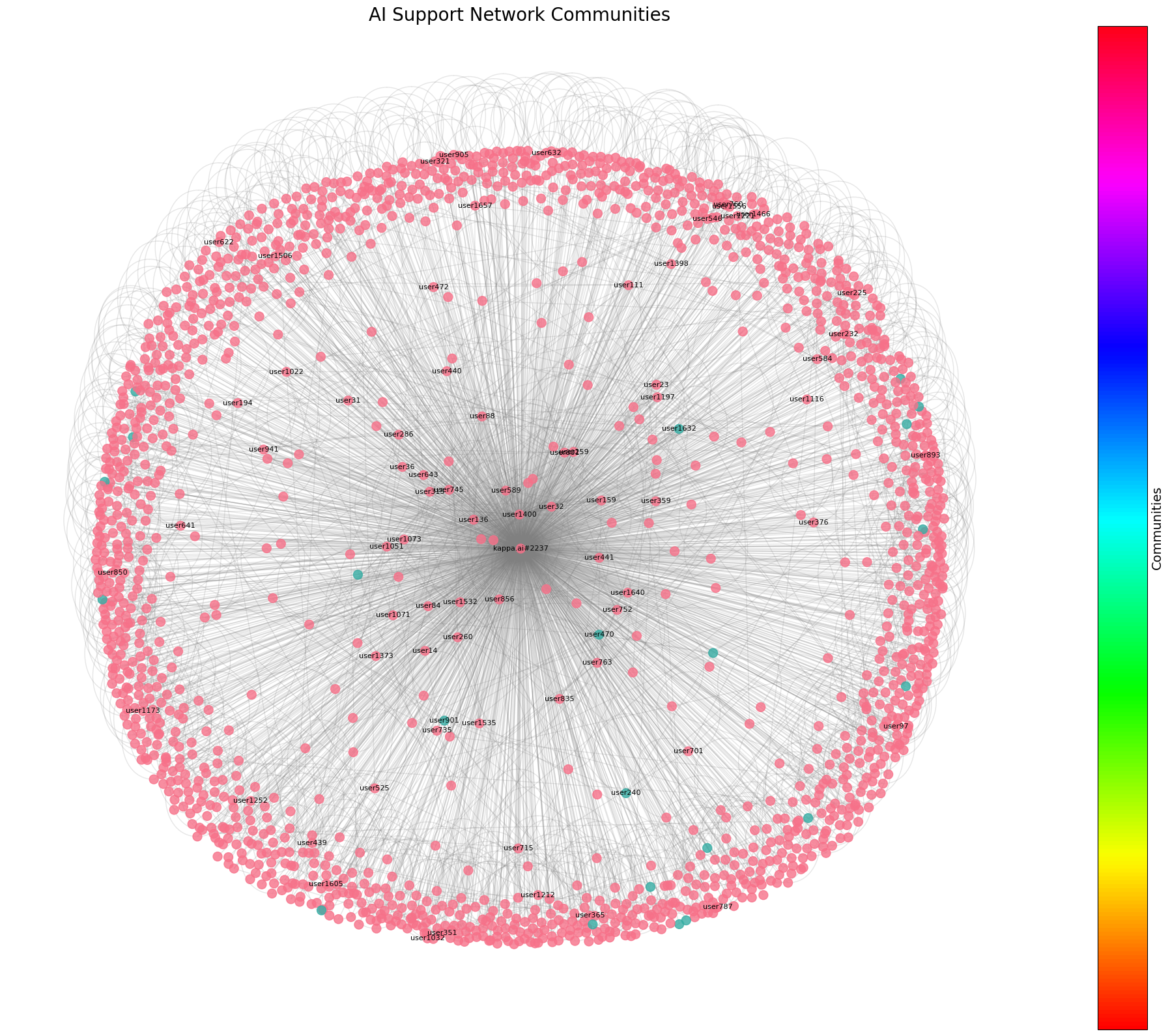}
  \caption{\textbf{Discord Community Network for AI-support Sub-channel. This figure shows the community network for the AI support channel, highlighting how users that interact frequently in AI-support Sub-channel. Compared to the user support channel, the AI support community shows fewer interactions but focuses on AI-centered interaction. This concludes that the AI support channel users seek answers from AI rather than from other users which leads to a lack of collaboration discussion.}}
  \label{fig.Discord-Community-Network-AI}
\end{figure}

In these visualizations, we highlighted specific nodes, including \textit{kappa.ai\#}-\textit{2237} and other highly connected users (with a degree greater than 5) in the case of AI-support sub-channel and top 10 connected users in the case of user-support sub-channel, to emphasize key actors within each network. This approach allowed us to visually compare the interaction dynamics between human-powered and AI-powered support systems. Fig.~\ref{fig.Discord-Community-Network-User} shows the User-support network communities, and Fig.~\ref{fig.Discord-Community-Network-AI} shows the AI-support network communities.

\subsubsection{\textbf{User-Support Network Communities vs. AI-Support Network Communities}}

The user-support network consists of numerous distinct communities, as indicated by the different node colors. This suggests that users are engaged in discussions on multiple topics or themes. Each community likely corresponds to a subset of users focused on specific issues or topics, which are relatively independent of other groups. For instance, one community might concentrate on troubleshooting software installation problems, while another could focus on feature usage or bug fixes. The variety of colors shows the diversity in the discussions taking place, indicating that users are not engaging with a single, central topic but are instead participating in several different conversations. This community-based structure highlights the collaborative nature of user support, where different users contribute their knowledge and experience to resolve a wide range of issues. Additionally, the central nodes with high degree centrality suggest that some users act as ``key responders'' or ``super users'', helping bridge different communities, facilitating the flow of information across multiple topics and resolving issues within the community.

In contrast, the AI-support network shows far fewer distinct communities, with much of the interaction centered around the AI bot, \textit{kappa.ai\#2237}. This centralized structure reflects a more transactional interaction model, where users seek specific answers from the AI rather than engaging in collaborative discussions with one another. The relative lack of community diversity compared to the user-support network suggests that AI interactions are generally more focused on singular, isolated questions rather than a broad array of themes. Since most users interact directly with the AI bot and not with each other, there is limited opportunity for different topic-based communities to form. This results in a more homogeneous network structure, where users are more likely to receive quick solutions from the AI without the extended back-and-forth discussions or topic differentiation seen in the user-driven support network.

Overall, the user-support network fosters a more collaborative, multi-user environment where participants can engage in rich, ongoing discussions. Meanwhile, the AI-support network operates more efficiently but lacks the interactive and community-building dynamics that characterize human-to-human support.

%
%
%
%

%

\section{Findings of Qualitative Analysis for RQ2. How do users interact with these different support systems? and What usage patterns emerge when users engage with these support systems?}
\subsection{Question and Response Mode in AI-support Sub-Channel}
\subsubsection{Questions posted in AI-support Sub-channel}
To explore RQ2, we examined how users interact with the AI-support and user-support sub-channels, uncovering key differences in the nature and style of interactions. The first difference is in the style of initial questions. In the AI-support sub-channel, users tend to ask simple, straightforward questions, such as ``How to'' questions, login issues, account problems, server-related errors, and queries about specific terms or concepts. Here are some examples:
\begin{quote}
``\textit{how i turn string into float array}'' [AI-support]\\
``\textit{i have a `new user' rank in vrchat, and i would like to upload my own avatar. how do i do that?}'' [AI-support]\\
``\textit{How do I change my costume height on pc?}'' [AI-support]\\
``\textit{Help i forgot my emaill pass word amd im out of uses the combination codes instead so will I ever get my VR account back}'' [AI-support]\\
\end{quote}
These questions reflect a preference for fast and direct responses. Rather than giving lengthy descriptions, users often focus on specific problems they need to solve quickly. In other cases, users may not ask a question directly but rather describe an issue or error they encountered, hoping the AI can identify the problem and provide a solution:
\begin{quote}
``\textit{On my steam account logging through there i kept getting ERROR 103?}'' [AI-support]\\
``\textit{Avatars are downloading incredibly slow and i have gig download speed}'' [AI-support]\\
``\textit{\seqsplit{Could not authenticate - Not logged in UnityEngine.Debug:LogError (object,UnityEngine.Object)}}'' [AI-support]\\
``\textit{Vrchat infinite loading startup} [AI-support]''
\end{quote}

Interestingly, users in the AI-support sub-channel often ask broader questions, which tend to be more vague, such as: ``\textit{How do I create avatars?}'' [AI-support]. These broad questions indicate that some users may be more focused on learning about general features rather than addressing a specific problem. This type of inquiry is less common in the user-support sub-channel, where users are typically more focused on resolving specific, immediate issues.

\subsubsection{Responses in AI-support Sub-channel}
\begin{figure*}[!h]
  \centering
  \includegraphics[width=.95\linewidth]{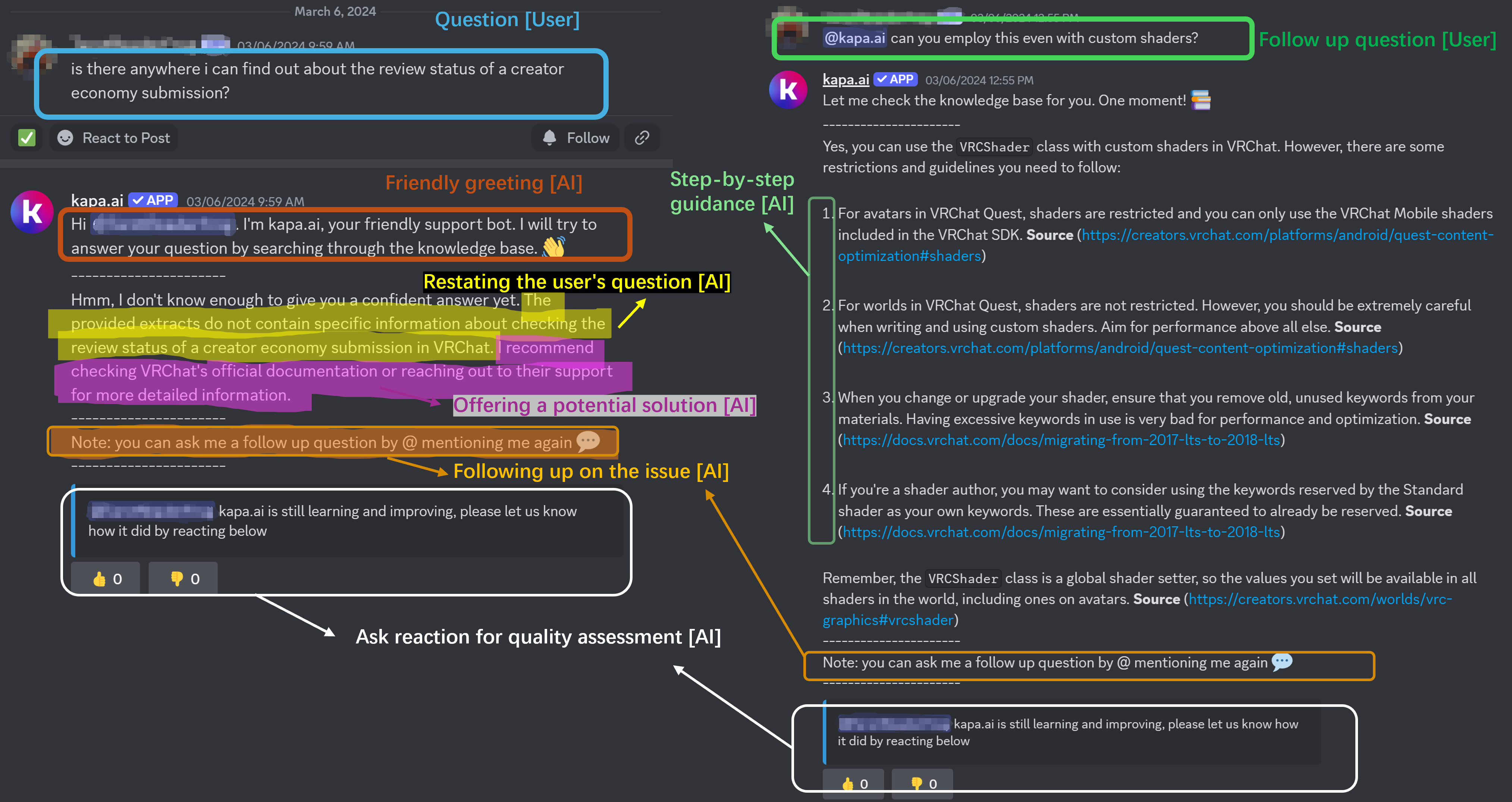}
  \caption{\textbf{Examples of Conversations in AI-support Sub-channel. The user asked where the user can find out about the review status of a creator economy submission (marked by light blue box). AI responded with a friendly greeting statement (marked by red box), restated the user’s question (marked by yellow marker), and gave a potential solution by recommending checking VRChat’s official documentation or reaching out to their support (marked by pink marker). The AI ended with a statement of following up on the issue and asking for a reaction for quality assessment (marked by orange marker). The user asks the follow-up question about whether the user can employ this even with custom shaders (marked by green box). The AI responds with step-by-step guidance but without restating the question like the first time (marked by light green). At the ends of each response of AI, it has a statement of following up on the issue and asking for a reaction for quality assessment (marked by white box).}}
  \label{fig.conversation-ai-support}
\end{figure*}

The AI bot's responses in the AI-support sub-channel follow a highly structured and formulaic pattern. The responses typically include several key elements: (1) Greeting: Starting with a friendly greeting (using a fixed format), (2) Empathy: In cases where users mention difficulties, the AI expresses empathy, for instance, ``\textit{Sorry to hear about the issue}''. This step is optional and depends on the content of the user's query. (3) Restating the Question: The AI restates the user's question to confirm understanding. (4) Providing a Solution: The AI suggests a potential solution to the user's problem. (5) Step-by-Step Guidance: For more complex issues, the AI provides detailed, step-by-step instructions to help users resolve the problem. (6) Asking for Clarification: If the AI cannot fully understand the question, it encourages users to clarify their questions. This iterative interaction process is especially common in technical questions involving code or troubleshooting. (7) Reaction Request: The AI asks for user feedback or a rating on the quality of the response to learn and improve future interactions. Fig.~\ref{fig.conversation-ai-support} shows a conversation between a user and the AI bot in the AI-support channel. 

This structured framework helps standardize AI support responses, but it also reveals some limitations, particularly in addressing more nuanced or contextual issues that might benefit from human judgment.

\subsection{Question and Response Mode in User-support Sub-Channel}

\subsubsection{Questions posted in User-support Sub-channel}
In the user-support sub-channel, interactions are noticeably more collaborative. Users often ask detailed questions, typically providing additional background information, screenshots, or error codes to help others better understand their issues (Figure~\ref{fig.coherence-scores-user}). Unlike the typically brief and direct questions found in the AI-support sub-channel, users in the user-support sub-channel tend to elaborate on their issues, with questions often being longer and more descriptive. This is because users frequently need to explain complex technical or in-game problems. Additionally, users may include existing solutions or methods they have already tried in their user-support posts, or even provide personal context or further explanation about why they are asking the question. Also, when posing questions, users also tend to be more humble compared to those in the AI-support sub-channel.
\begin{quote}
``\textit{To provide additional context: [more details] ... Forgive the long post but we've hit a point where we think it best to ask for any possible outside help to narrow down the cause of the crashing and find a work around to fix it, if it's something we can fix...} [User-support]''
\end{quote}

\begin{figure*}[!h]
  \centering
  \includegraphics[width=.95\linewidth]{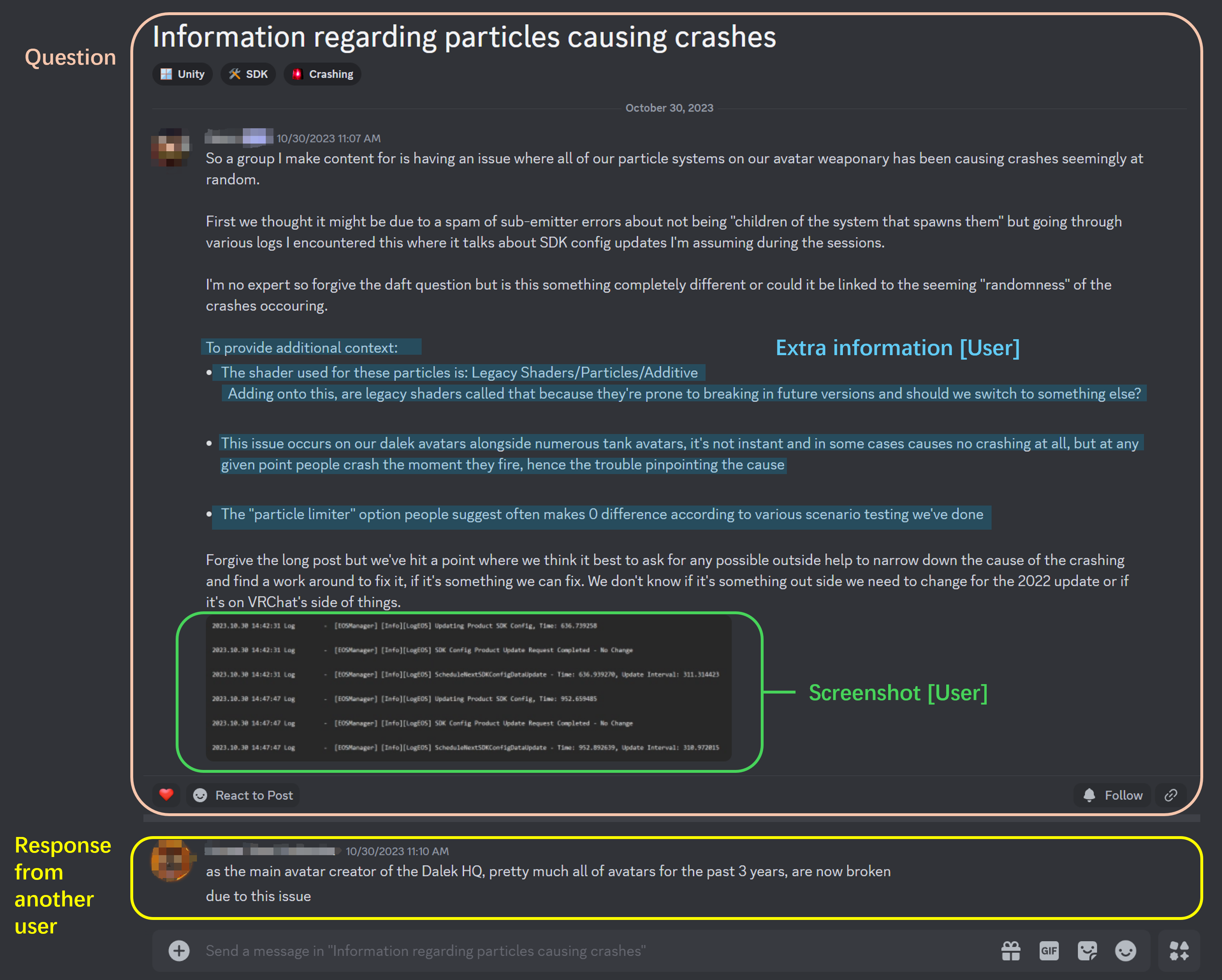}
  \caption{\textbf{Example of a conversation in user-support sub-channel. The light pink box contains the question posed by the user, within which the green box highlights a screenshot reference provided by the user to illustrate the issue. The questioner has described the encountered problem in detail (a random crash issue with all particle systems on an avatar weapon) and included additional background information about the problem (marked in light blue on the screenshot). The content in the yellow box consists of responses from other users, with one reply offering an answer from the perspective of an ``avatar'' creator.}}
  \label{fig.conversation-user-support}
\end{figure*}

Another example shows different attitudes and descriptions in different channels. Users may describe their problem more thoroughly in one channel and ask concisely in the other. User Anonymous 2 expressed urgency in the user-support channel: \begin{quote}
    ``\textit{Hello! I am a [Country Name] user playing VRChat! I sent my inquiry 5 times, but I didn't get a reply, so I sent it to Discord! I deleted the email linked to my VRChat account and am unable to change my password! What should I do? I can authenticate myself!}''
\end{quote} 
While in the AI-support channel, they briefly stated:
\begin{quote}
    ``\textit{I'm trying to upload a map from VRChat. However, I have forgotten my account password and need to change it, but I have deleted the email linked to VRChat and am unable to change my password. Can you change my email?}''
\end{quote}

On the other hand, we would like to particularly highlight the advantages of multimodal interaction in user support, especially when compared to AI support where only text interactions are possible. From the data, we found that although, overall, problem descriptions in user support are longer and more detailed, there still exists a large number of issues described only with a ``single sentence.'' However, unlike in AI support where the AI bot has to make additional inquiries in such cases, in user support, users can convey more information through multimedia files (screenshots or videos as attachments). Based on the rich content contained in images, users can describe the problems they encounter using only a few characters. This method also more intuitively showcases the issues they face, as some problems may be difficult to describe through natural language. Fig.~\ref{fig.user-support-screenshot1} provides examples of users asking questions using minimal text supplemented with multimedia files. Additionally, in an ongoing users' Q\&A dialogue, users may provide more supplementary information through screenshots to receive more targeted assistance (as shown in Fig.~\ref{fig.user-support-screenshot2}).

\aptLtoX[graphic=no,type=html]{\begin{figure}[htbp]
    \centering
        \includegraphics[width=\linewidth]{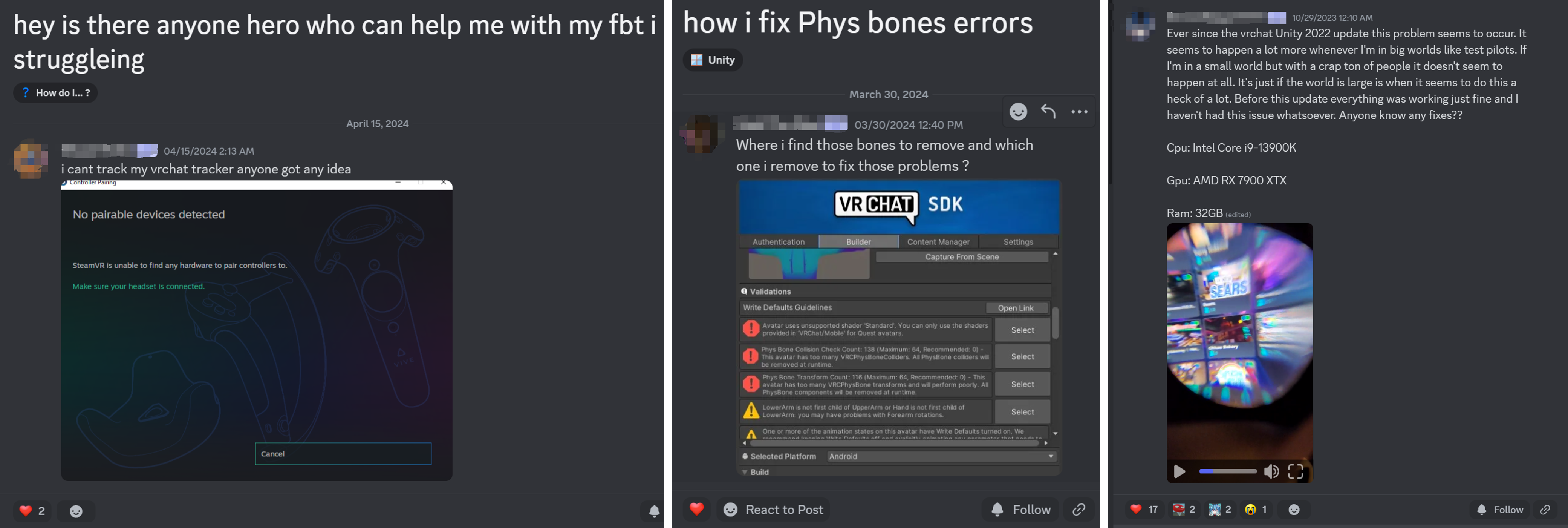}
        \caption{\textbf{Example of Users Use Screenshots to Give Extra Explanations for Questions. The question on the left shows the error page saying that no pairable devices were detected which is related to the question. The middle question shows a page where the user is having trouble with finding bones to remove. The question on the right shows the video of the problem that the user is having.}}
        \label{fig.user-support-screenshot1}
    \end{figure}

    \begin{figure}
        \centering
        \includegraphics[width=\linewidth]{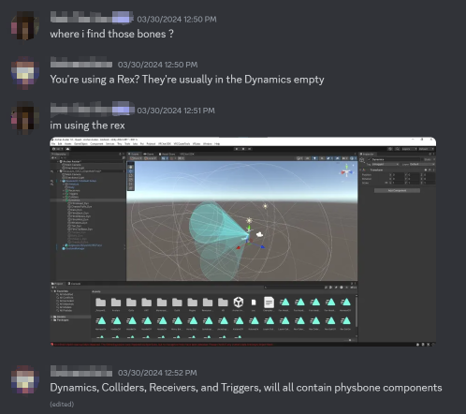}
        \caption{\textbf{Example of Users Use Screenshot to Give Additional Details in a Follow-up Conversations. Specifically, in this example, the questioner first posed the question, followed by the respondent asking for clarification. The questioner then provided a screenshot of the software (Rex) as an example of the issue encountered. Based on this screenshot, the respondent offered a more detailed solution.}}
        \label{fig.user-support-screenshot2}
\end{figure}}{\begin{figure*}[htbp]
    \centering
    \begin{minipage}{0.95\textwidth}
        \centering
        \includegraphics[width=\linewidth]{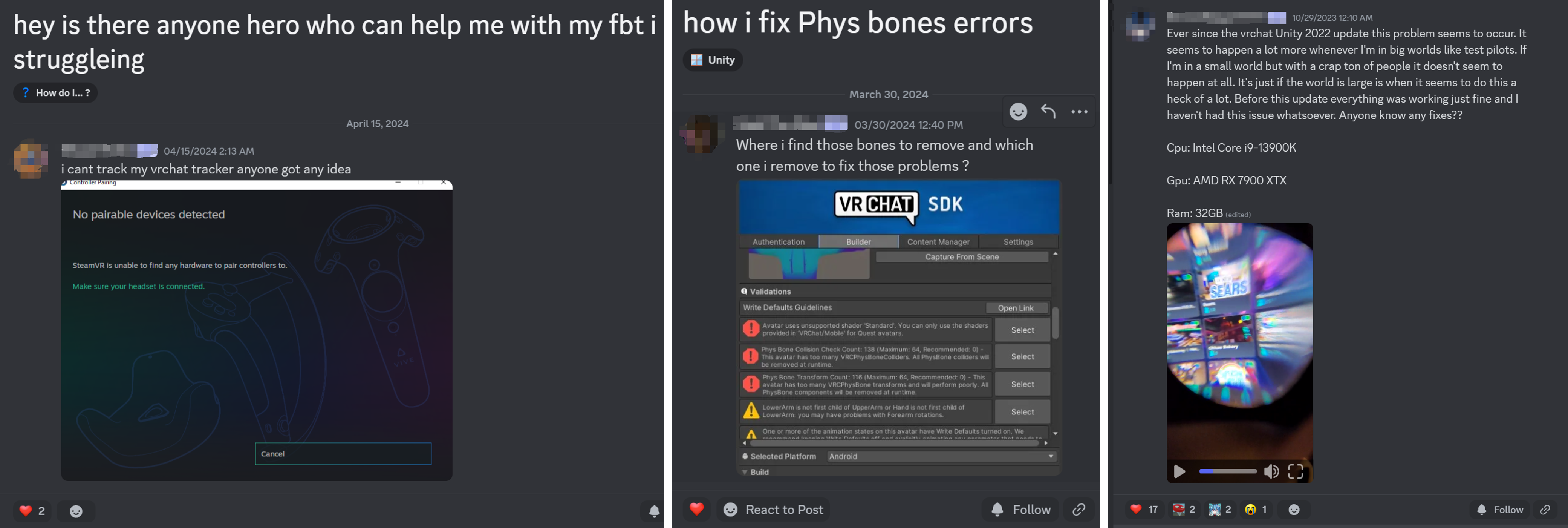}
        \caption{\textbf{Example of Users Use Screenshots to Give Extra Explanations for Questions. The question on the left shows the error page saying that no pairable devices were detected which is related to the question. The middle question shows a page where the user is having trouble with finding bones to remove. The question on the right shows the video of the problem that the user is having.}}
        \label{fig.user-support-screenshot1}
    \end{minipage}
    \hfill
    \begin{minipage}{0.7\textwidth}
        \centering
        \includegraphics[width=\linewidth]{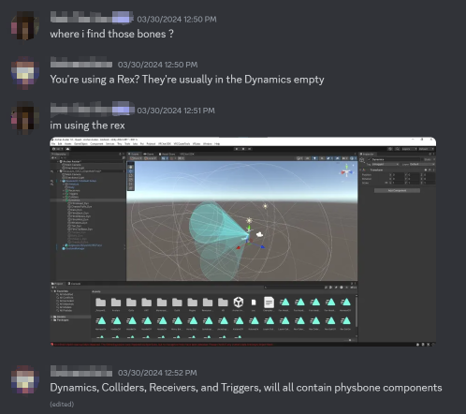}
        \caption{\textbf{Example of Users Use Screenshot to Give Additional Details in a Follow-up Conversations. Specifically, in this example, the questioner first posed the question, followed by the respondent asking for clarification. The questioner then provided a screenshot of the software (Rex) as an example of the issue encountered. Based on this screenshot, the respondent offered a more detailed solution.}}
        \label{fig.user-support-screenshot2}
    \end{minipage}

\end{figure*}}

\subsubsection{Responses in User-support}

In the user-support sub-channel, responses are primarily characterized by collaborative problem-solving. Specifically, while both our quantitative analysis results and previous examples indicate that questions in user-support are generally more detailed than those in AI-support, we still found some cases where questions lacked sufficient information, making them difficult to fully understand. In such cases, respondents often ask the original poster (OP) about any previous attempts they’ve made and gradually provide potential solutions through multiple rounds of dialogue. Additionally, in a single question thread, as the conversation evolves, the participants are not limited to just the original questioner and responder. Different users may join the conversation at various points, illustrating the more open and dynamic discussion trends in the user-support sub-channel. For example, in the following case, two repliers (A and B) each replied to the OP's initial question and a follow-up question, respectively. At the same time, we can observe that when focused on discussing solutions, some negative emotions are diffused over the course of multiple rounds of dialogue, without the need for the AI bot's empathy strategies. This may be attributed to community norms and moderator oversight. However, this could also make conversations in the user-support channel more unpredictable, as the development of discussions depends entirely on the participants' moods rather than a neutral framework.
\begin{quote}
``\textit{'Vrchat. Fix your god damn game. This stupid ass Unity error keeps popping up and it won't run.' [OP]\\
‘That's not a Unity error, that's a crash window. It happens when your client crashes.’ [replier A]\\
‘How do I fix it?’ [OP]\\
'try holding escape on the loading screen until you load in. it could be your current avatar'[replier B]}''
\end{quote}

Additionally, an important aspect of the user-support sub-channel is the presence of ``expert users (or experienced users),'' who play a crucial role in responding to inquiries. For example, in the aforementioned conversation, replier A and B are great examples of such users. They provide effective solutions with concise responses, unlike the AI bot, which typically offers lengthy, structured replies. These expert users are a key strength of the community’s mutual support. They may not necessarily be highly technical experts, but within the small user-support community, they are able to help others based on their own experiences. Since their solutions are derived from past learning (having likely encountered similar situations), they often eliminate unnecessary steps, making problem-solving more efficient.

However, gaining such expert knowledge comes with significant costs, especially when dealing with specific problems that may appear more randomly. In some cases, no one may know the solution to a particular issue, and only a few people might have encountered a similar problem. In such cases, the problem persists but gets ignored by the community. For instance, a user posted the following question in September 2023:
\begin{quote}
``\textit{Okay, not really sure where to ask this question, figure I might start here. I use a Quest 2 with Quest Pro Controllers, primarily on PC with Virtual Desktop. I love the Quest Pro controllers, but I have a consistent problem with the index finger posing in game. The Quest Pro controllers have very sensitive capacitive sensors on the triggers that allow it to detect your index finger even when it's not in contact with the trigger. If I'm in the Quest Home, it works fine and makes sense. I can see fairly accurate posing of my index finger, changing smoothly from pointed straight out, to touching the button, to fully depressed. Unfortunately, when I get into VRChat, what seems to be happening is that, even when my finger is off the trigger, the Quest is reporting some very tiny amount of detection, or movement, on the finger, so I can't actually point my finger in VRChat. Sometimes it's fine, other times it gets stuck down, and I can't point my finger again without going through a bunch of hand poses to kind of 'reset' it. This happens on both controllers. I've reset the controllers, updated them, etc. A simple fix for this would be a very simple 'dead zone' on the index finger positioning, either in VRChat or Virtual Desktop, but it doesn't seem like there is such an option anywhere. Has anyone else run into this? I don't know who to bug about this! Could use a point in the right direction.}'' [User-support]
\end{quote}
This post has still not received any responses, leaving the user stuck in a ``dead zone'' regarding what they described as a ``simple fix'' for setting a dead zone.

In contrast, user-support benefits from another positive effect: user aggregation. When different users remain active and comment on a particular unresolved issue in the user-support channel, it increases the exposure of the problem and fosters collective efforts to find a solution. Specifically, users often reply to unresolved posts with ``same problem here'' to inform others that they are experiencing a similar issue. These problems might not have an immediate solution, and the replies, in such cases, serve not to solve the issue but to search for a method of resolving it. Over time, as more users join the discussion, the issue might be resolved through collective input or attract the attention of official support and be addressed accordingly. In the process, the users experiencing the issue, future users, and the application itself all stand to benefit from the post, potentially contributing to community development and the formation of a new user base~\cite{10.1145/985921.986080}.

However, one downside is that the emergence of lengthy discussions often means the problem may go unresolved for a significant period. As the number of replies and the amount of information increase, it may become challenging for users to keep track of the issue's progress. Fig.~\ref{fig.Long1}-\ref{fig.Long6} illustrates a post that resonated with the community, showing a progression from initial user support (``same issue here,'' Fig.~\ref{fig.Long1}) to attempts at finding a solution (Fig.~\ref{fig.Long2}), sharing information (Fig.~\ref{fig.Long3}), follow-ups on progress (Fig.~\ref{fig.Long4}), the emergence of new issues (Fig.~\ref{fig.Long5}), and eventually the problem being largely resolved (Fig.~\ref{fig.Long6}). This process spanned approximately six months and garnered over 650 follow-up posts.

During this period, the solution to the problem was gradually advanced, and users engaged in extensive discussions to attempt to resolve it. In fact, in the early stages of the problem, users offered potential solutions, though these suggestions were not always applicable. Meanwhile, although at one point the issue seemed to be resolved, ``\textit{omg bro the jolting has just about completely stopped!!! it’s working!!!!!!!!!}'' [User-support], a new, related issue emerged some time later. Several versions of the software and the unique circumstances of different users further complicated the content of the post. Additionally, the heavy use of colloquial language in the replies made the thread difficult to follow, and as the discussion progressed, the emotions expressed were dynamic, with some debates inevitably arising within the conversation.


\subsection{Other Findings from Comparative Analyses}
We also discovered an interesting result from comparing the two channels: the same users often ask similar questions in both the AI-support and user-support subchannels. Based on the timing of these questions, user feedback, and follow-up comments, the reasons why users ask the same questions in both subchannels are mainly due to the following
\begin{itemize}

\item[\textbf{(1)}] Not receiving a timely response in one channel, so they ask in the other channel. For example, one user asked in the user-support channel: 
\begin{quote}``\textit{Ive apparently got my age entered wrong on my acc and didnt know it bc i cant disablr the new content filter any help would be great thanks}'' [User-support]
\end{quote} 
but did not receive a reply. Therefore, this user asked again in the AI-support channel: 
\begin{quote}
``\textit{Is it possible to update age info on my account}'' [AI-support] 
\end{quote}
\item[\textbf{(2)}] Hoping to get more professional or faster answers. Some users may think the AI-support channel can provide quicker help or hope to leverage AI capabilities to solve their problems. User asked in the AI-support channel: 
\begin{quote}``\textit{Is there an alternative method to \\`EventSystem.current.currentSelectedGameObject.name' to get the name of the button that has been pressed?}'' [AI-support] 
\end{quote} 
Since the user did not get a satisfactory reply, the user elaborated the question in the user-support channel: 
\begin{quote}
``Hello, I'm wondering if there's a way so that when a button calls a function in a script, it passes itself with it, in other words: get the name of the button that's been pressed. Asked Kappa and it mentioned the OnButtonClicked which does just that, but it isn't a supported event in Udon, I think? And there was another result talking about \texttt{\seqsplit{EventSystem.current.currentSelectedGameObject.name}} but that function isn't exposed to Udon. Any ideas? Thanks!'' [User-support]
\end{quote}
\item[\textbf{(3)}] Their question was not resolved in one channel, so they try seeking answers in the other channel. For example, User Anonymous 1 asked in the user-support channel, but the problem was not solved. They then asked again in the AI-support channel: 
\begin{quote}
    ``\textit{If I upload a world to VRChat in Unity 2022 can I downgrade it to 2019}'' [AI-support]
\end{quote}
\end{itemize}

In summary, users ask similar questions in both the AI-support and user-support sub-channels mainly because they did not receive satisfactory replies in one channel, hope to get faster or more professional help, and try different communication methods in different channels to resolve their issues.

\section{Discussion}
This study contributes to the literature on human-AI collaboration by focusing on individual and collaborative efforts to support online community management. Through a series of analyses to explore interaction dynamics within a Discord community, we revealed a collaborative dynamic in which human and AI support contributed significantly to community engagement and growth. Both humans and AI played crucial roles in addressing the diverse needs of community members, enabling them to find potential solutions and strengthen community ties.

Specifically, human support provided a nuanced contextual understanding of questions through varying levels of interaction, fostering discussions that could lead to higher engagement and constructive, civil conversations. Meanwhile, AI support offered detailed, structured answers, enhancing efficiency and outcome-oriented discussions. These two forms of support complemented each other, allowing community members to leverage each other's strengths effectively. Members could seamlessly transition between human and AI assistance to maximize their options for finding solutions.

Our study presents an exemplary case of a human-bot collaborative support system for online community management, highlighting design opportunities for future work in this area. We reflect on how humans and AI can assist each other in this domain and propose that AI can reduce the human workload to boost community engagement in several directions: 1. AI bot can generate non-technical content to foster interaction and promote engagement; 2. AI bot can learn from user support to provide empathetic and emotional responses; 3. AI bot can adopt user support strategies to enhance its problem-solving abilities. This would enable AI to intervene appropriately during collaborative discussions.

\subsection{Non-technical solidarity support is an integral part of online communities}
Our findings reveal that the user-support sub-channel often involves non-expert users attempting to help each other, which, while not always providing immediate solutions, contributes significantly to community engagement and growth. For instance, we observed cases where users replied with ``same problem here'' to unresolved issues, not necessarily offering solutions but creating a sense of shared experience. This phenomenon, while potentially viewed as unhelpful in terms of immediate problem-solving, serves several important functions in community building, such as (1) it increases the visibility of issues, potentially attracting more knowledgeable users or official support. (2) It fosters a sense of community among users facing similar challenges. (3) It contributes to the collective knowledge base, as these threads may eventually lead to solutions through continued discussion. 

Interestingly, this aspect of community interaction is absent in the AI-support sub-channel, where responses are always aimed at providing direct solutions. While AI support is designed to be consistently helpful, our findings suggest that a perfect AI solution system might inadvertently hinder the development of user-help communities. The imperfect nature of human support, with its mix of helpful and seemingly non-technical solidarity support, appears to be crucial for sustaining community engagement and fostering long-term problem-solving capabilities within the user base~\cite{doi:10.1080/0144929X.2020.1818828}. \rev{In addition, we also acknowledge that part of the so called ``unhelpful'' content in user-support takes the form of humor, sarcasm, or colloquial language. While these responses may not provide technical solutions, they play a vital role in shaping the community social dynamics. Prior research has shown that humor and playful interactions can reduce tension, increase participation, and strengthen relational bonds in online communities~\cite{Kraut2012Building}. From this perspective, the prevalence of colloquial or joking responses should not be dismissed as noise, instead it reflects the inherently social nature of peer support environments. Our sentiment analysis approach may not fully capture these nuances, but our qualitative inspection suggests that such posts contribute positively to sustaining engagement and maintaining a supportive atmosphere.} Prior research indicates that the process of mutual help and even off-topic interactions can strengthen community ties and user satisfaction~\cite{Kraut2012Building}.

Prior work highlights AI's functional roles for different contexts, such as providing accurate answers with faster, clear explanations and transparency~\cite{10.1145/3274426,10.1145/3613904.3642135,doi:10.1080/0144929X.2020.1818828}. AI can provide more social roles for community management to foster interaction. In this sense, AI bots can be designed to generate explanatory (non-direct answer) content, providing assistance in organizing, managing, and tracking progress~\cite{WIJENAYAKE2020106302} for user support to some extent, thereby preventing unresolved issues from falling into a dead zone. For example, if no one responds to a comment within a certain period, the AI bot can generate a post to promote discussion.


\subsection{Leveraging users' different expectations for helpful feedback to enhance the design of AI-assisted robots} 

Our analysis of user interactions in both AI-support and user-support sub-channels reveals distinct patterns of user expectations and behaviors. In the user-support channel, users tend to be more emotionally expressive, both positively and negatively, while also demonstrating greater tolerance for diverse responses. For example, we noted instances where initial frustration was diffused through multiple rounds of dialogue, without requiring the kind of structured empathy responses provided by the AI bot.

This observation suggests an opportunity to enhance AI-assisted support systems by incorporating elements of the dynamic, emotionally nuanced interactions found in user-support communities. Introducing AI agents that can engage in socially and emotionally intelligent interactions has been shown to improve user satisfaction and trust~\cite{10.1145/1067860.1067867,chi2021developing}. For instance, semi-anonymous AI bots could be integrated into user-support communities to increase user engagement and the number of interaction rounds. These bots could mimic the casual, emotionally varied communication style of human users, encouraging users to provide more detailed feedback and engage in longer conversations.

However, developing AI bots capable of recognizing and responding to emotional cues requires careful implementation. While empathetic responses can foster a more supportive environment, maintaining balance is crucial, especially when AI chatbots assist end users by providing specific content~\cite{10.1145/3613904.3642135}. Striking the right balance between user satisfaction and the need for precise information, and adopting appropriate response strategies~\cite{10.1145/3613904.3642135},  can help prevent responses from appearing insincere or overshadowing the technical support aspect~\cite{akdilek2024influence,10.1145/3686852.3687069}. The design of such AI agents should consider the appropriate level of emotional intelligence and transparency to maintain user trust~\cite{10.1145/2901790.2901842}.

An enhanced AI bot could potentially recognize the user's frustration, acknowledge it appropriately, and then proceed to offer technical support, mimicking the more natural flow of human interactions observed in the user-support channel. This approach aligns with findings that users prefer AI assistants that can adapt their communication style to the user's emotional state~\cite{10.1145/3563659.3563666}.

\subsection{Learning from the Crowd: Enhancing AI Support Through Human Problem-Solving Strategies} \label{subsec:dicussion-system}

\begin{figure*}[!h]
  \centering
  \includegraphics[width=.88\linewidth]{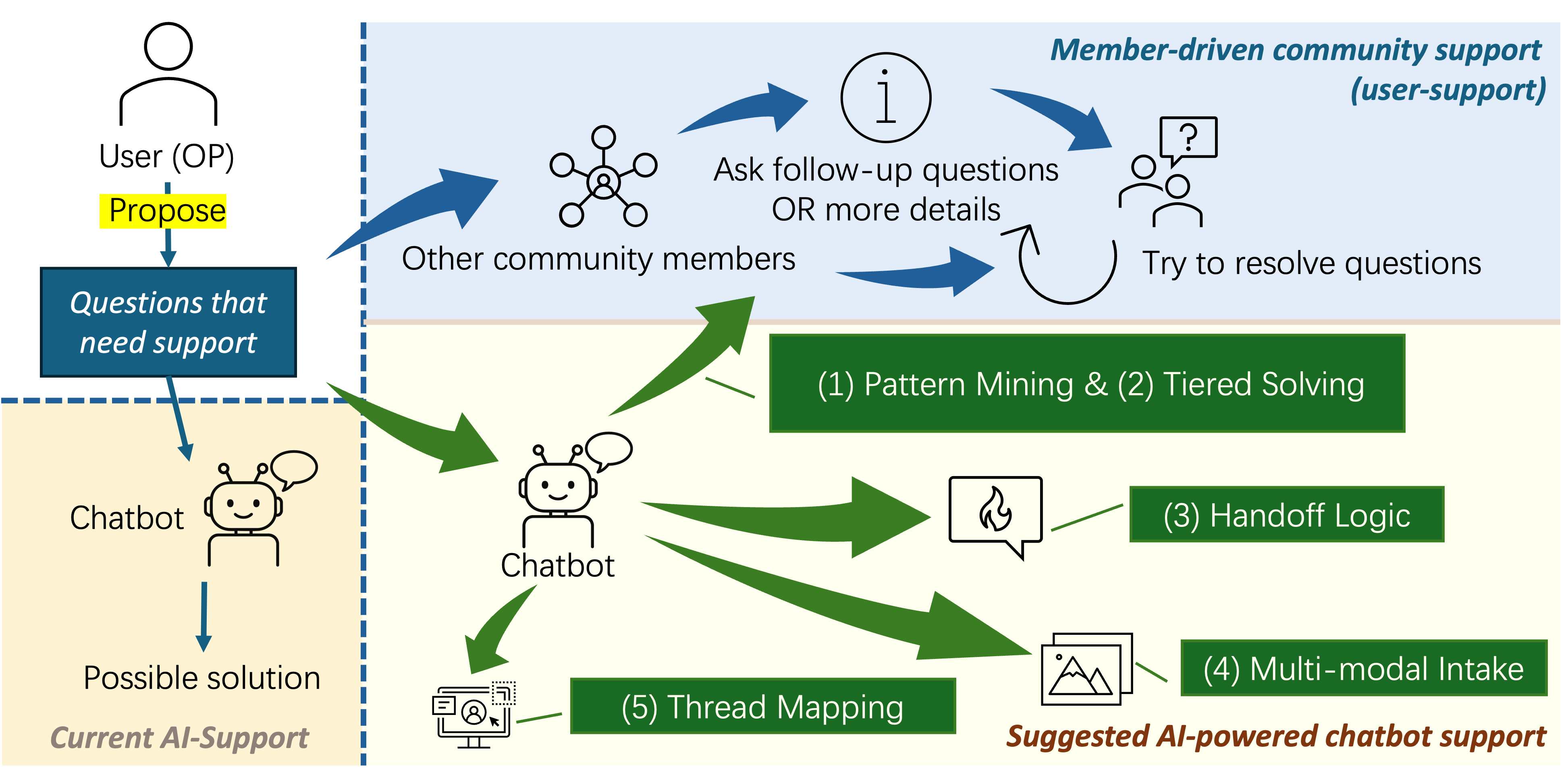}
  \caption{\textbf{Comparison of user-driven support workflows versus AI chatbot-driven support workflows, with proposed AI chatbot enhancements.}}
  \label{fig.suggested-ai}
\end{figure*}

Our findings highlight significant differences in problem-solving approaches between the AI-support and user-support sub-channels. The user-support channel often demonstrates a more nuanced, context-aware approach to problem-solving, leveraging collective knowledge and experience. This suggests an opportunity for AI systems to learn from and incorporate human problem-solving strategies to enhance their support capabilities. Specifically, we propose a system of referral learning (see in Fig.~\ref{fig.suggested-ai}), where the AI bot could be trained on successful problem-solving interactions from the user-support sub-channel. This concept is similar to approaches in crowd-powered systems that combine human intelligence with machine/AI efficiency~\cite{10.1145/1866029.1866078} or Human-AI bidirectional alignment~\cite{shen2024bidirectionalhumanaialignmentsystematic}. This could include (1) \textbf{Human-like Query Pattern Analysis}: Analyzing patterns in how experienced users ask for additional information, especially for unclear questions. (2) \textbf{Adaptive Step-by-Step Resolution}: Learning to provide step-by-step solutions that are tailored to the user's level of expertise, similar to how human experts in the user-support channel often break down complex solutions into simple manageable steps, and more directly tell people how to process it. (3) \textbf{Community-AI Handoff Intelligence}: Incorporating the ability to recognize when a problem might benefit from community input. In cases where the AI is uncertain, it could suggest posting the question in the user-support channel, effectively combining the strengths of both support systems. (4) \textbf{Multimodal Context Processing}: Developing the capability to handle multi-modal inputs, such as screenshots, which are frequently used in the user-support channel to provide richer context about issues. (5) \textbf{Discussion Thread Dynamics Tracking}: Developing the ability to track community events, such as the capability to record and follow special events and community dynamics, especially for posts in user-support that generate extensive, long-term discussions. This could assist users in tracking specific topics.

By implementing these strategies, AI support systems could evolve to provide more contextually relevant, user-friendly assistance, bridging the gap between the efficiency of AI and the nuanced understanding often demonstrated in human-to-human support interactions. In conclusion, while AI support systems offer consistency and immediacy~\cite{MALLICK2024103355}, our findings suggest that the path to truly effective support lies in a hybrid approach. This approach would combine the structured efficiency of AI with the dynamic, context-aware, and community-driven aspects of human support, ultimately leading to a more robust and user-centric support ecosystem.

\subsubsection{\rev{\textbf{Ethical Considerations of Introducing Community-AI Handoff Intelligence}}}

\rev{While integrating AI bots into human-centered support channels offers potential benefits, it also raises ethical and practical concerns. A key issue lies in the risk of AI bots disseminating AI-generated content under the guise of human authorship, which can lead to identity forgery, misinformation, conspiracy theories, and broader challenges for public discourse~\cite{Yang_nbcnews2025}. To mitigate such risks, we put forward several design recommendations and examine how the \textit{Community-AI Handoff Intelligence model} may be applied in Discord and similar online communities.}

\rev{First, the origin of posted questions should remain with human users. Specifically, when an AI bot poses a question in a user-support channel, the content should derive from unresolved queries in the AI-support channel rather than being generated without referencing. This ensures that AI contributes by summarizing unsolved problems and their context instead of introducing hallucinatory or fictitious issues.}

\rev{Second, the distribution of responsibility is also an important issue. Given the importance of user-centered design and the nature of community platforms, assigning full accountability to end-users for AI-generated questions is neither fair nor feasible. On the one hand, users may lack the knowledge to validate all AI outputs, and on the other hand, comprehensive human checking would add extra cognitive load, which may not benefit the development of social media platforms.We therefore recommend distributing responsibility among three parties: the user, the AI bot, and the platform. For the users, especially askers/questioners, they can make a quick check and grant authorization to decide whether the AI bot should post the relevant AI-generated/summarized question in the user support channel (similar to the concept of human-in-the-loop~\cite{mckay2024realizing}), and whether to self-disclose the identity of the original human asker~\cite{10.1145/2818048.2820010,10.1145/2858036.2858414}. For the AI bot, an additional AI checker~\cite{zhang2025harnessingpoweraiqualitative} or a self-evaluatior with reflection~\cite{agashe2025agent} could be introduced before publishing (similar to the concept of LLM-as-a-judge~\cite{NEURIPS2023_91f18a12}), and at the time of posting, it can include a label such as ``\textit{This question was generated by AI from unresolved queries in the AI support channel}'' to clarify the AI-generated status of the content. Since the AI bot itself may not be able to take responsibility for real-world interactions, such responsibility may ultimately need to be transferred to the platform's supervisory role. However, this can greatly alleviate users' misunderstandings of AI-generated content, such as assuming that it was a direct question from a real person. For the platform, it should take on the supervision of AI-generated content to prevent potentially harmful information. Although this introduces extra workload, such a trade-off may be worthwhile for sustaining healthy community development.}

\rev{In addition, role restrictions are essential. In our scenario (the VRChat Discord community), the AI bot should function only as a questioner in the user-support channel, not as an answerer. This helps prevent identity impersonation and the uncontrolled proliferation of AI-generated responses. While individual users may still employ AI to generate posts manually, such practices fall outside the scope of this model and our current discussion.}

\rev{With these design considerations in place, the AI bot can be positioned as an intermediary agent facilitating information exchange between channels. Its primary value lies in bridging knowledge gaps, retrieving relevant information, and reducing interaction overhead for community members.}

\rev{Finally, it is important to note that our discussion here focuses specifically on Section~\ref{subsec:dicussion-system} (3), rather than (1) or other parts of the system of referral learning. Other components are primarily concerned with technical aspects (e.g., reinforcement learning, multimodal interaction) and functional implementations, rather than the ethical implications of AI in online communities.}

\section{Limitations}
While our study provides valuable insights into the interactions between users and support systems in online communities, it is important to acknowledge several limitations: (1) Limited Scope of Data Source: Our dataset is derived from a single Discord community, specifically the VRChat community. Although this community boasts a large user base, it may not be fully representative of all online communities that utilize AI-assisted support systems. The dynamics and effectiveness of AI bots may vary in different communities or platforms, potentially limiting the generalizability of our findings. (2) Static Analysis of a Dynamic Environment: Our analysis provides a snapshot of user interactions at a specific point in time. However, online communities and AI technologies are rapidly evolving. The patterns and behaviors we observed may have changed since data collection, and the AI bot's capabilities may have been updated. A longitudinal study could provide more comprehensive insights into the evolving nature of these interactions. (3) Limited Analysis of AI Bot Learning: While we observed the AI bot's responses, our study did not have access to data on how the bot learns or improves over time. Understanding the bot's learning process could provide valuable insights into the development of more effective AI support systems. \rev{(4) Sensitivity of Sentiment Analysis to Colloquial Language: Our sentiment analysis relies on a BERT-based model that is not optimized for humor, sarcasm, or colloquial language. To mitigate this limitation, we complemented model outputs with qualitative inspection to contextualize polarity estimates, and we plan to explore models and annotations that explicitly target humor and sarcasm in future work.} (5) The underlying model, system architecture, and knowledge sources used by ``\textit{kappa.ai\#2237}'' in the VRChat Discord channel remain unknown, as its creator and owner have not released technical details or open-sourced the system. As a result, it is unclear whether the chatbot's responses are generated by LLMs, retrieval-based systems, or predefined templates. Nevertheless, the primary focus of this study is on users' interaction behaviors and engagement patterns with the AI bot rather than its internal technical design. Therefore, this uncertainty is unlikely to substantially affect the main findings related to user behavior and interaction dynamics. These limitations present important opportunities for future research to build upon our findings and develop a more comprehensive understanding of AI-assisted support in online communities. In particular, future work could incorporate controlled experiments to more systematically compare user responses to AI and human support under comparable conditions, apply inferential statistical methods to test the robustness and significance of observed differences, and examine the longitudinal evolution of these sub-channels to better understand how support practices, engagement patterns, and user expectations change over time.

\section{Conclusion}
This study presents a comprehensive analysis of user interactions within AI-assisted and human-based support systems in the context of the VRChat online community. By examining data from the ``user-support'' and ``ai-support'' sub-channels over a significant period (from very beginning day of AI bot released in VRChat Discord community), we have uncovered distinct patterns of user engagement and problem-solving approaches in these two support environments. Our mixed-methods approach, combining quantitative analyses with qualitative inductive coding, revealed significant differences in how users interact with AI and human support systems. Importantly, we found that seemingly ``unhelpful'' content in user-support channels plays a crucial role in community building and collective problem-solving, a factor absent in AI-support interactions. This highlights the importance of human interaction in fostering a supportive community environment. These findings suggest that an ideal support ecosystem might combine the efficiency of AI with the nuanced, community-driven aspects of human support. Such a hybrid approach could lead to more effective problem resolution and a more engaged user community. 

\begin{acks}
We sincerely thank the anonymous reviewers for their insightful comments and constructive feedback, which greatly helped strengthen this work. We would also like to clarify that we used ChatGPT and Claude solely for spell-checking and grammar revisions to enhance content clarity. Our research team assumes full responsibility for the content of this publication. We did not use generative AI for data collection, analysis, or image generation. The authors declare that they have no known competing financial interests or personal relationships that could have appeared to influence the work reported in this paper.
\end{acks}



\balance
\bibliographystyle{ACM-Reference-Format}
\bibliography{sample-base}

@article{10.1145/3432938,
author = {Freeman, Guo and Maloney, Divine},
title = {Body, Avatar, and Me: The Presentation and Perception of Self in Social Virtual Reality},
year = {2021},
issue_date = {December 2020},
publisher = {Association for Computing Machinery},
address = {New York, NY, USA},
volume = {4},
number = {CSCW3},
url = {https://doi.org/10.1145/3432938},
doi = {10.1145/3432938},
journal = {Proc. ACM Hum.-Comput. Interact.},
month = {jan},
articleno = {239},
numpages = {27}
}

@inproceedings{10.1145/3313831.3376606,
author = {Tanenbaum, Theresa Jean and Hartoonian, Nazely and Bryan, Jeffrey},
title = {"How do I make this thing smile?": An Inventory of Expressive Nonverbal Communication in Commercial Social Virtual Reality Platforms},
year = {2020},
isbn = {9781450367080},
publisher = {Association for Computing Machinery},
address = {New York, NY, USA},
url = {https://doi.org/10.1145/3313831.3376606},
doi = {10.1145/3313831.3376606},
booktitle = {Proceedings of the 2020 CHI Conference on Human Factors in Computing Systems},
pages = {1–13},
numpages = {13},
location = {Honolulu, HI, USA},
series = {CHI '20}
}

@inproceedings{10.1145/3197391.3205451,
author = {McVeigh-Schultz, Joshua and M\'{a}rquez Segura, Elena and Merrill, Nick and Isbister, Katherine},
title = {What's It Mean to "Be Social" in VR? Mapping the Social VR Design Ecology},
year = {2018},
isbn = {9781450356312},
publisher = {Association for Computing Machinery},
address = {New York, NY, USA},
url = {https://doi.org/10.1145/3197391.3205451},
doi = {10.1145/3197391.3205451},
booktitle = {Proceedings of the 2018 ACM Conference Companion Publication on Designing Interactive Systems},
pages = {289–294},
numpages = {6},
location = {Hong Kong, China},
series = {DIS '18 Companion}
}

@inproceedings{10.1145/3173574.3173863,
author = {Smith, Harrison Jesse and Neff, Michael},
title = {Communication Behavior in Embodied Virtual Reality},
year = {2018},
isbn = {9781450356206},
publisher = {Association for Computing Machinery},
address = {New York, NY, USA},
url = {https://doi.org/10.1145/3173574.3173863},
doi = {10.1145/3173574.3173863},
booktitle = {Proceedings of the 2018 CHI Conference on Human Factors in Computing Systems},
pages = {1–12},
numpages = {12},
location = {Montreal QC, Canada},
series = {CHI '18}
}

@INPROCEEDINGS{5428431,
  author={Mennecke, Brian E. and Triplett, Janea L. and Hassall, Lesya M. and Conde, Zayira Jordan},
  booktitle={2010 43rd Hawaii International Conference on System Sciences}, 
  title={Embodied Social Presence Theory}, 
  year={2010},
  volume={},
  number={},
  pages={1-10},
  doi={10.1109/HICSS.2010.179}}

@INPROCEEDINGS{9417636,
  author={Sykownik, Philipp and Graf, Linda and Zils, Christoph and Masuch, Maic},
  booktitle={2021 IEEE Virtual Reality and 3D User Interfaces (VR)}, 
  title={The Most Social Platform Ever? A Survey about Activities \& Motives of Social VR Users}, 
  year={2021},
  volume={},
  number={},
  pages={546-554},
  doi={10.1109/VR50410.2021.00079}}

@inproceedings{10.1145/3411764.3445426,
author = {Saffo, David and Di Bartolomeo, Sara and Yildirim, Caglar and Dunne, Cody},
title = {Remote and Collaborative Virtual Reality Experiments via Social VR Platforms},
year = {2021},
isbn = {9781450380966},
publisher = {Association for Computing Machinery},
address = {New York, NY, USA},
url = {https://doi.org/10.1145/3411764.3445426},
doi = {10.1145/3411764.3445426},
booktitle = {Proceedings of the 2021 CHI Conference on Human Factors in Computing Systems},
articleno = {523},
numpages = {15},
location = {Yokohama, Japan},
series = {CHI '21}
}

@inproceedings{wang2020social,
  title={Social VR: a new form of social communication in the future or a beautiful illusion?},
  author={Wang, Minhan},
  booktitle={Journal of Physics: Conference Series},
  volume={1518},
  number={1},
  pages={012032},
  year={2020},
  doi={10.1088/1742-6596/1518/1/012032},
  organization={IOP Publishing}
}

@article{10.1145/3492836,
author = {Freeman, Guo and Acena, Dane and McNeese, Nathan J. and Schulenberg, Kelsea},
title = {Working Together Apart through Embodiment: Engaging in Everyday Collaborative Activities in Social Virtual Reality},
year = {2022},
issue_date = {January 2022},
publisher = {Association for Computing Machinery},
address = {New York, NY, USA},
volume = {6},
number = {GROUP},
url = {https://doi.org/10.1145/3492836},
doi = {10.1145/3492836},
journal = {Proc. ACM Hum.-Comput. Interact.},
month = {jan},
articleno = {17},
numpages = {25}
}

@article{dwivedi2022metaverse,
  title={Metaverse beyond the hype: Multidisciplinary perspectives on emerging challenges, opportunities, and agenda for research, practice and policy},
  author={Dwivedi, Yogesh K and Hughes, Laurie and Baabdullah, Abdullah M and Ribeiro-Navarrete, Samuel and Giannakis, Mihalis and Al-Debei, Mutaz M and Dennehy, Denis and Metri, Bhimaraya and Buhalis, Dimitrios and Cheung, Christy MK and others},
  journal={International journal of information management},
  volume={66},
  pages={102542},
  year={2022},
  dor={https://doi.org/10.1016/j.ijinfomgt.2022.102542},
  publisher={Elsevier}
}

@inproceedings{10.1145/3411764.3445335,
author = {Krau\ss{}, Veronika and Boden, Alexander and Oppermann, Leif and Reiners, Ren\'{e}},
title = {Current Practices, Challenges, and Design Implications for Collaborative AR/VR Application Development},
year = {2021},
isbn = {9781450380966},
publisher = {Association for Computing Machinery},
address = {New York, NY, USA},
url = {https://doi.org/10.1145/3411764.3445335},
doi = {10.1145/3411764.3445335},
booktitle = {Proceedings of the 2021 CHI Conference on Human Factors in Computing Systems},
articleno = {454},
numpages = {15},
location = {Yokohama, Japan},
series = {CHI '21}
}

@inproceedings{zhang2024exploring,
  title={Exploring Virtual Reality Through Ihde’s Instrumental Realism},
  author={Zhang, He and Carroll, John M},
  booktitle={International Conference on Information},
  pages={82--93},
  year={2024},
  doi={https://doi.org/10.1007/978-3-031-57860-1_6},
  organization={Springer}
}

@article{10.1145/2133806.2133826,
author = {Blei, David M.},
title = {Probabilistic Topic Models},
year = {2012},
issue_date = {April 2012},
publisher = {Association for Computing Machinery},
address = {New York, NY, USA},
volume = {55},
number = {4},
issn = {0001-0782},
url = {https://doi.org/10.1145/2133806.2133826},
doi = {10.1145/2133806.2133826},
journal = {Commun. ACM},
month = {apr},
pages = {77–84},
numpages = {8}
}

@article{jelodar2019latent,
  title={Latent Dirichlet allocation (LDA) and topic modeling: models, applications, a survey},
  author={Jelodar, Hamed and Wang, Yongli and Yuan, Chi and Feng, Xia and Jiang, Xiahui and Li, Yanchao and Zhao, Liang},
  journal={Multimedia tools and applications},
  volume={78},
  pages={15169--15211},
  year={2019},
  doi={https://doi.org/10.1007/s11042-018-6894-4},
  publisher={Springer}
}

@inproceedings{10.5555/2390948.2391052,
author = {Stevens, Keith and Kegelmeyer, Philip and Andrzejewski, David and Buttler, David},
title = {Exploring Topic Coherence over Many Models and Many Topics},
year = {2012},
publisher = {Association for Computational Linguistics},
address = {USA},
booktitle = {Proceedings of the 2012 Joint Conference on Empirical Methods in Natural Language Processing and Computational Natural Language Learning},
pages = {952–961},
numpages = {10},
location = {Jeju Island, Korea},
series = {EMNLP-CoNLL '12}
}

@inproceedings{stevens2012exploring,
  title={Exploring topic coherence over many models and many topics},
  author={Stevens, Keith and Kegelmeyer, Philip and Andrzejewski, David and Buttler, David},
  booktitle={Proceedings of the 2012 joint conference on empirical methods in natural language processing and computational natural language learning},
  pages={952--961},
  year={2012}
}

@inproceedings{zhao2011comparing,
  title={Comparing twitter and traditional media using topic models},
  author={Zhao, Wayne Xin and Jiang, Jing and Weng, Jianshu and He, Jing and Lim, Ee-Peng and Yan, Hongfei and Li, Xiaoming},
  booktitle={Advances in Information Retrieval: 33rd European Conference on IR Research, ECIR 2011, Dublin, Ireland, April 18-21, 2011. Proceedings 33},
  pages={338--349},
  year={2011},
  doi={https://doi.org/10.1007/978-3-642-20161-5_34},
  organization={Springer}
}

@inproceedings{10.1145/3613904.3642787,
author = {Cai, Jie and Lin, Ya-Fang and Zhang, He and Carroll, John M.},
title = {Third-Party Developers and Tool Development For Community Management on Live Streaming Platform Twitch},
year = {2024},
isbn = {9798400703300},
publisher = {Association for Computing Machinery},
address = {New York, NY, USA},
url = {https://doi.org/10.1145/3613904.3642787},
doi = {10.1145/3613904.3642787},
booktitle = {Proceedings of the CHI Conference on Human Factors in Computing Systems},
articleno = {926},
numpages = {18},
location = {Honolulu, HI, USA},
series = {CHI '24}
}

@article{egger2022topic,
  title={A topic modeling comparison between lda, nmf, top2vec, and bertopic to demystify twitter posts},
  author={Egger, Roman and Yu, Joanne},
  journal={Frontiers in sociology},
  volume={7},
  pages={886498},
  year={2022},
  doi ={https://doi.org/10.3389/fsoc.2022.886498},
  publisher={Frontiers Media SA}
}

@ARTICLE{5416713,
  author={Phan, Xuan-Hieu and Nguyen, Cam-Tu and Le, Dieu-Thu and Nguyen, Le-Minh and Horiguchi, Susumu and Ha, Quang-Thuy},
  journal={IEEE Transactions on Knowledge and Data Engineering}, 
  title={A Hidden Topic-Based Framework toward Building Applications with Short Web Documents}, 
  year={2011},
  volume={23},
  number={7},
  pages={961-976},
  doi={10.1109/TKDE.2010.27}}

@article{likhitha2019detailed,
  title={A detailed survey on topic modeling for document and short text data},
  author={Likhitha, S and Harish, BS and Kumar, HM Keerthi},
  journal={International Journal of Computer Applications},
  volume={178},
  number={39},
  pages={1--9},
  year={2019},
  doi={10.5120/ijca2019919265}
}

@article{albalawi2020using,
  title={Using topic modeling methods for short-text data: A comparative analysis},
  author={Albalawi, Rania and Yeap, Tet Hin and Benyoucef, Morad},
  journal={Frontiers in artificial intelligence},
  volume={3},
  pages={42},
  year={2020},
  doi = {https://doi.org/10.3389/frai.2020.00042},
  publisher={Frontiers Media SA}
}

@article{girvan2002community,
  title={Community structure in social and biological networks},
  author={Girvan, Michelle and Newman, Mark EJ},
  journal={Proceedings of the national academy of sciences},
  volume={99},
  number={12},
  pages={7821--7826},
  year={2002},
  doi={https://doi.org/10.1073/pnas.122653799},
  publisher={National Acad Sciences}
}

@article{brin1998anatomy,
  title={The anatomy of a large-scale hypertextual web search engine},
  author={Brin, Sergey and Page, Lawrence},
  journal={Computer networks and ISDN systems},
  volume={30},
  number={1-7},
  pages={107--117},
  year={1998},
  doi={https://doi.org/10.1016/S0169-7552(98)00110-X},
  publisher={Elsevier}
}

@misc{devlin2019bertpretrainingdeepbidirectional,
      title={BERT: Pre-training of Deep Bidirectional Transformers for Language Understanding}, 
      author={Jacob Devlin and Ming-Wei Chang and Kenton Lee and Kristina Toutanova},
      year={2019},
      eprint={1810.04805},
      archivePrefix={arXiv},
      primaryClass={cs.CL},
      url={https://arxiv.org/abs/1810.04805}, 
}

@article{holton2007coding,
  title={The coding process and its challenges},
  author={Holton, Judith A},
  journal={The Sage handbook of grounded theory},
  volume={3},
  pages={265--289},
  ISBN={9781446275726},
  year={2007}
}

@article{thomas2003general,
  title={A general inductive approach for qualitative data analysis},
  author={Thomas, David R},
  year={2003},
  epring={https://d1wqtxts1xzle7.cloudfront.net/84611396/Inductive_20Content_20Analysis-libre.pdf?1650539704=&response-content-disposition=inline%3B+filename%3DQualitative_Data_Analysis.pdf&Expires=1726010819&Signature=AwoQbH~3B6GoCjog01Z9hoKbNgl2Gy9wiaOr244mh93oEtqD4HWr65kheHhHrJuZdZaD1X4m5EOkTE6A3fhfAQA0n4qdeBe0k-2GmD0~BKY3FFJcO7qLR0JVS24QirciNCotYKR8TZnwrZKffkbAe5it-HxTQ5QzQLduxCMZQXuJr6oK-5d4PkLd7MvFcvaoazHLGiey48wnLb1iPOyfurWM-YoAFpXlryr-5xKLhFKY~LlPdAGq2MdENhe3q3dU91kJpgQgk-iuMLxmuyScgmBy2PmhZk0vKPbkgg439wWt9MRaHLv7Y7STQfdTZMmEDT~mdao8E3HOWM6NEnmZWw__&Key-Pair-Id=APKAJLOHF5GGSLRBV4ZA},
  publisher={SAGE}
}

@inproceedings{10.1145/3411764.3445363,
author = {A. Sparrow, Lucy and Gibbs, Martin and Arnold, Michael},
title = {The Ethics of Multiplayer Game Design and Community Management: Industry Perspectives and Challenges},
year = {2021},
isbn = {9781450380966},
publisher = {Association for Computing Machinery},
address = {New York, NY, USA},
url = {https://doi.org/10.1145/3411764.3445363},
doi = {10.1145/3411764.3445363},
booktitle = {Proceedings of the 2021 CHI Conference on Human Factors in Computing Systems},
articleno = {325},
numpages = {13},
location = {Yokohama, Japan},
series = {CHI '21}
}

@ARTICLE{10535527,
  author={Zhang, He and Li, Xinyang and Fu, Xinyi and Qiu, Christine and Zhang, Jiyuan and Carroll, John M.},
  journal={IEEE Transactions on Games}, 
  title={Understanding Fear Responses and Coping Mechanisms in VR Horror Gaming: Insights From Semi-Structured Interviews}, 
  year={2024},
  volume={},
  number={},
  pages={1-14},
  doi={10.1109/TG.2024.3403768}}

@article{ZHANG2025100144,
title = {Harnessing the power of AI in qualitative research: Exploring, using and redesigning ChatGPT},
journal = {Computers in Human Behavior: Artificial Humans},
volume = {4},
pages = {100144},
year = {2025},
issn = {2949-8821},
doi = {https://doi.org/10.1016/j.chbah.2025.100144},
url = {https://www.sciencedirect.com/science/article/pii/S2949882125000283},
author = {He Zhang and Chuhao Wu and Jingyi Xie and Yao Lyu and Jie Cai and John M. Carroll}
}

@article{glaser1965constant,
  title={The constant comparative method of qualitative analysis},
  author={Glaser, Barney G},
  journal={Social problems},
  volume={12},
  number={4},
  pages={436--445},
  year={1965},
  doi = {https://doi.org/10.2307/798843},
  publisher={Oxford University Press Oxford, UK}
}

@article{faraj2015leading,
  title={Leading collaboration in online communities},
  author={Faraj, Samer and Kudaravalli, Srinivas and Wasko, Molly},
  journal={MIS quarterly},
  volume={39},
  number={2},
  pages={393--412},
  year={2015},
  url={https://www.jstor.org/stable/26628359},
  publisher={JSTOR}
}

@article{10.1145/3651990,
author = {Cheng, Ruijia and Wang, Ruotong and Zimmermann, Thomas and Ford, Denae},
title = {“It would work for me too”: How Online Communities Shape Software Developers’ Trust in AI-Powered Code Generation Tools},
year = {2024},
issue_date = {June 2024},
publisher = {Association for Computing Machinery},
address = {New York, NY, USA},
volume = {14},
number = {2},
issn = {2160-6455},
url = {https://doi.org/10.1145/3651990},
doi = {10.1145/3651990},
journal = {ACM Trans. Interact. Intell. Syst.},
month = {may},
articleno = {11},
numpages = {39}
}

@inproceedings{10.1145/985921.986080,
author = {Seay, A. Fleming and Jerome, William J. and Lee, Kevin Sang and Kraut, Robert E.},
title = {Project massive: a study of online gaming communities},
year = {2004},
isbn = {1581137036},
publisher = {Association for Computing Machinery},
address = {New York, NY, USA},
url = {https://doi.org/10.1145/985921.986080},
doi = {10.1145/985921.986080},
booktitle = {CHI '04 Extended Abstracts on Human Factors in Computing Systems},
pages = {1421–1424},
numpages = {4},
location = {Vienna, Austria},
series = {CHI EA '04}
}

@article{10.1145/3579505,
author = {Meng, Jingbo and Rheu, Minjin (MJ) and Zhang, Yue and Dai, Yue and Peng, Wei},
title = {Mediated Social Support for Distress Reduction: AI Chatbots vs. Human},
year = {2023},
issue_date = {April 2023},
publisher = {Association for Computing Machinery},
address = {New York, NY, USA},
volume = {7},
number = {CSCW1},
url = {https://doi.org/10.1145/3579505},
doi = {10.1145/3579505},
journal = {Proc. ACM Hum.-Comput. Interact.},
month = {apr},
articleno = {72},
numpages = {25}
}

@article{hill2015real,
  title={Real conversations with artificial intelligence: A comparison between human--human online conversations and human--chatbot conversations},
  author={Hill, Jennifer and Ford, W Randolph and Farreras, Ingrid G},
  journal={Computers in human behavior},
  volume={49},
  pages={245--250},
  year={2015},
  publisher={Elsevier}
}

@inproceedings{cui2017superagent,
  title={Superagent: A customer service chatbot for e-commerce websites},
  author={Cui, Lei and Huang, Shaohan and Wei, Furu and Tan, Chuanqi and Duan, Chaoqun and Zhou, Ming},
  booktitle={Proceedings of ACL 2017, system demonstrations},
  pages={97--102},
  year={2017}
}

@inproceedings{10.1145/2858036.2858288,
author = {Luger, Ewa and Sellen, Abigail},
title = {"Like Having a Really Bad PA": The Gulf between User Expectation and Experience of Conversational Agents},
year = {2016},
isbn = {9781450333627},
publisher = {Association for Computing Machinery},
address = {New York, NY, USA},
url = {https://doi.org/10.1145/2858036.2858288},
doi = {10.1145/2858036.2858288},
booktitle = {Proceedings of the 2016 CHI Conference on Human Factors in Computing Systems},
pages = {5286–5297},
numpages = {12},
location = {San Jose, California, USA},
series = {CHI '16}
}

@book{kraut2012building,
  title={Building Successful Online Communities: Evidence-based Social Design},
  author={Kraut, RE},
  year={2012},
  publisher={MIT Press}
}

@article{10.1145/1067860.1067867,
author = {Bickmore, Timothy W. and Picard, Rosalind W.},
title = {Establishing and maintaining long-term human-computer relationships},
year = {2005},
issue_date = {June 2005},
publisher = {Association for Computing Machinery},
address = {New York, NY, USA},
volume = {12},
number = {2},
issn = {1073-0516},
url = {https://doi.org/10.1145/1067860.1067867},
doi = {10.1145/1067860.1067867},
journal = {ACM Trans. Comput.-Hum. Interact.},
month = {jun},
pages = {293–327},
numpages = {35}
}

@article{chi2021developing,
  title={Developing a formative scale to measure consumers’ trust toward interaction with artificially intelligent (AI) social robots in service delivery},
  author={Chi, Oscar Hengxuan and Jia, Shizhen and Li, Yafang and Gursoy, Dogan},
  journal={Computers in Human Behavior},
  volume={118},
  pages={106700},
  year={2021},
  publisher={Elsevier}
}

@article{akdilek2024influence,
  title={The Influence of Generative AI on Interpersonal Communication Dynamics},
  author={Akdilek, Sumeyya and Akdilek, Ibrahim and Punyanunt-Carter, Narissra Maria},
  journal={The Role of Generative AI in the Communication Classroom},
  pages={167--190},
  year={2024},
  publisher={IGI Global}
}

@inproceedings{10.1145/3686852.3687069,
author = {Zhang, He and Xie, Jingyi and Wu, Chuhao and Cai, Jie and Kim, Chanmin and Carroll, John M.},
title = {The Future of Learning: Large Language Models through the Lens of Students},
year = {2024},
isbn = {9798400711060},
publisher = {Association for Computing Machinery},
address = {New York, NY, USA},
url = {https://doi.org/10.1145/3686852.3687069},
doi = {10.1145/3686852.3687069},
booktitle = {Proceedings of the 25th Annual Conference on Information Technology Education},
pages = {12–18},
numpages = {7},
location = {El Paso, TX, USA},
series = {SIGITE '24}
}

@inproceedings{10.1145/2901790.2901842,
author = {Liao, Q. Vera and Davis, Matthew and Geyer, Werner and Muller, Michael and Shami, N. Sadat},
title = {What Can You Do? Studying Social-Agent Orientation and Agent Proactive Interactions with an Agent for Employees},
year = {2016},
isbn = {9781450340311},
publisher = {Association for Computing Machinery},
address = {New York, NY, USA},
url = {https://doi.org/10.1145/2901790.2901842},
doi = {10.1145/2901790.2901842},
booktitle = {Proceedings of the 2016 ACM Conference on Designing Interactive Systems},
pages = {264–275},
numpages = {12},
location = {Brisbane, QLD, Australia},
series = {DIS '16}
}

@inbook{10.1145/3563659.3563666,
author = {Janowski, Kathrin and Ritschel, Hannes and Andr\'{e}, Elisabeth},
title = {Adaptive Artificial Personalities},
year = {2022},
isbn = {9781450398961},
publisher = {Association for Computing Machinery},
address = {New York, NY, USA},
edition = {1},
url = {https://doi.org/10.1145/3563659.3563666},
booktitle = {The Handbook on Socially Interactive Agents: 20 Years of Research on Embodied Conversational Agents, Intelligent Virtual Agents, and Social Robotics Volume 2: Interactivity, Platforms, Application},
pages = {155–194},
numpages = {40}
}

@inproceedings{10.1145/1866029.1866078,
author = {Bernstein, Michael S. and Little, Greg and Miller, Robert C. and Hartmann, Bj\"{o}rn and Ackerman, Mark S. and Karger, David R. and Crowell, David and Panovich, Katrina},
title = {Soylent: a word processor with a crowd inside},
year = {2010},
isbn = {9781450302715},
publisher = {Association for Computing Machinery},
address = {New York, NY, USA},
url = {https://doi.org/10.1145/1866029.1866078},
doi = {10.1145/1866029.1866078},
booktitle = {Proceedings of the 23nd Annual ACM Symposium on User Interface Software and Technology},
pages = {313–322},
numpages = {10},
location = {New York, New York, USA},
series = {UIST '10}
}

@article{haythornthwaite2007social,
  title={Social networks and online community},
  author={Haythornthwaite, Caroline},
  journal={The Oxford handbook of Internet psychology},
  pages={121--137},
  year={2007}
}

@article{WIJENAYAKE2020106302,
title = {Impact of contextual and personal determinants on online social conformity},
journal = {Computers in Human Behavior},
volume = {108},
pages = {106302},
year = {2020},
issn = {0747-5632},
doi = {https://doi.org/10.1016/j.chb.2020.106302},
url = {https://www.sciencedirect.com/science/article/pii/S074756322030056X},
author = {Senuri Wijenayake and Niels {van Berkel} and Vassilis Kostakos and Jorge Goncalves}
}

@inproceedings{10.1145/3613904.3642135,
author = {Wester, Joel and Schrills, Tim and Pohl, Henning and van Berkel, Niels},
title = {“As an AI language model, I cannot”: Investigating LLM Denials of User Requests},
year = {2024},
isbn = {9798400703300},
publisher = {Association for Computing Machinery},
address = {New York, NY, USA},
url = {https://doi.org/10.1145/3613904.3642135},
doi = {10.1145/3613904.3642135},
booktitle = {Proceedings of the 2024 CHI Conference on Human Factors in Computing Systems},
articleno = {979},
numpages = {14},
location = {Honolulu, HI, USA},
series = {CHI '24}
}

@article{10.1145/3274426,
author = {Seering, Joseph and Flores, Juan Pablo and Savage, Saiph and Hammer, Jessica},
title = {The Social Roles of Bots: Evaluating Impact of Bots on Discussions in Online Communities},
year = {2018},
issue_date = {November 2018},
publisher = {Association for Computing Machinery},
address = {New York, NY, USA},
volume = {2},
number = {CSCW},
url = {https://doi.org/10.1145/3274426},
doi = {10.1145/3274426},
journal = {Proc. ACM Hum.-Comput. Interact.},
month = nov,
articleno = {157},
numpages = {29}
}

@article{doi:10.1080/0144929X.2020.1818828,
author = {Niels van Berkel, Benjamin Tag, Jorge Goncalves and Simo Hosio},
title = {Human-centred artificial intelligence: a contextual morality perspective},
journal = {Behaviour \& Information Technology},
volume = {41},
number = {3},
pages = {502--518},
year = {2022},
publisher = {Taylor \& Francis},
doi = {10.1080/0144929X.2020.1818828},
URL = {https://doi.org/10.1080/0144929X.2020.1818828},
eprint = {https://doi.org/10.1080/0144929X.2020.1818828}
}

@article{ferrara2023social,
  title={Social bot detection in the age of ChatGPT: Challenges and opportunities},
  author={Ferrara, Emilio},
  journal={First Monday},
  doi={https://doi.org/10.5210/fm.v28i6.13185},
  year={2023}
}

@article{zhu2022me,
  title={It is me, chatbot: working to address the COVID-19 outbreak-related mental health issues in China. User experience, satisfaction, and influencing factors},
  author={Zhu, Yonghan and Janssen, Marijn and Wang, Rui and Liu, Yang},
  journal={International Journal of Human--Computer Interaction},
  volume={38},
  number={12},
  pages={1182--1194},
  year={2022},
  doi={https://doi.org/10.1080/10447318.2021.1988236},
  publisher={Taylor \& Francis}
}

@book{kempt2020chatbots,
  title={Chatbots and the domestication of AI: a relational approach},
  author={Kempt, Hendrik},
  year={2020},
  publisher={Springer}
}

@book{davenport2018ai,
  title={The AI advantage: How to put the artificial intelligence revolution to work},
  author={Davenport, Thomas H},
  year={2018},
  publisher={mit Press}
}

@article{blut2021understanding,
  title={Understanding anthropomorphism in service provision: a meta-analysis of physical robots, chatbots, and other AI},
  author={Blut, Markus and Wang, Cheng and W{\"u}nderlich, Nancy V and Brock, Christian},
  journal={Journal of the Academy of Marketing Science},
  volume={49},
  pages={632--658},
  year={2021},
doi={https://doi.org/10.1007/s11747-020-00762-y},
  publisher={Springer}
}

@inproceedings{10.1145/3593013.3594013,
author = {Neumann, Terrence and Wolczynski, Nicholas},
title = {Does AI-Assisted Fact-Checking Disproportionately Benefit Majority Groups Online?},
year = {2023},
isbn = {9798400701924},
publisher = {Association for Computing Machinery},
address = {New York, NY, USA},
url = {https://doi.org/10.1145/3593013.3594013},
doi = {10.1145/3593013.3594013},
booktitle = {Proceedings of the 2023 ACM Conference on Fairness, Accountability, and Transparency},
pages = {480–490},
numpages = {11},
location = {Chicago, IL, USA},
series = {FAccT '23}
}

@article{mouronte2024patterns,
  title={Patterns of human and bots behaviour on Twitter conversations about sustainability},
  author={Mouronte-L{\'o}pez, Mary Luz and G{\'o}mez S{\'a}nchez-Seco, Javier and Benito, Rosa M},
  journal={Scientific Reports},
  volume={14},
  number={1},
  pages={3223},
  doi={https://doi.org/10.1038/s41598-024-52471-z},
  year={2024},
  publisher={Nature Publishing Group UK London}
}

@article{SEITZ2022102848,
title = {Can we trust a chatbot like a physician? A qualitative study on understanding the emergence of trust toward diagnostic chatbots},
journal = {International Journal of Human-Computer Studies},
volume = {165},
pages = {102848},
year = {2022},
issn = {1071-5819},
doi = {https://doi.org/10.1016/j.ijhcs.2022.102848},
url = {https://www.sciencedirect.com/science/article/pii/S1071581922000751},
author = {Lennart Seitz and Sigrid Bekmeier-Feuerhahn and Krutika Gohil}
}

@article{MALLICK2024103355,
title = {What you say vs what you do: Utilizing positive emotional expressions to relay AI teammate intent within human–AI teams},
journal = {International Journal of Human-Computer Studies},
volume = {192},
pages = {103355},
year = {2024},
issn = {1071-5819},
doi = {https://doi.org/10.1016/j.ijhcs.2024.103355},
url = {https://www.sciencedirect.com/science/article/pii/S1071581924001381},
author = {Rohit Mallick and Christopher Flathmann and Wen Duan and Beau G. Schelble and Nathan J. McNeese}
}

@article{HOORN2024103142,
title = {The media inequality, uncanny mountain, and the singularity is far from near: Iwaa and Sophia robot versus a real human being},
journal = {International Journal of Human-Computer Studies},
volume = {181},
pages = {103142},
year = {2024},
issn = {1071-5819},
doi = {https://doi.org/10.1016/j.ijhcs.2023.103142},
url = {https://www.sciencedirect.com/science/article/pii/S1071581923001519},
author = {Johan F. Hoorn and Ivy S. Huang}
}

@article{RAPP2021102630,
title = {The human side of human-chatbot interaction: A systematic literature review of ten years of research on text-based chatbots},
journal = {International Journal of Human-Computer Studies},
volume = {151},
pages = {102630},
year = {2021},
issn = {1071-5819},
doi = {https://doi.org/10.1016/j.ijhcs.2021.102630},
url = {https://www.sciencedirect.com/science/article/pii/S1071581921000483},
author = {Amon Rapp and Lorenzo Curti and Arianna Boldi}
}

@article{HOBERT2023103108,
title = {Chatbots for active learning: A case of phishing email identification},
journal = {International Journal of Human-Computer Studies},
volume = {179},
pages = {103108},
year = {2023},
issn = {1071-5819},
doi = {https://doi.org/10.1016/j.ijhcs.2023.103108},
url = {https://www.sciencedirect.com/science/article/pii/S1071581923001179},
author = {Sebastian Hobert and Asbjørn Følstad and Effie Lai-Chong Law}
}

@article{JANSSEN2022102921,
title = {How to Make chatbots productive – A user-oriented implementation framework},
journal = {International Journal of Human-Computer Studies},
volume = {168},
pages = {102921},
year = {2022},
issn = {1071-5819},
doi = {https://doi.org/10.1016/j.ijhcs.2022.102921},
url = {https://www.sciencedirect.com/science/article/pii/S1071581922001410},
author = {Antje Janssen and Davinia {Rodríguez Cardona} and Jens Passlick and Michael H. Breitner}
}

@article{10.1145/3610191,
author = {Cai, Jie and Chowdhury, Sagnik and Zhou, Hongyang and Wohn, Donghee Yvette},
title = {Hate Raids on Twitch: Understanding Real-Time Human-Bot Coordinated Attacks in Live Streaming Communities},
year = {2023},
issue_date = {October 2023},
publisher = {Association for Computing Machinery},
address = {New York, NY, USA},
volume = {7},
number = {CSCW2},
url = {https://doi.org/10.1145/3610191},
doi = {10.1145/3610191},
journal = {Proc. ACM Hum.-Comput. Interact.},
month = oct,
articleno = {342},
numpages = {28}
}

@inproceedings{10.1145/3678884.3681875,
author = {Li, Xian and Han, Yuanning and Liu, Di and An, Pengcheng and Niu, Shuo},
title = {FlowGPT: Exploring Domains, Output Modalities, and Goals of Community-Generated AI Chatbots},
year = {2024},
isbn = {9798400711145},
publisher = {Association for Computing Machinery},
address = {New York, NY, USA},
url = {https://doi.org/10.1145/3678884.3681875},
doi = {10.1145/3678884.3681875},
booktitle = {Companion Publication of the 2024 Conference on Computer-Supported Cooperative Work and Social Computing},
pages = {355–361},
numpages = {7},
location = {San Jose, Costa Rica},
series = {CSCW Companion '24}
}

@article{10.1145/3449083,
author = {Wang, Liuping and Wang, Dakuo and Tian, Feng and Peng, Zhenhui and Fan, Xiangmin and Zhang, Zhan and Yu, Mo and Ma, Xiaojuan and Wang, Hongan},
title = {CASS: Towards Building a Social-Support Chatbot for Online Health Community},
year = {2021},
issue_date = {April 2021},
publisher = {Association for Computing Machinery},
address = {New York, NY, USA},
volume = {5},
number = {CSCW1},
url = {https://doi.org/10.1145/3449083},
doi = {10.1145/3449083},
journal = {Proc. ACM Hum.-Comput. Interact.},
month = apr,
articleno = {9},
numpages = {31}
}

@inproceedings{10.1145/3290605.3300680,
author = {Seering, Joseph and Luria, Michal and Kaufman, Geoff and Hammer, Jessica},
title = {Beyond Dyadic Interactions: Considering Chatbots as Community Members},
year = {2019},
isbn = {9781450359702},
publisher = {Association for Computing Machinery},
address = {New York, NY, USA},
url = {https://doi.org/10.1145/3290605.3300680},
doi = {10.1145/3290605.3300680},
booktitle = {Proceedings of the 2019 CHI Conference on Human Factors in Computing Systems},
pages = {1–13},
numpages = {13},
location = {Glasgow, Scotland Uk},
series = {CHI '19}
}

@article{10.1145/3686941,
author = {Seering, Joseph and Khadka, Manas and Haghighi, Nava and Yang, Tanya and Xi, Zachary and Bernstein, Michael},
title = {Chillbot: Content Moderation in the Backchannel},
year = {2024},
issue_date = {November 2024},
publisher = {Association for Computing Machinery},
address = {New York, NY, USA},
volume = {8},
number = {CSCW2},
url = {https://doi.org/10.1145/3686941},
doi = {10.1145/3686941},
journal = {Proc. ACM Hum.-Comput. Interact.},
month = nov,
articleno = {402},
numpages = {26}
}

@article{faraj2011knowledge,
  title={Knowledge collaboration in online communities},
  author={Faraj, Samer and Jarvenpaa, Sirkka L and Majchrzak, Ann},
  journal={Organization science},
  volume={22},
  number={5},
  pages={1224--1239},
  year={2011},
  doi={https://doi.org/10.1287/orsc.1100.0614},
  publisher={INFORMS}
}

@article{TSAI2013475,
title = {Explaining members' proactive participation in virtual communities},
journal = {International Journal of Human-Computer Studies},
volume = {71},
number = {4},
pages = {475-491},
year = {2013},
issn = {1071-5819},
doi = {https://doi.org/10.1016/j.ijhcs.2012.12.002},
url = {https://www.sciencedirect.com/science/article/pii/S1071581912002121},
author = {Hsien-Tung Tsai and Peiyu Pai}
}

@article{BARCELLINI2008558,
title = {User and developer mediation in an Open Source Software community: Boundary spanning through cross participation in online discussions},
journal = {International Journal of Human-Computer Studies},
volume = {66},
number = {7},
pages = {558-570},
year = {2008},
note = {Collaborative and social aspects of software development},
issn = {1071-5819},
doi = {https://doi.org/10.1016/j.ijhcs.2007.10.008},
url = {https://www.sciencedirect.com/science/article/pii/S1071581907001437},
author = {Flore Barcellini and Françoise Détienne and Jean-Marie Burkhardt}
}

@article{HSU2007153,
title = {Knowledge sharing behavior in virtual communities: The relationship between trust, self-efficacy, and outcome expectations},
journal = {International Journal of Human-Computer Studies},
volume = {65},
number = {2},
pages = {153-169},
year = {2007},
issn = {1071-5819},
doi = {https://doi.org/10.1016/j.ijhcs.2006.09.003},
url = {https://www.sciencedirect.com/science/article/pii/S1071581906001431},
author = {Meng-Hsiang Hsu and Teresa L. Ju and Chia-Hui Yen and Chun-Ming Chang}
}

@article{MALINEN2015228,
title = {Understanding user participation in online communities: A systematic literature review of empirical studies},
journal = {Computers in Human Behavior},
volume = {46},
pages = {228-238},
year = {2015},
issn = {0747-5632},
doi = {https://doi.org/10.1016/j.chb.2015.01.004},
url = {https://www.sciencedirect.com/science/article/pii/S0747563215000163},
author = {Sanna Malinen}
}

@inproceedings{10.1145/503376.503446,
author = {Kelly, Sean Uberoi and Sung, Christopher and Farnham, Shelly},
title = {Designing for improved social responsibility, user participation and content in on-line communities},
year = {2002},
isbn = {1581134533},
publisher = {Association for Computing Machinery},
address = {New York, NY, USA},
url = {https://doi.org/10.1145/503376.503446},
doi = {10.1145/503376.503446},
booktitle = {Proceedings of the SIGCHI Conference on Human Factors in Computing Systems},
pages = {391–398},
numpages = {8},
location = {Minneapolis, Minnesota, USA},
series = {CHI '02}
}

@article{10.1145/3068777.3068781,
author = {Shu, Kai and Wang, Suhang and Tang, Jiliang and Zafarani, Reza and Liu, Huan},
title = {User Identity Linkage across Online Social Networks: A Review},
year = {2017},
issue_date = {December 2016},
publisher = {Association for Computing Machinery},
address = {New York, NY, USA},
volume = {18},
number = {2},
issn = {1931-0145},
url = {https://doi.org/10.1145/3068777.3068781},
doi = {10.1145/3068777.3068781},
journal = {SIGKDD Explor. Newsl.},
month = mar,
pages = {5–17},
numpages = {13}
}

@inproceedings{10.1145/2187836.2187907,
author = {Bakshy, Eytan and Rosenn, Itamar and Marlow, Cameron and Adamic, Lada},
title = {The role of social networks in information diffusion},
year = {2012},
isbn = {9781450312295},
publisher = {Association for Computing Machinery},
address = {New York, NY, USA},
url = {https://doi.org/10.1145/2187836.2187907},
doi = {10.1145/2187836.2187907},
booktitle = {Proceedings of the 21st International Conference on World Wide Web},
pages = {519–528},
numpages = {10},
location = {Lyon, France},
series = {WWW '12}
}

@article{GHANI2019417,
title = {Social media big data analytics: A survey},
journal = {Computers in Human Behavior},
volume = {101},
pages = {417-428},
year = {2019},
issn = {0747-5632},
doi = {https://doi.org/10.1016/j.chb.2018.08.039},
url = {https://www.sciencedirect.com/science/article/pii/S074756321830414X},
author = {Norjihan Abdul Ghani and Suraya Hamid and Ibrahim Abaker {Targio Hashem} and Ejaz Ahmed}
}

@article{10.1145/3274319,
author = {Faas, Travis and Dombrowski, Lynn and Young, Alyson and Miller, Andrew D.},
title = {Watch Me Code: Programming Mentorship Communities on Twitch.tv},
year = {2018},
issue_date = {November 2018},
publisher = {Association for Computing Machinery},
address = {New York, NY, USA},
volume = {2},
number = {CSCW},
url = {https://doi.org/10.1145/3274319},
doi = {10.1145/3274319},
journal = {Proc. ACM Hum.-Comput. Interact.},
month = nov,
articleno = {50},
numpages = {18}
}

@inproceedings{10.1145/3478431.3499385,
author = {Bridson, Kathryn and Atkinson, Jeffrey and Fleming, Scott D.},
title = {Delivering Round-the-Clock Help to Software Engineering Students Using Discord: An Experience Report},
year = {2022},
isbn = {9781450390705},
publisher = {Association for Computing Machinery},
address = {New York, NY, USA},
url = {https://doi.org/10.1145/3478431.3499385},
doi = {10.1145/3478431.3499385},
booktitle = {Proceedings of the 53rd ACM Technical Symposium on Computer Science Education - Volume 1},
pages = {759–765},
numpages = {7},
location = {Providence, RI, USA},
series = {SIGCSE 2022}
}

@article{10.1145/3359157,
author = {Jiang, Jialun Aaron and Kiene, Charles and Middler, Skyler and Brubaker, Jed R. and Fiesler, Casey},
title = {Moderation Challenges in Voice-based Online Communities on Discord},
year = {2019},
issue_date = {November 2019},
publisher = {Association for Computing Machinery},
address = {New York, NY, USA},
volume = {3},
number = {CSCW},
url = {https://doi.org/10.1145/3359157},
doi = {10.1145/3359157},
journal = {Proc. ACM Hum.-Comput. Interact.},
month = nov,
articleno = {55},
numpages = {23}
}

@article{10.1145/3610053,
author = {Choi, Frederick and Bajpai, Tanvi and Pratipati, Sowmya and Chandrasekharan, Eshwar},
title = {ConvEx: A Visual Conversation Exploration System for Discord Moderators},
year = {2023},
issue_date = {October 2023},
publisher = {Association for Computing Machinery},
address = {New York, NY, USA},
volume = {7},
number = {CSCW2},
url = {https://doi.org/10.1145/3610053},
doi = {10.1145/3610053},
journal = {Proc. ACM Hum.-Comput. Interact.},
month = oct,
articleno = {262},
numpages = {30}
}

@article{ABRAMCZUK2023103104,
title = {Meet me in VR! Can VR space help remote teams connect: A seven-week study with Horizon Workrooms},
journal = {International Journal of Human-Computer Studies},
volume = {179},
pages = {103104},
year = {2023},
issn = {1071-5819},
doi = {https://doi.org/10.1016/j.ijhcs.2023.103104},
url = {https://www.sciencedirect.com/science/article/pii/S1071581923001131},
author = {Katarzyna Abramczuk and Zbigniew Bohdanowicz and Bartosz Muczyński and Kinga H. Skorupska and Daniel Cnotkowski}
}

@article{10.1145/3579529,
author = {Alcala, Katrina and D'Achille, Anthony and Bruckman, Amy},
title = {The Stage and the Theatre: AltspaceVR and its Relationship to Discord},
year = {2023},
issue_date = {April 2023},
publisher = {Association for Computing Machinery},
address = {New York, NY, USA},
volume = {7},
number = {CSCW1},
url = {https://doi.org/10.1145/3579529},
doi = {10.1145/3579529},
journal = {Proc. ACM Hum.-Comput. Interact.},
month = {April},
articleno = {96},
numpages = {21}
}

@article{ABURUMMAN2022102819,
title = {Nonverbal communication in virtual reality: Nodding as a social signal in virtual interactions},
journal = {International Journal of Human-Computer Studies},
volume = {164},
pages = {102819},
year = {2022},
issn = {1071-5819},
doi = {https://doi.org/10.1016/j.ijhcs.2022.102819},
url = {https://www.sciencedirect.com/science/article/pii/S1071581922000489},
author = {Nadine Aburumman and Marco Gillies and Jamie A. Ward and Antonia F.de C. Hamilton}
}

@misc{zhang2024redefiningqualitativeanalysisai,
      title={Redefining Qualitative Analysis in the AI Era: Utilizing ChatGPT for Efficient Thematic Analysis}, 
      author={He Zhang and Chuhao Wu and Jingyi Xie and Yao Lyu and Jie Cai and John M. Carroll},
      year={2024},
      eprint={2309.10771},
      archivePrefix={arXiv},
      primaryClass={cs.HC},
      url={https://arxiv.org/abs/2309.10771}, 
}

@article{10.1145/3637410,
author = {Wester, Joel and Pohl, Henning and Hosio, Simo and van Berkel, Niels},
title = {"This Chatbot Would Never...": Perceived Moral Agency of Mental Health Chatbots},
year = {2024},
issue_date = {April 2024},
publisher = {Association for Computing Machinery},
address = {New York, NY, USA},
volume = {8},
number = {CSCW1},
url = {https://doi.org/10.1145/3637410},
doi = {10.1145/3637410},
journal = {Proc. ACM Hum.-Comput. Interact.},
month = apr,
articleno = {133},
numpages = {28}
}

@inproceedings{10.1145/3640457.3688069,
author = {Ariza-Casabona, Alejandro and Boratto, Ludovico and Salam\'{o}, Maria},
title = {A Comparative Analysis of Text-Based Explainable Recommender Systems},
year = {2024},
isbn = {9798400705052},
publisher = {Association for Computing Machinery},
address = {New York, NY, USA},
url = {https://doi.org/10.1145/3640457.3688069},
doi = {10.1145/3640457.3688069},
booktitle = {Proceedings of the 18th ACM Conference on Recommender Systems},
pages = {105–115},
numpages = {11},
location = {Bari, Italy},
series = {RecSys '24}
}

@article{
doi:10.1126/science.1165821,
author = {Stephen P. Borgatti  and Ajay Mehra  and Daniel J. Brass  and Giuseppe Labianca },
title = {Network Analysis in the Social Sciences},
journal = {Science},
volume = {323},
number = {5916},
pages = {892-895},
year = {2009},
doi = {10.1126/science.1165821},
URL = {https://www.science.org/doi/abs/10.1126/science.1165821},
eprint = {https://www.science.org/doi/pdf/10.1126/science.1165821}}

@article{otte2002social,
  title={Social network analysis: a powerful strategy, also for the information sciences},
  author={Otte, Evelien and Rousseau, Ronald},
  journal={Journal of information Science},
  volume={28},
  number={6},
  pages={441--453},
  year={2002},
  doi={https://doi.org/10.1177/016555150202800601},
  publisher={Sage Publications Sage CA: Thousand Oaks, CA}
}

@article{10.1145/3392863,
author = {Wijenayake, Senuri and van Berkel, Niels and Kostakos, Vassilis and Goncalves, Jorge},
title = {Quantifying the Effect of Social Presence on Online Social Conformity},
year = {2020},
issue_date = {May 2020},
publisher = {Association for Computing Machinery},
address = {New York, NY, USA},
volume = {4},
number = {CSCW1},
url = {https://doi.org/10.1145/3392863},
doi = {10.1145/3392863},
journal = {Proc. ACM Hum.-Comput. Interact.},
month = may,
articleno = {55},
numpages = {22}
}

@inproceedings{10.1145/3706599.3720120,
author = {Zhang, He and Zha, Siyu and Cai, Jie and Wohn, Donghee Yvette and Carroll, John M.},
title = {Generative AI in Virtual Reality Communities: A Preliminary Analysis of the VRChat Discord Community},
year = {2025},
isbn = {9798400713958},
publisher = {Association for Computing Machinery},
address = {New York, NY, USA},
url = {https://doi.org/10.1145/3706599.3720120},
doi = {10.1145/3706599.3720120},
booktitle = {Proceedings of the Extended Abstracts of the CHI Conference on Human Factors in Computing Systems},
articleno = {305},
numpages = {11},
location = {
},
series = {CHI EA '25}
}

@article{rzeszewski2024social,
  title={Social relations and spatiality in VR-Making spaces meaningful in VRChat},
  author={Rzeszewski, Michal and Evans, Leighton},
  journal={Emotion, Space and Society},
  volume={53},
  pages={101038},
  year={2024},
  doi={https://doi.org/10.1016/j.emospa.2024.101038},
  publisher={Elsevier}
}

@misc{shen2024bidirectionalhumanaialignmentsystematic,
      title={Towards Bidirectional Human-AI Alignment: A Systematic Review for Clarifications, Framework, and Future Directions}, 
      author={Hua Shen and Tiffany Knearem and Reshmi Ghosh and Kenan Alkiek and Kundan Krishna and Yachuan Liu and Ziqiao Ma and Savvas Petridis and Yi-Hao Peng and Li Qiwei and Sushrita Rakshit and Chenglei Si and Yutong Xie and Jeffrey P. Bigham and Frank Bentley and Joyce Chai and Zachary Lipton and Qiaozhu Mei and Rada Mihalcea and Michael Terry and Diyi Yang and Meredith Ringel Morris and Paul Resnick and David Jurgens},
      year={2024},
      eprint={2406.09264},
      archivePrefix={arXiv},
      primaryClass={cs.HC},
      url={https://arxiv.org/abs/2406.09264}, 
}

@article{heise2019internet,
  title={Internet research: Ethical guidelines 3.0},
  author={Heise, Anne Hove Henriksen and Hongladarom, Soraj and Jobin, Anna and Kinder-Kurlanda, Katharina and Sun, Sun and Lim, Elisabetta Locatelli and Markham, Annette and Reilly, Paul J and Tiidenberg, Katrin and Wilhelm, Carsten},
  journal={WildApricot},
  year={2019},
  url={https://aoir.org/reports/ethics3.pdf}
}

@misc{zhang2025harnessingpoweraiqualitative,
      title={Harnessing the Power of AI in Qualitative Research: Role Assignment, Engagement, and User Perceptions of AI-Generated Follow-Up Questions in Semi-Structured Interviews}, 
      author={He Zhang and Yueyan Liu and Xin Guan and Jie Cai and John M. Carroll},
      year={2025},
      eprint={2509.12709},
      archivePrefix={arXiv},
      primaryClass={cs.HC},
      url={https://arxiv.org/abs/2509.12709}, 
}

@inproceedings{
agashe2025agent,
title={Agent S: An Open Agentic Framework that Uses Computers Like a Human},
author={Saaket Agashe and Jiuzhou Han and Shuyu Gan and Jiachen Yang and Ang Li and Xin Eric Wang},
booktitle={Towards Agentic AI for Science: Hypothesis Generation, Comprehension, Quantification, and Validation},
year={2025},
url={https://openreview.net/forum?id=43XMKuTTK0}
}

@inproceedings{NEURIPS2023_91f18a12,
 author = {Zheng, Lianmin and Chiang, Wei-Lin and Sheng, Ying and Zhuang, Siyuan and Wu, Zhanghao and Zhuang, Yonghao and Lin, Zi and Li, Zhuohan and Li, Dacheng and Xing, Eric and Zhang, Hao and Gonzalez, Joseph E and Stoica, Ion},
 booktitle = {Advances in Neural Information Processing Systems},
 editor = {A. Oh and T. Naumann and A. Globerson and K. Saenko and M. Hardt and S. Levine},
 pages = {46595--46623},
 publisher = {Curran Associates, Inc.},
 title = {Judging LLM-as-a-Judge with MT-Bench and Chatbot Arena},
 url = {https://proceedings.neurips.cc/paper_files/paper/2023/file/91f18a1287b398d378ef22505bf41832-Paper-Datasets_and_Benchmarks.pdf},
 volume = {36},
 year = {2023}
}

@inproceedings{mckay2024realizing,
  title={Realizing the Promise of AI Governance Involving Humans-in-the-Loop},
  author={McKay, Margaret H},
  booktitle={International Conference on Human-Computer Interaction},
  pages={107--123},
  year={2024},
  doi={https://doi.org/10.1007/978-3-031-76827-9_7},
  organization={Springer}
}

@inproceedings{10.1145/2818048.2820010,
author = {Wang, Yi-Chia and Burke, Moira and Kraut, Robert},
title = {Modeling Self-Disclosure in Social Networking Sites},
year = {2016},
isbn = {9781450335928},
publisher = {Association for Computing Machinery},
address = {New York, NY, USA},
url = {https://doi.org/10.1145/2818048.2820010},
doi = {10.1145/2818048.2820010},
booktitle = {Proceedings of the 19th ACM Conference on Computer-Supported Cooperative Work \& Social Computing},
pages = {74–85},
numpages = {12},
location = {San Francisco, California, USA},
series = {CSCW '16}
}

@inproceedings{10.1145/2858036.2858414,
author = {Ma, Xiao and Hancock, Jeff and Naaman, Mor},
title = {Anonymity, Intimacy and Self-Disclosure in Social Media},
year = {2016},
isbn = {9781450333627},
publisher = {Association for Computing Machinery},
address = {New York, NY, USA},
url = {https://doi.org/10.1145/2858036.2858414},
doi = {10.1145/2858036.2858414},
booktitle = {Proceedings of the 2016 CHI Conference on Human Factors in Computing Systems},
pages = {3857–3869},
numpages = {13},
location = {San Jose, California, USA},
series = {CHI '16}
}

@misc{Yang_nbcnews2025, title={Researchers secretly infiltrated a popular Reddit Forum with AI bots, causing outrage}, url={https://www.nbcnews.com/tech/tech-news/reddiit-researchers-ai-bots-rcna203597}, journal={NBCNews.com}, publisher={NBCUniversal News Group}, author={Yang, Angela}, year={2025}, month={Apr}}

@article{10.1145/3449161,
author = {Kim, Soomin and Eun, Jinsu and Seering, Joseph and Lee, Joonhwan},
title = {Moderator Chatbot for Deliberative Discussion: Effects of Discussion Structure and Discussant Facilitation},
year = {2021},
issue_date = {April 2021},
publisher = {Association for Computing Machinery},
address = {New York, NY, USA},
volume = {5},
number = {CSCW1},
url = {https://doi.org/10.1145/3449161},
doi = {10.1145/3449161},
journal = {Proc. ACM Hum.-Comput. Interact.},
month = apr,
articleno = {87},
numpages = {26}
}

@misc{cai2026twitchthirdpartydeveloperssupport,
      title={Twitch Third-Party Developers' Support Seeking and Provision Practices on Discord}, 
      author={Jie Cai and He Zhang and Yueyan Liu and John M. Carroll and Chun Yu},
      year={2026},
      eprint={2604.07732},
      archivePrefix={arXiv},
      primaryClass={cs.HC},
      url={https://arxiv.org/abs/2604.07732}, 
}

@misc{chen2026understandingnewcomerpersistencesocial,
      title={Understanding Newcomer Persistence in Social VR: A Case Study of VRChat}, 
      author={Qijia Chen and Andrea Bellucci and Giulio Jacucci},
      year={2026},
      eprint={2603.25223},
      archivePrefix={arXiv},
      primaryClass={cs.HC},
      url={https://arxiv.org/abs/2603.25223}, 
}

@inproceedings{10.1145/3706598.3713561,
author = {Hu, Yang and Freeman, Guo and Panchanadikar, Ruchi},
title = {``Grab the Chat and Stick It to My Wall": Understanding How Social VR Streamers Bridge Immersive VR Experiences with Streaming Audiences Outside VR},
year = {2025},
isbn = {9798400713941},
publisher = {Association for Computing Machinery},
address = {New York, NY, USA},
url = {https://doi.org/10.1145/3706598.3713561},
doi = {10.1145/3706598.3713561},
booktitle = {Proceedings of the 2025 CHI Conference on Human Factors in Computing Systems},
articleno = {459},
numpages = {19},
location = {
},
series = {CHI '25}
}

@inproceedings{10.1145/3715336.3735825,
author = {Hu, Yang and Freeman, Guo},
title = {Understanding Social VR Streamers’ Unique Challenges in Managing Cross-Reality Social Interactions Through Multi-dimensional VR Interfaces},
year = {2025},
isbn = {9798400714856},
publisher = {Association for Computing Machinery},
address = {New York, NY, USA},
url = {https://doi.org/10.1145/3715336.3735825},
doi = {10.1145/3715336.3735825},
booktitle = {Proceedings of the 2025 ACM Designing Interactive Systems Conference},
pages = {1725–1739},
numpages = {15},
location = {
},
series = {DIS '25}
}

@inproceedings{10.1145/3744736.3749361,
author = {Hu, Yang and Panchanadikar, Ruchi and Freeman, Guo},
title = {Beyond Traditional Gaming: Understanding How Social VR Streaming Creates New Forms of Play},
year = {2025},
isbn = {9798400720239},
publisher = {Association for Computing Machinery},
address = {New York, NY, USA},
url = {https://doi.org/10.1145/3744736.3749361},
doi = {10.1145/3744736.3749361},
booktitle = {Companion Proceedings of the Annual Symposium on Computer-Human Interaction in Play},
pages = {15–22},
numpages = {8},
location = {
},
series = {CHI PLAY Companion '25}
}

@inproceedings{10.1145/3706598.3713503,
author = {Pan, Shuyi and de Graaf, Maartje M.A.},
title = {Developing a Social Support Framework: Understanding the Reciprocity in Human-Chatbot Relationship},
year = {2025},
isbn = {9798400713941},
publisher = {Association for Computing Machinery},
address = {New York, NY, USA},
url = {https://doi.org/10.1145/3706598.3713503},
doi = {10.1145/3706598.3713503},
booktitle = {Proceedings of the 2025 CHI Conference on Human Factors in Computing Systems},
articleno = {182},
numpages = {13},
location = {
},
series = {CHI '25}
}

@article{kim2025enhancing,
  title={Enhancing user experience with a generative AI chatbot},
  author={Kim, Jeong Soo and Kim, Minseong and Baek, Tae Hyun},
  journal={International Journal of Human--Computer Interaction},
  volume={41},
  number={1},
  pages={651--663},
  year={2025},
doi={https://doi.org/10.1080/10447318.2024.2311971},
  publisher={Taylor \& Francis}
}

\appendix
\section{Appendix}
\subsection{More Figures}

\aptLtoX[graphic=no,type=html]{\begin{figure}[!h]
    \centering
        \includegraphics[width=\linewidth]{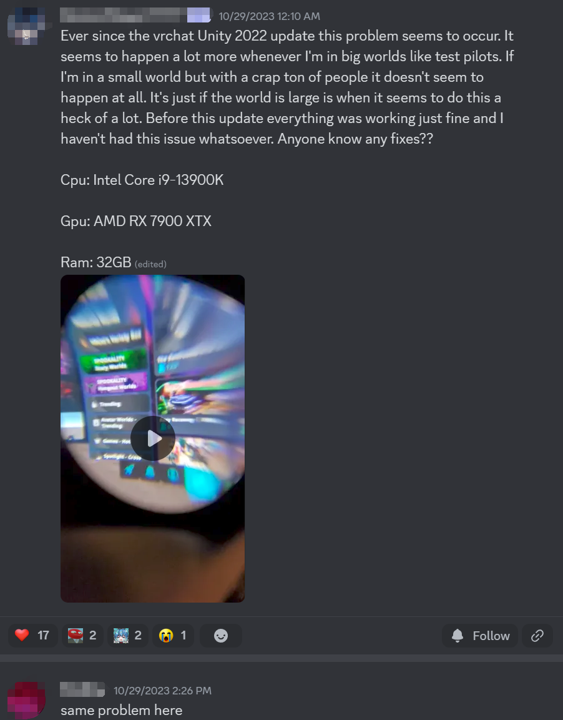}
        \caption{Example of Initial User Support. It shows a discussion thread on Discord where a user is reporting a technical issue after a VRChat Unity 2022 update. The user describes encountering problems primarily in large worlds, such as ``test pilots,'' but not in smaller worlds with fewer people. The issue did not exist prior to the update, and the user is asking for help with fixes.}
        \label{fig.Long1}
    \end{figure}
    \begin{figure}
        \centering
        \includegraphics[width=\linewidth]{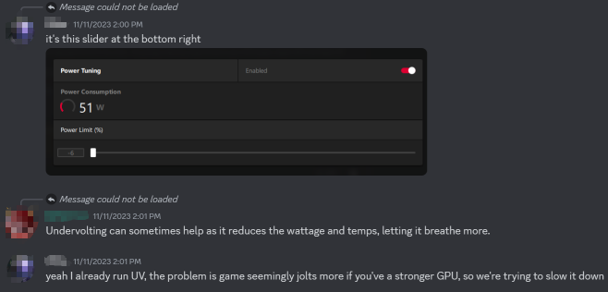}
        \caption{Example of Users Attempt at Finding a Solution. It shows a conversation in Discord about optimizing GPU performance to address a gaming issue. The discussion revolves around power tuning and undervolting a graphics card.}
        \label{fig.Long2}
    \end{figure}}{\begin{figure*}[!h]
    \centering
    \begin{minipage}{0.35\textwidth}
        \centering
        \includegraphics[width=\linewidth]{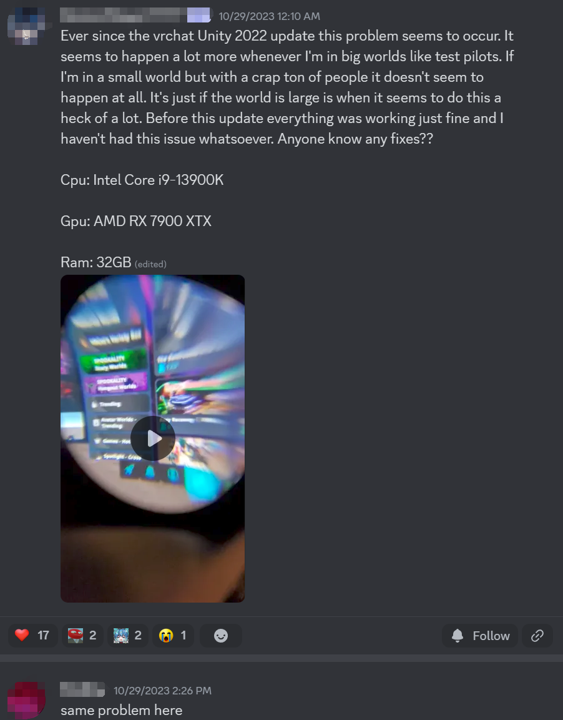}
        \caption{Example of Initial User Support. It shows a discussion thread on Discord where a user is reporting a technical issue after a VRChat Unity 2022 update. The user describes encountering problems primarily in large worlds, such as ``test pilots,'' but not in smaller worlds with fewer people. The issue did not exist prior to the update, and the user is asking for help with fixes.}
        \label{fig.Long1}
    \end{minipage}
    \hfill
    \begin{minipage}{0.58\textwidth}
        \centering
        \includegraphics[width=\linewidth]{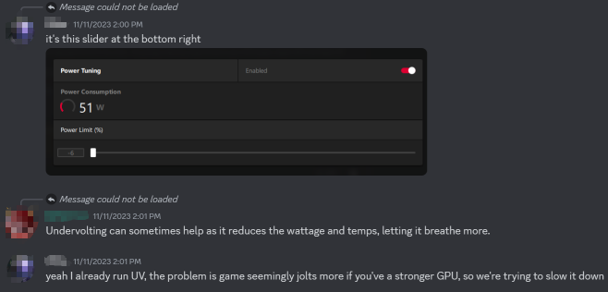}
        \caption{Example of Users Attempt at Finding a Solution. It shows a conversation in Discord about optimizing GPU performance to address a gaming issue. The discussion revolves around power tuning and undervolting a graphics card.}
        \label{fig.Long2}
    \end{minipage}
\end{figure*}}

\aptLtoX[graphic=no,type=html]{\begin{figure}[!h]
    \centering
        \includegraphics[width=\linewidth]{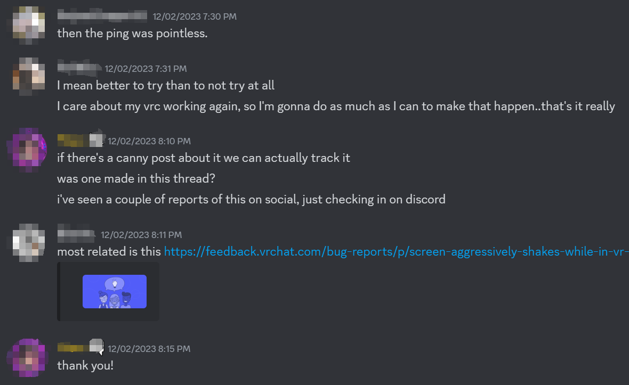}
        \caption{Example of Sharing Information. It shows an example of information sharing in a conversation on Discord. The participants are discussing an issue with VRChat, and they collaboratively work to identify and share resources related to the problem.}
        \label{fig.Long3}
    \end{figure}
    \begin{figure}
        \centering
        \includegraphics[width=\linewidth]{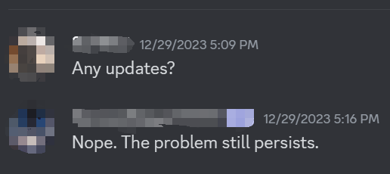}
        \caption{Example of Follow-up on Progress. It shows an example of a follow-up on progress in a conversation on Discord. It illustrates users checking on the status of an unresolved issue.}
        \label{fig.Long4}
\end{figure}}{\begin{figure*}[!h]
    \centering
    \begin{minipage}{0.58\textwidth}
        \centering
        \includegraphics[width=\linewidth]{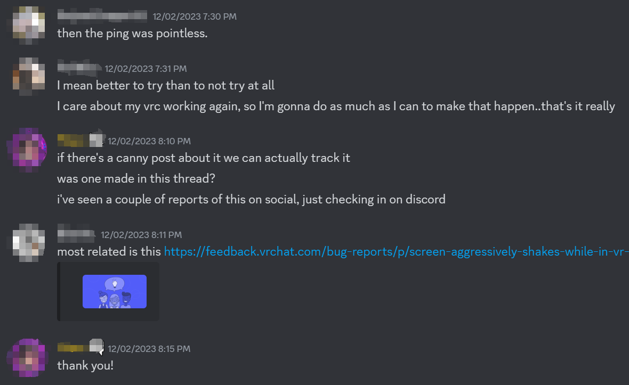}
        \caption{Example of Sharing Information. It shows an example of information sharing in a conversation on Discord. The participants are discussing an issue with VRChat, and they collaboratively work to identify and share resources related to the problem.}
        \label{fig.Long3}
    \end{minipage}
    \hfill
    \begin{minipage}{0.38\textwidth}
        \centering
        \includegraphics[width=\linewidth]{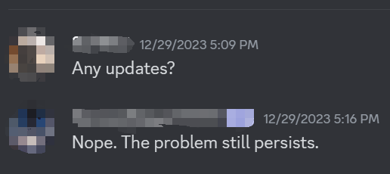}
        \caption{Example of Follow-up on Progress. It shows an example of a follow-up on progress in a conversation on Discord. It illustrates users checking on the status of an unresolved issue.}
        \label{fig.Long4}
    \end{minipage}
\end{figure*}}

\aptLtoX[graphic=no,type=html]{\begin{figure}[!h]
    \centering
        \includegraphics[width=\linewidth]{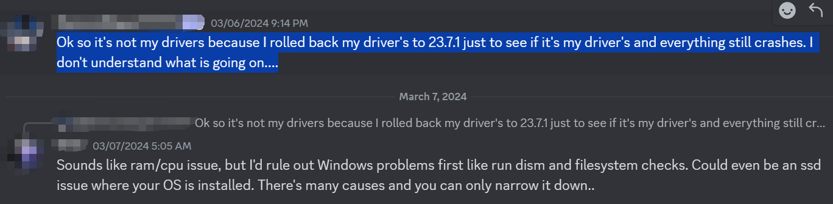}
        \caption{Example of Emergence of New Issues. It shows an example of the emergence of new issues in a discussion on Discord. The conversation involves troubleshooting a persistent problem and considering potential new causes.}
        \label{fig.Long5}
    \end{figure}
    \begin{figure}
        \centering
        \includegraphics[width=\linewidth]{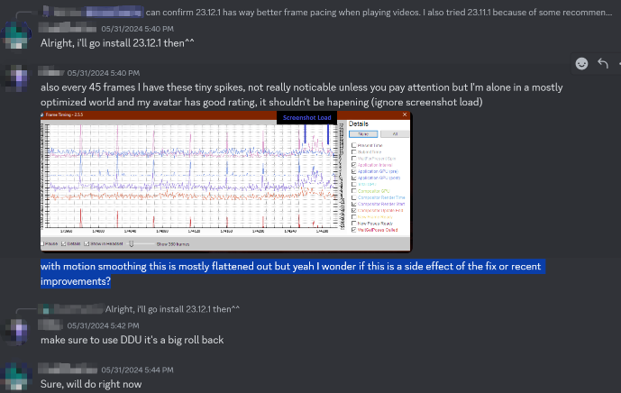}
        \caption{Example of Final Stage. It showcases a conversation representing the final stage of troubleshooting in a technical support discussion on Discord. It illustrates users sharing solutions, observations, and actions to resolve a problem.}
        \label{fig.Long6}
\end{figure}}{\begin{figure*}[!h]
    \centering
    \begin{minipage}{0.48\textwidth}
        \centering
        \includegraphics[width=\linewidth]{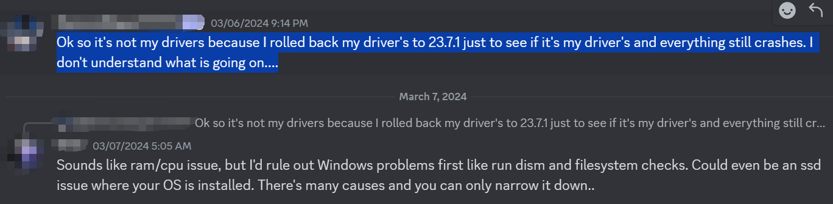}
        \caption{Example of Emergence of New Issues. It shows an example of the emergence of new issues in a discussion on Discord. The conversation involves troubleshooting a persistent problem and considering potential new causes.}
        \label{fig.Long5}
    \end{minipage}
    \hfill
    \begin{minipage}{0.48\textwidth}
        \centering
        \includegraphics[width=\linewidth]{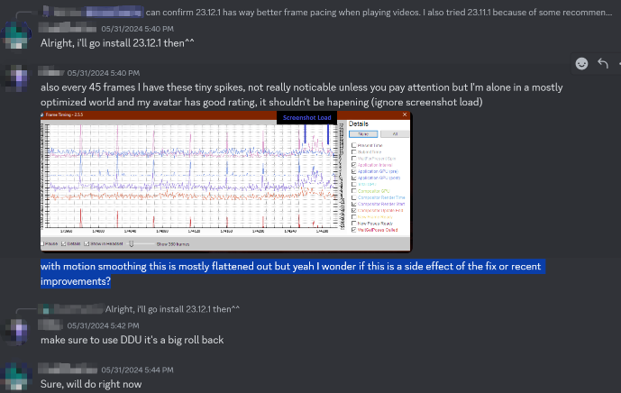}
        \caption{Example of Final Stage. It showcases a conversation representing the final stage of troubleshooting in a technical support discussion on Discord. It illustrates users sharing solutions, observations, and actions to resolve a problem.}
        \label{fig.Long6}
    \end{minipage}
\end{figure*}}

\end{document}